\documentclass[fleqn,usenatbib]{mnras}

\usepackage{newtxtext,newtxmath}
\usepackage{orcidlink}
\usepackage[T1]{fontenc}

\usepackage{anyfontsize} 

\DeclareRobustCommand{\VAN}[3]{#2}
\let\VANthebibliography\thebibliography
\def\thebibliography{\DeclareRobustCommand{\VAN}[3]{##3}\VANthebibliography}

\usepackage{graphicx}	
\usepackage{amsmath}	
\usepackage{xcolor}

\title[PAH correlations and IMF- PAH Variations in the SFCs of NGC 628]{Spatial Correlations of PAH, UV, H$\alpha$ Emission and IMF- PAH Variations in the Star-Forming Complexes of NGC 628}

\author[S Amrutha]{
S Amrutha\orcidlink{0009-0005-6072-9252},$^{1,2}$\thanks{E-mail:amrutha.s@iiap.res.in, amrutharao99@gmail.com}
Dimitra Rigopoulou\orcidlink{0000-0001-6854-7545},$^{3,4}$
Mousumi Das\orcidlink{0000-0001-8996-6474},$^{1}$
Jyoti Yadav\orcidlink{0000-0002-5641-8102},$^{5,6}$
\\
$^{1}$Indian Institute of Astrophysics, Koramangala II Block, Bangalore 560034, India\\
$^{2}$Pondicherry University, R.V. Nagar, Kalapet, 605014, Puducherry, India\\
$^{3}$Oxford Astrophysics, Denys Wilkinson Building, University of Oxford, Keble Road, Oxford OX1 3RH, UK\\
$^{4}$School of Sciences, European University Cyprus, Diogenes street, Engomi, 1516 Nicosia, Cyprus\\
$^{5}$Instituto de Astrofísica de Canarias, Vía Láctea s/n, E-38205 La
Laguna, Spain\\
$^{6}$Departamento de Astrofísica, Universidad de La Laguna, E-38206
La Laguna, Spain\\
}

\date{Accepted XXX. Received YYY; in original form ZZZ}

\pubyear{\the\year{}}

\begin{document}
\label{firstpage}
\pagerange{\pageref{firstpage}--\pageref{lastpage}}
\maketitle

\begin{abstract}
We examine the spatial correlation of emission from polycyclic aromatic hydrocarbons (PAH) (3.3~$\mathrm{\mu m}$, 7.7~$\mathrm{\mu m}$, and 11.3~$\mathrm{\mu m}$) with the far-ultraviolet (FUV), near-ultraviolet (NUV), and H$\alpha$ emission from star-forming complexes (SFCs) in different regions of NGC 628. We use the James Webb Space Telescope to detect PAH emission, along with the Ultraviolet Imaging Telescope and Multi Unit Spectroscopic Explorer observations to sample the UV and H$\alpha$ emission, respectively. We investigate the correlation of PAH luminosities with FUV, NUV, and H$\alpha$ luminosities for the extracted SFCs. We find that the arm and spur SFCs show bright PAH emission, except for those with only NUV emission, located primarily in bubbles and superbubbles. We also examine these correlations in the phantom void, the largest superbubble in NGC 628. We investigate the trend for FUV-NUV and the Initial mass function (IMF) index $\alpha_2$ derived from the ratio of B stars with PAH band ratios (F335M/F1130W, F770W/F1130W, and F335M/F770W). We find that PAH band ratios increase as the IMF becomes more top-heavy and the SFCs become bluer, consistent with enhanced PAH excitation and ionisation in stronger UV radiation fields. However, $\alpha_2$ of spur SFCs shows a flat trend with PAH band ratios compared to arms. Upon closer examination, two SFCs in the shell region of the phantom void showed a flatter IMF compared to the SFC located near the elongated bubble. This is likely due to gas accumulation, possibly through feedback mechanisms in the shell region, while the SFC near the elongated bubble may have formed in a region affected by shear.
\end{abstract}

\begin{keywords}
galaxies: individual(NGC 628)-- galaxies: star-forming regions-- infrared: galaxies--infrared: ISM
\end{keywords}



\section{Introduction} \label{sec:intro}

Star formation is crucial for understanding the evolution of galaxies \citep{Kennicutt_1998}. The most commonly used star formation tracers are UV and H$\alpha$, however there are others found throughout the electromagnetic spectrum \citep{calzetti.book.2013}. Although both UV and H$\alpha$ trace recent star formation, UV emission arises directly from massive stars and traces star formation for around 200 Myr \citep{2007ApJS..173..538T}, whereas H$\alpha$ is the recombination line emission from the ionised regions around massive stars, hence tracing star formation for around 10 Myr. However, both can experience extinction, and these tracers alone will not provide complete information on the star formation rate (SFR). On the other hand, mid-infrared (MIR) and near-infrared (NIR) emissions are less affected by extinction \citep{Belfiore}. Polycyclic aromatic hydrocarbons (PAHs) have also been proposed as effective SFR tracers \citep{Rigopoulou_1999, 2004ApJ...613..986P}.

PAHs are organic molecules with multiple aromatic rings containing 20-1000 carbon atoms \citep{1989ApJS...71..733A, 2021MNRAS.504.5287R}. These molecules are an essential part of the Interstellar medium (ISM), and up to 20$\%$ of the total IR emission in star-forming galaxies can come from these molecules \citep{Smith_2007, 2020NatAs...4..339L}. They emit in distinctive IR bands at 3.3, 6.2, 7.7, 11.3, 12.7, and 17 $\mu$m \citep{1984A&A...137L...5L, 1989ApJS...71..733A}. These molecules can be vibrationally excited with a single UV photon, and they de-excite emitting the distinctive PAH features \citep{2008ARA&A..46..289T}.
Both stretching and bending vibrational modes contribute to these features. The 3.3 $\mu$m feature is seen when C-H bonds stretch, whereas C-C bending and C-H out-of-plane bending contribute to the features at 6-9$\mu$m and features beyond 11$\mu$m, respectively \citep{1984A&A...137L...5L, 1989ApJS...71..733A}.

PAH molecules can be excited by the wide range of radiation fields \citep{2001ApJ...554..778L}. PAH emission is observed to remain relatively constant in regions of a galaxy with uniform metallicity and weak radiation fields, tracing the diffuse ISM \citep{2023ApJ...944L...8S}. However, it is particularly bright in regions illuminated by UV and visible photons from massive stars in star-forming regions tracing the front of the photodissociation regions (PDRs) around stars, which separates the ionised gas from the surrounding molecular gas \citep{Tielens1993}. Hence, their bright emission is closely associated with star formation. Their band ratios can be used to investigate the properties of the molecules, such as size and charge, which in turn helps us understand the source of UV photons \citep{2021ApJ...917....3D, 2022ApJ...931...38M, 2023ApJ...944L..12C, 2024MNRAS.532.1598R}. Small PAH molecules emit strongly at 3.3 $\mathrm{\mu m}$. Larger molecules emit at longer wavelengths, specifically around 11.3 and 17 $\mathrm{\mu m}$. Ionised PAHs are associated with C-C vibrational modes, leading to 6.2 and 7.7 $\mathrm{\mu m}$ emissions. Meanwhile, neutral PAHs emit at 3.3, 11.3, and 17 $\mathrm{\mu m}$, which are primarily linked to C-H vibrational modes. Hence, emission from the various PAH bands traces the size and charge (or degree of ionisation) of PAH molecules \citep{Allamandola_1999}.

PAHs have been extensively studied in various astrophysical regions, from planetary, protoplanetary, and reflection nebulae to the H II regions and ISM of galaxies \citep{2008ARA&A..46..289T}. The 3.3, 6.2, 7.7, and 11.3 $\mu$m features have been mostly used to study the PAH properties in galaxies \citep{2021MNRAS.504.5287R, Sandstrom_2023, 2024MNRAS.532.1598R, Lyu_2025}. Early-type and active galaxies show an abundance of neutral larger molecules, whereas star-forming galaxies have more ionised molecules \citep{Kaneda_2008,2010ApJ...721.1090V, García2022, garcia2024}. Smaller molecules may undergo preferential destruction from a harsh environment \citep{Smith_2007} or their abundance could also increase due to the shattering of large grains \citep{2012A&A...541A..10Y}. There can also be the destruction of PAHs in regions with strong emission from Active Galactic Nuclei (AGNs). However, some studies have shown that there is no clear evidence for the preferential destruction of smaller PAH molecules \citep{2024MNRAS.532.1598R}. Rather, ionisation appears to be the dominant factor governing PAH emission. Furthermore, PAHs have been detected in the immediate vicinity of AGNs \citep{García2022, Lai2022, Armus_2023}. In addition to this, properties such as metallicity affect the production and emission of PAH molecules \cite[]{madden2006, 2012ApJ...744...20S, 2019ApJ...876...62C}. With the James Webb Space Telescope (JWST), several studies are dedicated to understanding PAH molecules and their emissions in detail. These include spectroscopic \citep{ Lai_2023, 10.1093/mnras/stac3729, García2022, garcia2024} and photometric studies of nearby galaxies \citep{2023ApJ...944L...8S, 2023ApJ...944L..16E, 2024ApJ...971...32P, 2025ApJ...983..137R, Yadav2026A&A...709A.172Y} and also those focusing on high redshift (z) galaxies \citep{2023ApJ...950....7S, Young_2023, 2024A&A...690A..89S, 2025arXiv251007365M}.

Our study uses the JWST filters covering the PAH bands (F335M, F770W, and F1130W) of NGC 628 to compare the spatial distribution of PAH with UV, using observations from the Ultraviolet Imaging Telescope (UVIT) as well as the H$\alpha$ distribution obtained from the Multi Unit Spectroscopic Explorer (MUSE) instrument. 
Our aim is to understand how the physical properties of the PAHs in star-forming complexes (SFCs) vary with different stellar initial mass functions (IMFs). The latter have been obtained from the catalogue of star-formation properties of NGC 628 by \cite{Amrutha_2025} (hereafter A\&D25). We investigated how the PAH properties vary over the different regions (arms and spurs). We also closely examined the PAH distribution around the phantom void, which is prominent in the PAH bands of NGC 628 (also known as the Phantom galaxy). 

NGC 628 is a well-known grand design spiral galaxy, at a distance of 9.84 Mpc \citep{10.1093/mnras/staa3668}. This massive galaxy \cite[log($M_{*}/M_{\odot}$)=10.34;][] {2019ApJS..244...24L} is nearly face-on \cite[i=8$^\circ$.9;][] {Lang_2020} and shows young and massive star formation \citep{Yadav_2021}. Owing to its favourable orientation and active star formation, NGC 628 has been a common target for spatially resolved studies of star formation, including investigations of PAH emission using the broadband and medium-band filters \cite[e.g.,][] {Sandstrom_2023} of JWST. Most studies have focused on understanding variation of the PAH fraction as a function of the properties of the ISM, along with their variation with properties such as star formation rate, stellar age, metallicity, and gas ionisation \citep{Chastenet_2023, Dale_2023, ujjwal, Sutter_2024, Baron_2024}. \cite{Thilker_2023} examined the diffuse and filamentary emissions seen in PAH emission and their relationship to star formation. One of the striking features observed in the JWST is the bubbles and superbubbles formed due to feedback from massive stars and supernovae (SNe) explosions. This has also been a key research topic for NGC 628 \citep{Mayya2023, Barnes_2023, Watkins_2023}.



In the following sections, we first present the data in Section \ref{Data}, and then in Section \ref{sec:Analysis and results} we describe the analysis of our data, along with the results. In Section \ref{sec:discussion}, we discuss our results and compare them with previous studies. Finally, we summarise all our findings in Section \ref{sec:Summary}.

\section{Data} \label{Data}

\begin{figure*}
\includegraphics[scale=0.32]{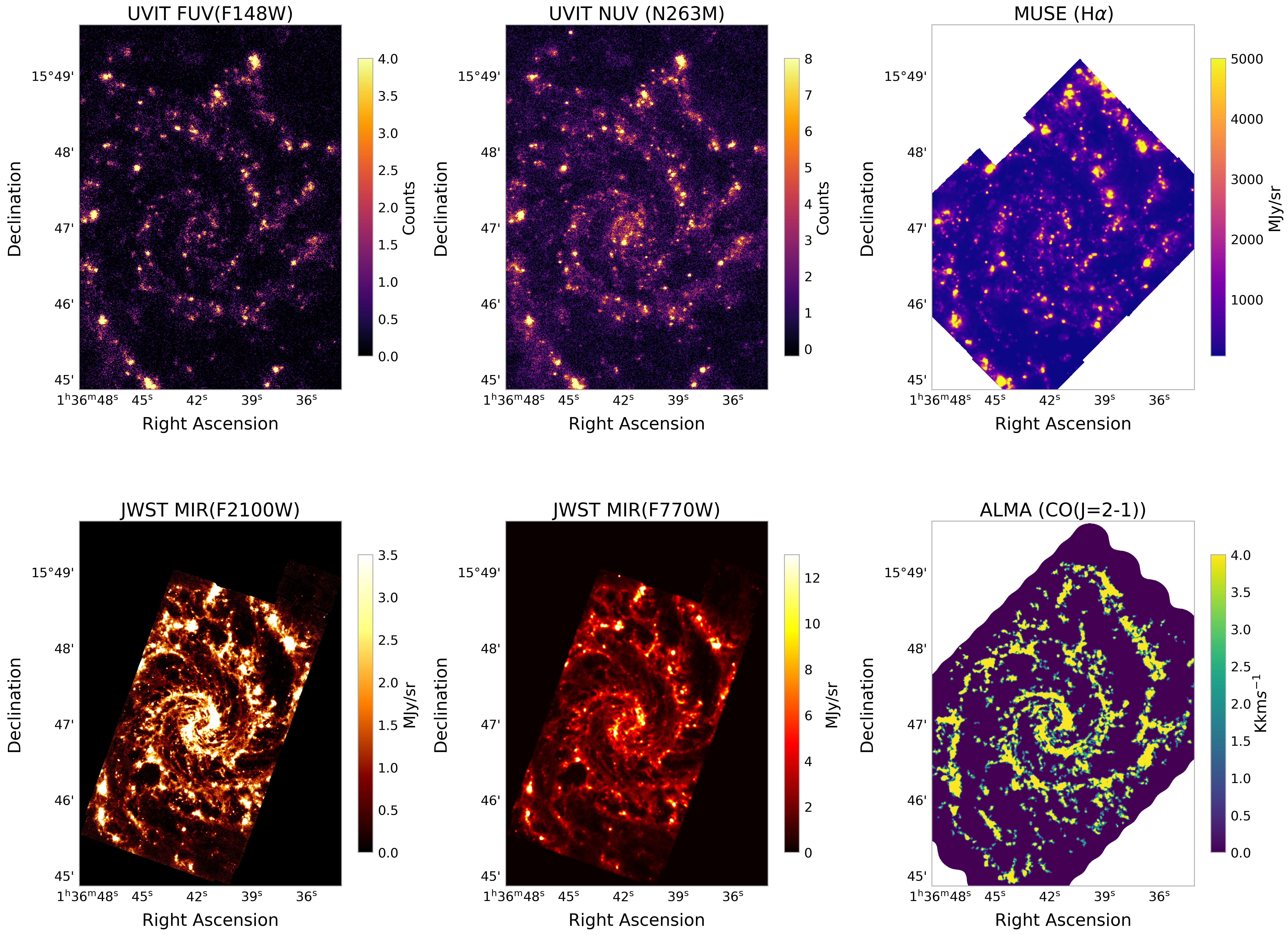}
\caption{Multiwavelength images of NGC 628.}
\label{fiG:multi-wavelength_images}
\end{figure*}

We used archival fully calibrated level 3 JWST Mid-Infrared Instrument \cite[MIRI;][]{Rieke2015PASP..127..584R} F770W ($\Delta \lambda =1.95 \mu m$), F1130W ($\Delta \lambda =0.73 \mu m$) images, and the Near Infrared Camera \cite[NIRcam;][]{Rieke2005SPIE.5904....1R} F335M image ($\Delta \lambda =0.38 \mu m$) to study the PAH emission in NGC 628. To estimate and remove the underlying stellar and dust continuum, we additionally used the NIRCam F300M and F360M images for the F335M band, and the MIRI F560W and F2100W images for the F770W and F1130W bands. We also used MIRI F2100W for extinction correction. These images (except F560W) are part of the Physics at High Angular resolution in Nearby GalaxieS  \cite[PHANGS-JWST;] []{Lee_2023, Williams_2024} treasury survey. The image F560W is part of the Feedback in Emerging extrAgalactic Star clusTers \cite[FEAST;] []{2024ApJ...971..118C, 2024ApJ...971...32P, 2026MNRAS.545f2025B} survey. These high-level science products were retrieved from the Mikulski Archive for Space Telescopes portal. The F770W and F1130W images have an angular resolution of 0$^{\prime \prime}$.269 and 0$^{\prime \prime}$.375, respectively. The NIRCAM filters F300M, F335M, and F360M have an approximate angular resolution of 0$^{\prime \prime}$.11. 

We used the archival MUSE H$\alpha$ emission-line map (PHANGS-MUSE; \cite{Eric}). The angular resolution of the MUSE H$\alpha$ map is around 0$^{\prime \prime}$.6. We used reduced UVIT's far-ultraviolet (FUV) and near-ultraviolet (NUV) observations from A\&D25. The UVIT is a twin telescope with FUV (1300–1800\r{A}) as one instrument, and the other instrument has the NUV (2000–3000\r{A}) and visible (VIS) bands. We used the CaF2 F148W ($\Delta \lambda =500$\r{A}), and the NUVB4 N263M ($\Delta \lambda =275$\r{A}) filter images for our analysis. The angular resolution of the FUV and NUV bands is around 1$^{\prime \prime}$.2-1$^{\prime \prime}$.5. By looking at the field stars in the FUV and NUV images, we obtained a mean PSF of those stars to be around 1$^{\prime \prime}$.2. The information on the reduction and other analyses to get a science-ready UVIT image of NGC 628 is explained in A\&D25. 

We convolved all JWST and MUSE images with appropriate Gaussian kernels using AstroPy’s \texttt{convolve\_fft} to match the UVIT resolution, with a Full Width at Half Maximum of approximately 1$^{\prime \prime}$.2. For JWST images, we approximated the native PSF of each image as a Gaussian and computed the required Gaussian convolution kernel from the difference between the native and target resolutions. The kernel width was determined separately for each filter to match all images to a common final resolution. We utilise these convolved images for further analysis. With this resolution, we are able to resolve OB associations that are approximately 57 parsecs in size.

To characterise star-forming regions, we make use of an existing catalogue from A\&D25, which lists the star formation properties of SFCs in NGC 628, including UV and H$\alpha$ emission. These SFCs were extracted using SExtractor \citep{1996A&AS..117..393B} with a threshold of 5$\sigma$ in all three bands for a confident detection. The NUV image reaches a 5$\sigma$ limiting magnitude of $m_{\mathrm{NUV}} = 23.36$, yielding 690 SFCs, while the FUV image is shallower, with a 5$\sigma$ limiting magnitude of $m_{\mathrm{FUV}} = 22.78$, resulting in 195 SFCs. The SFCs were then classified based on their location (arm, spurs, and interarm regions), and the associated emission in the H$\alpha$, FUV, and NUV bands. Arms and interarm SFCs were classified based on their location in the NUV image. However, spurs are features that extend outward from the spiral arms and are evenly spaced in azimuth. They can extend up to kiloparsec scale lengths as per simulations \citep{Williams_2022}. Spurs are observed better in molecular gas maps as seen in the ALMA image in Figure \ref{fiG:multi-wavelength_images}. Hence, the SFCs were plotted on the ALMA CO map \cite[PHANGS-ALMA;] []{Leroy_2021a, Leroy_2021}  to identify those SFCs associated with spur regions (See Figure 2 in paper A\&D25).  

\subsection{Continuum subtraction from JWST PAH bands} \label{Continuum subtraction}

The JWST F335M, F770W, and F1130W filters are dominated by PAH emission features. However, these medium- and broad-band filters also contain significant contributions from the underlying stellar and dust continuum. Therefore, it is necessary to separate the PAH emission from the continuum components to obtain accurate measurements of the PAH fluxes.

The starlight continuum from F335M is removed by linearly interpolating between F300M and F360M with a varying slope as mentioned in \cite{Lai_2020}. However, with this method, the F335M continuum is over-subtracted in regions where PAH emission is dominant. This is due to the presence of PAH contribution in the F360M band in such regions \citep{Sandstrom_2023}. Hence, we use the following relation from \cite{Sandstrom_2023} for optimised continuum subtraction, which is given below.

\begin{multline} \label{eq1}
\mathrm{F335M_{Cont}= (\frac{B_{Lai}}{B_{PAH}-B_{Lai}})(B_{PAH}F360M-F335M}\\
\mathrm{+A_{Lai}F300M)+A_{Lai}F300M }
\end{multline} 

\begin{equation} \label{eqPAH}
\mathrm{F335M_{PAH}=F335M-F335M_{Cont}}
\end{equation}

\noindent
The coefficients A$\mathrm{_{Lai}}$ and B$\mathrm{_{Lai}}$ in Equation \ref{eq1} are 0.35 and 0.65, respectively, as mentioned in \cite{Lai_2020}. B$\mathrm{_{PAH}=}$1.6 is the observed slope of the colours for PAH-dominated regions as obtained for NGC 628 in \cite{Sandstrom_2023}.

We used the spectroscopically calibrated method developed by \cite{2025ApJ...983...79D} to derive 7.7 and 11.3 $\mu$m PAH fluxes from the broadband MIRI F770W and F1130W filters. This method calibrates photometrically derived PAH fluxes against spectroscopic measurements using synthetic MIRI photometry. The thermal dust continuum is approximated by a single power law with index $\alpha$, allowing the continuum contribution to be estimated from pairs of MIRI filters consisting of a blue (B) and red (R) continuum band selected from F560W, F1000W, F1500W, and F2100W. For NGC 628, observations are available in F560W, F1000W, and F2100W, enabling the application of several filter combinations. The resulting expressions for the 7.7 and 11.3 $\mu$m PAH fluxes are given by Equations \ref{eq4} and \ref{eq5}, respectively.

\begin{equation} \label{eq4}
F_{\mathrm{PAH 7.7}}(\mathrm{Wm^{-1})}= (9.00 \pm 0.21) \times10^{-14} \times (f_{F770}-g_{con}f_{B}^{1-\alpha}f_{R}^{\alpha})
\end{equation}

\begin{equation} \label{eq5}
F_{\mathrm{PAH 11.3}}(\mathrm{Wm^{-1})}= (1.77 \pm 0.04) \times10^{-14} \times (f_{F1130}-g_{con}f_{B}^{1-\alpha}f_{R}^{\alpha})
\end{equation}

\noindent
where the flux densities $f$ are given in Jy, while $f_{B}$ and $f_{R}$ are flux densities from blue and red filters, respectively, and filter combinations are shown in Table \ref{tab: continuum parameters}. The adopted filter combinations and the corresponding values of $g_{con}$ and $\alpha$ for particular combinations are listed in Table \ref{tab: continuum parameters}.

\begin{figure}
\includegraphics[scale=0.37]{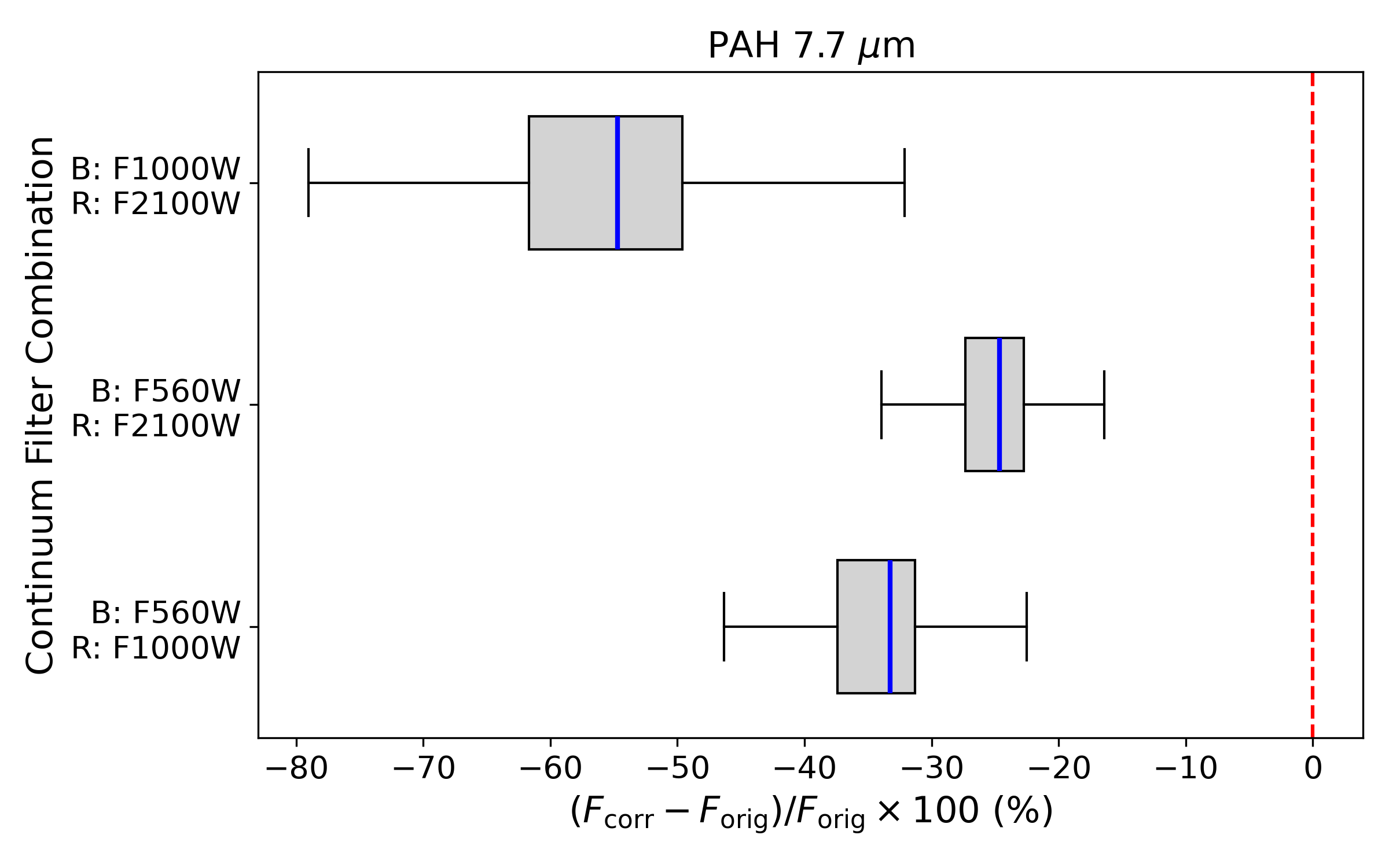}
\includegraphics[scale=0.37]{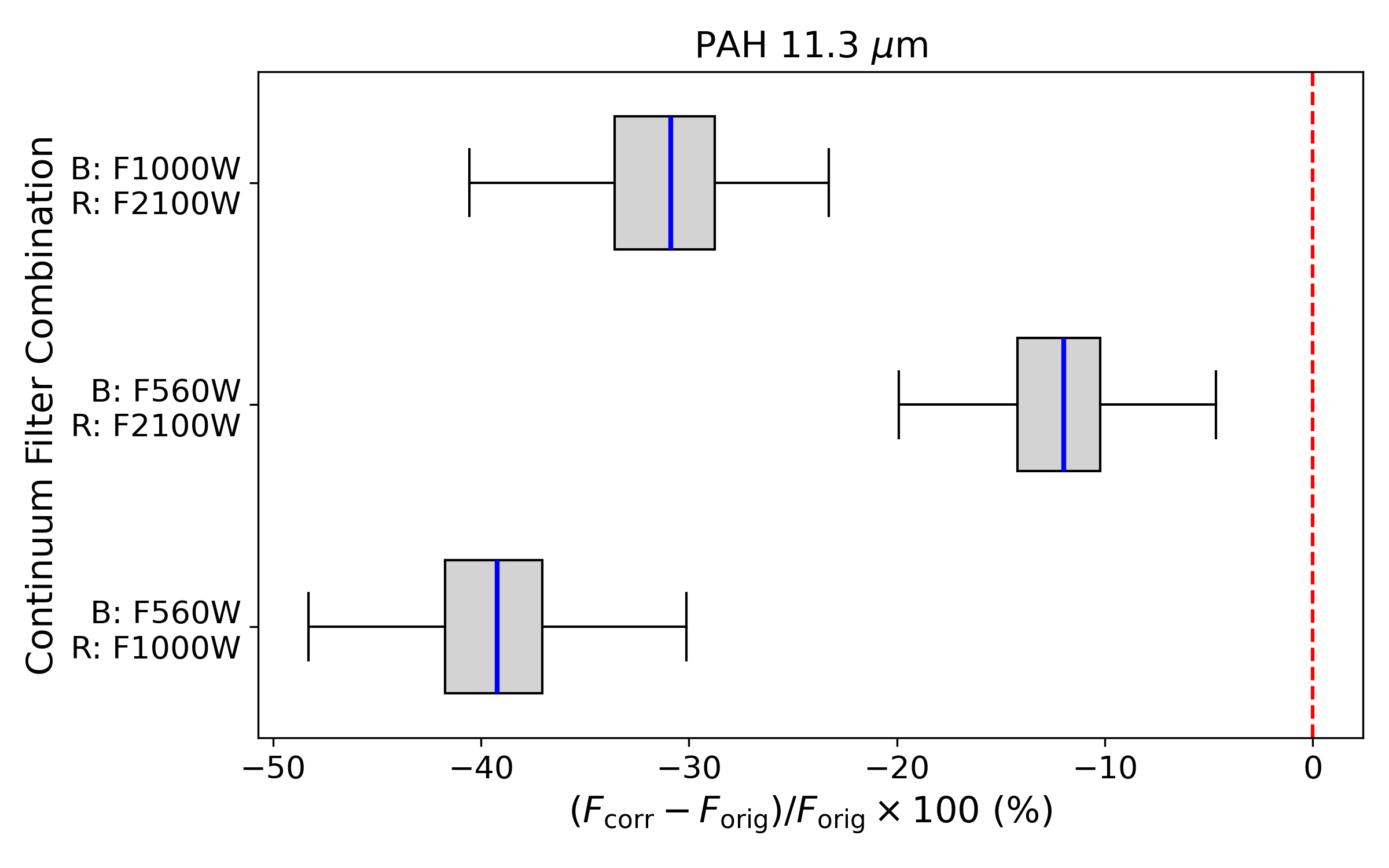}
\caption{Comparisons between corrected photometric PAH flux and uncorrected PAH flux from F770W (top panel )and F1130W (bottom panel) filters for various combinations of
JWST/MIRI filters using Equation \ref{eq4} and \ref{eq5}. The blue lines are median values, the ends of the boxes are 25th and 75th percentiles, respectively, and the whiskers indicate the largest percent difference within 1.5 times the 25th–75th percentile range on either side.}
\label{fig: continuum}
\end{figure}

We applied these relations to all SFCs using each available filter combination listed in Table \ref{tab: continuum parameters}. To assess the sensitivity of the derived PAH fluxes to the continuum estimate, we compared the fractional continuum contribution removed from the F770W and F1130W fluxes for each filter combination. The resulting distributions are shown in Figure \ref{fig: continuum}.

We favour the correction involving F560W and F2100W filters for further analysis, as the continuum covered by these two filters is relatively flat without prominent emission lines compared to the F1000W, which contains the silicate absorption feature and the pure rotational $\mathrm{H}_2$ 0--0 S(5) line. In addition, the F560W-F2100W combination produces the smallest scatter in the continuum-corrected PAH fluxes for both F770W and F1130W bands (Figure \ref{fig: continuum}), indicating a more stable continuum subtraction.

\begin{table}
\caption{Prescription Parameters for MIRI PAH extraction}
    \centering
    \begin{tabular}{cccccc}
    \hline
        &7.7 $\mu$m& & &11.3 $\mu$m&\\
        \hline
        Filters & $g_{con}$ & $\alpha$ &  & $g_{con}$ & $\alpha$  \\
       (B, R)  &  & &&\\
    \hline
       F560W, F1000W  & 0.53 &0.91 & &1.22&1.18\\
       F560W, F2100W  &  0.23 & 0.62&&0.53& 0.49\\
       F1000W, F2100W &  -0.36 &1.51 &&0.17&1\\ 
    \hline
    \end{tabular}
    \label{tab: continuum parameters}
    \begin{minipage}{70 mm}
    \textit{Note: $g_{con}$ and $\alpha$ are parameters for equations \ref{eq4} and \ref{eq5}. These equations are from \cite{2025ApJ...983...79D}. } 
    \end{minipage}%
\end{table}

\section{Analysis and Results } \label{sec:Analysis and results} 

\subsection{PAH correlation with the UV and H\texorpdfstring{$\alpha$}{Lg} emission} \label{PAH correlation with the UV and Halpha emission}
The correlation of PAH with SFR has been established in previous studies \citep{Calzetti_2007, Shipley_2016}. Earlier studies have used the Infrared Space Observatory (ISO) and Spitzer low-resolution data and broadband MIR emission to establish the relation \citep{Calzetti_2007, Battisti_2015}. These have used integrated luminosities of galaxies to obtain this correlation. Therefore, it is important to verify whether this correlation also applies to individual SFCs. Recent studies using JWST have further explored PAH-based SFR calibrations using longer-wavelength recombination lines (such as P$\alpha$ and Br$\alpha$) at $\sim$40 pc scales in nearby galaxies \citep{2024ApJ...971..115G, 2026ApJ...997...20G}. In this work, we instead examine whether the established UV and H$\alpha$ correlations with PAH emission hold in spatially resolved regions such as SFCs.

\begin{figure*}
\centering
(i)\\
\includegraphics[scale=0.55]{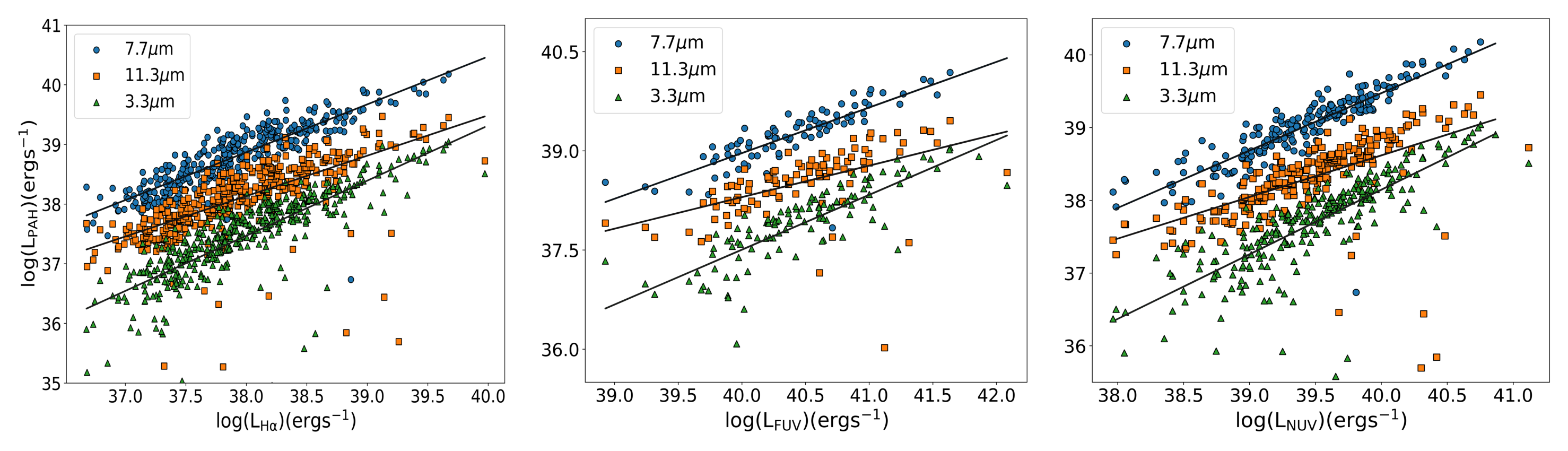}
\vspace{0.3cm}

\centering
(ii)\\
\includegraphics[scale=0.55]{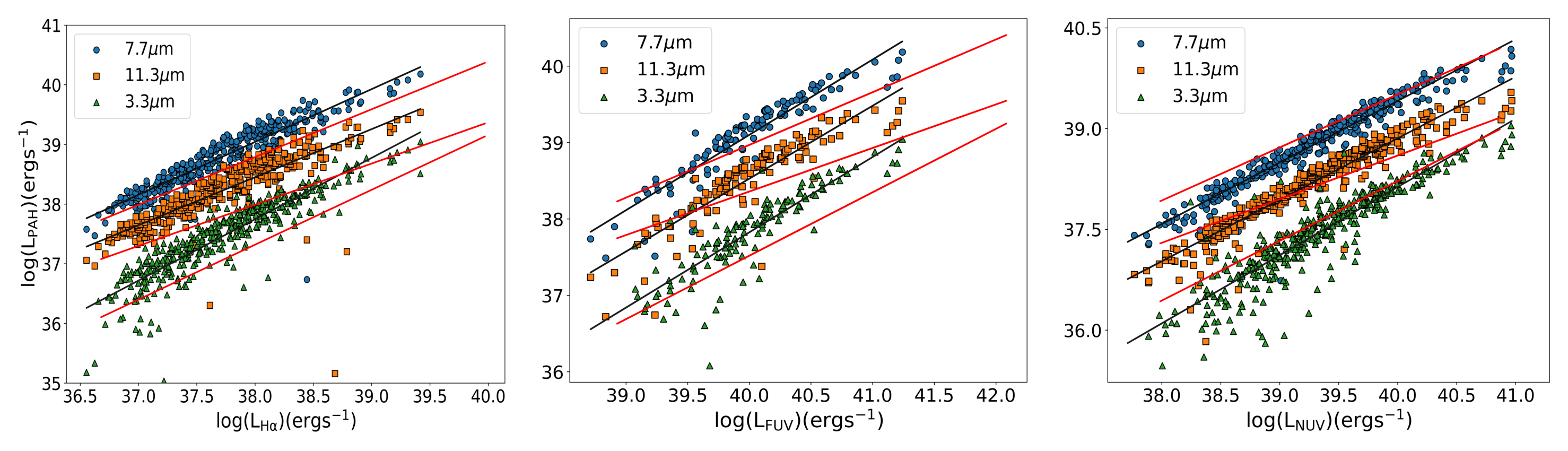}

\caption{Correlation of H$\alpha$, FUV and NUV luminosities with different PAH luminosities (7.7 $\mathrm{\mu m}$, 11.3 $\mathrm{\mu m}$ and 3.3 $\mathrm{\mu m}$). This correlation is for extracted SFCs, whose UV and H$\alpha$ luminosities are corrected for extinction with (i) the Balmer decrement method and (ii) MIRI 21$\mu$m. The black trend lines represent the best fit for each plot, while the red lines in panel (ii) show the trend lines from panel (i) overplotted for comparison.}
\label{fig: correlation}
\end{figure*}

Here we investigate how the PAH emission from JWST filters correlates with H$\alpha$, FUV, and NUV emission using SFCs from the A\&D25 catalogue. The H$\alpha$ and UV luminosities of SFCs are corrected for extinction using the Balmer decrement and MIR emission (here using JWST F2100W band), as detailed in A\&D25. For the Balmer decrement, we used the ratio of two nebular Balmer emission lines, H$\alpha$ and H$\beta$. To correct for extinction using MIR emission, we employed well-established extinction correction relations from \cite{2009ApJ...703.1672K} and \cite{Hao_2011}. These corrections are based on the Spitzer MIPS 24 µm luminosity. Hence, we identified a set of bright regions in the matched-resolution images and measured their fluxes in both bands. We calculated the ratio of the JWST 21 µm luminosity to the MIPS 24 µm luminosity for several bright regions and found a ratio of 4.54. We then multiplied this ratio by the 21 µm luminosity to apply it in the extinction correction relation. In our further analysis, we primarily used the luminosities corrected with the Balmer decrement; however, we also considered the JWST 21$\mathrm{\mu m}$ method for comparison.


We overlaid the SFCs on convolved PAH images (F770W, F1130W, and continuum-subtracted F335M) and measured luminosities in each band. However, we implemented signal-to-noise ($\mathrm{S/N}$) limits on the JWST data, as continuum-subtracted PAH maps, particularly the F335M map, can contain negative pixel values owing to oversubtraction and background fluctuations. To ensure robust measurements, we restricted our analysis to SFCs with $\mathrm{S/N}>3$ in all PAH maps. For the 7.7 and 11.3 $\mu$m PAH measurements, this criterion was imposed on the original F770W and F1130W fluxes before continuum subtraction. We plotted the correlation of different PAH bands with H$\alpha$, FUV, and NUV emission as shown in  Figure \ref{fig: correlation}. We fitted linear regression as given below,

\begin{equation} \label{eq2}
\mathrm{log(L(PAH))=Alog(L(K))+B}
\end{equation}

\noindent where L(PAH) is continuum-subtracted PAH luminosity (7.7 $\mathrm{\mu m}$, 11.3 $\mathrm{\mu m}$, 3.3 $\mathrm{\mu m}$) and the term L(K) corresponds to the extinction-corrected luminosity for H$\alpha$, FUV, or NUV, with K denoting any of these star formation tracers. Both L(PAH) and L(K) are in ergs$^{-1}$. A and B are fitting parameters as mentioned in Table \ref{table1}.

\begin{table*}
\caption{Fitting parameters for Equation \ref{eq2}.}
\begin{center}
\begin{tabular}{|c|c|cccc|cccc|}
\hline
Star formation & PAH bands & \multicolumn{4}{c|}{Balmer decrement method} & \multicolumn{4}{c|}{21 $\mu$m method} \\
Tracer &$\mu$m   &A &B&$\sigma$   & $\rho$  &A &B&$\sigma$   & $\rho$  \\
\hline
&3.3     & 0.92 $\pm$ 0.04       & 2.36 $\pm$ 1.39 & 0.46         & 0.86 & 1.03 $\pm$ 0.02        & -1.32 $\pm$ 0.94 & 0.27               & 0.93                   \\
H$\alpha$ &7.7     &0.80 $\pm$ 0.02         & 8.39 $\pm$ 0.96 & 0.27          & 0.89   &0.88 $\pm$ 0.02         & 5.42 $\pm$ 0.81 & 0.22          & 0.94                    \\
&11.3   & 0.69 $\pm$ 0.03          & 11.77 $\pm$ 1.29 & 0.37         & 0.84 & 0.80 $\pm$ 0.03          & 7.86 $\pm$ 1.16 & 0.32         & 0.90                \\
\hline
&3.3     & 0.83 $\pm$ 0.07        & 4.32 $\pm$ 2.76 & 0.29               & 0.85 & 0.99 $\pm$ 0.05        & -1.59 $\pm$ 2.04 & 0.28               & 0.90             \\
FUV& 7.7   & 0.69 $\pm$ 0.05        & 11.37 $\pm$ 2.08 & 0.24 & 0.88   & 0.99 $\pm$ 0.04        & -0.33 $\pm$ 1.47 & 0.20 & 0.96               \\
&11.3 & 0.57 $\pm$ 0.06 & 15.56 $\pm$ 2.51 & 0.29                & 0.80 & 0.95 $\pm$ 0.04 & 0.37 $\pm$ 1.77 & 0.25                & 0.94                 \\
\hline

&3.3     & 0.89 $\pm$ 0.04        & 2.62 $\pm$ 1.78 & 0.40         &  0.86 & 1.02 $\pm$ 0.02        & -2.49 $\pm$ 0.96 & 0.28         &  0.92      \\
NUV& 7.7   & 0.79 $\pm$ 0.03        & 7.91 $\pm$ 1.30 & 0.25         & 0.91    & 0.92 $\pm$ 0.01        & 2.73 $\pm$ 0.50 & 0.16         & 0.98                   \\
&11.3  & 0.65 $\pm$ 0.05          & 12.60 $\pm$ 2.03 &  0.40    & 0.83      & 0.92 $\pm$ 0.02          & 2.22 $\pm$ 0.61 &  0.19    & 0.97       \\
\hline

\end{tabular}
\end{center}
\label{table1}
\centering
 \begin{minipage}{150 mm}
    \textit{Note:} A and B are the fitting parameters for the Equation \ref{eq2}, which represents the correlation between PAH bands and star formation tracers as shown in the Figure \ref{fig: correlation}. $\sigma$ and $\rho$ represent the standard deviation and Spearman's correlation coefficient of these fits, respectively.
    \end{minipage}%
\end{table*}

\cite{Shipley_2016} derived the calibration relation between the extinction-corrected  H$\alpha$ luminosity and the individual PAH luminosities. The slopes of the linear relation for 7.7$ \mu m$ and 11.3 $\mathrm{\mu m}$ PAH luminosities are 1 and 0.94 with intercepts of 1.12 and 2.87, respectively. Certainly, there is a clear difference between the parameters of \cite{Shipley_2016} calibration relation and those mentioned in Table \ref{table1}. However, the relations reported in \cite{Shipley_2016} are based on integrated emission from the whole galaxy with very diverse samples, and therefore, there is a lot more scatter in their correlations.

There are very few calibration relations connecting $L({PAH})$ with UV luminosities. \cite{Zhang_2021}  derived calibration relations for integrated PAH luminosity with H$\alpha$, FUV and NUV luminosities, and found slopes of 0.849 $\pm$ 0.026, 0.842 $\pm$ 0.027, and 0.865 $\pm$ 0.029. However, when we examined our fitted parameters for individual PAH bands against H$\alpha$ and UV luminosities corrected for extinction using the Balmer decrement, we observed band-to-band variations, depending on the specific PAH band used. We find that the fitted slopes decrease with increasing PAH wavelength. 7.7 and 11.3 $\mu$m PAH luminosities exhibit flatter linear fits compared to the 3.3 $\mu$m luminosity, suggesting that the 3.3 $\mu$m feature is more sensitive to variations in recent star formation. We also find that the correlations involving the 3.3 $\mathrm{\mu m}$ PAH luminosity exhibit larger scatter than those involving the 7.7 and 11.3 $\mathrm{\mu m}$ PAH luminosities, likely reflecting the intrinsically weaker nature of the 3.3 $\mathrm{\mu m}$ PAH feature. In addition, uncertainties associated with continuum subtraction in the F335M band may contribute to the observed scatter.

The fitted relations involving H$\alpha$ and NUV luminosities are broadly similar across the PAH bands. In contrast, the FUV correlations show significant variation, as indicated in Table~\ref{table1}, with shallower slopes and larger uncertainties than those of the other tracers. For example, the fitted slope for H$\alpha$ and NUV with 7.7 $\mu$m PAH is 0.80 $\pm$ 0.02 and 0.79 $\pm$ 0.03, respectively. Whereas, the slope for the correlations with FUV has a value of 0.69 $\pm$ 0.05. Similarly, the fitted parameters are shallow for the FUV and other PAH bands. One plausible explanation for this is that the FUV data is intrinsically shallower than the other datasets, resulting in lower completeness at faint luminosities. Consequently, under a fixed detection threshold, a larger fraction of low-luminosity FUV-emitting SFCs may remain undetected, which can contribute to both the reduced number of identified FUV SFCs and the increased uncertainty in the fitted relations.

We also examined correlations between the PAH-band luminosities and extinction-corrected luminosities derived using MIR-based extinction corrections. In this case, the correlations are stronger (Spearman's correlation coefficient, $\rho \geq 0.9$), and the fitted slopes are steeper than those obtained using Balmer-decrement extinction corrections (Figure~\ref{fig: correlation}(ii) and Table~\ref{table1}). This indicates that the derived calibration relations are sensitive to the adopted extinction-correction method.

\subsection{ PAH emission across the different regions of NGC 628} \label{Association of PAH emission across the different regions of NGC 628}

\begin{figure*}
\includegraphics[scale=0.43]{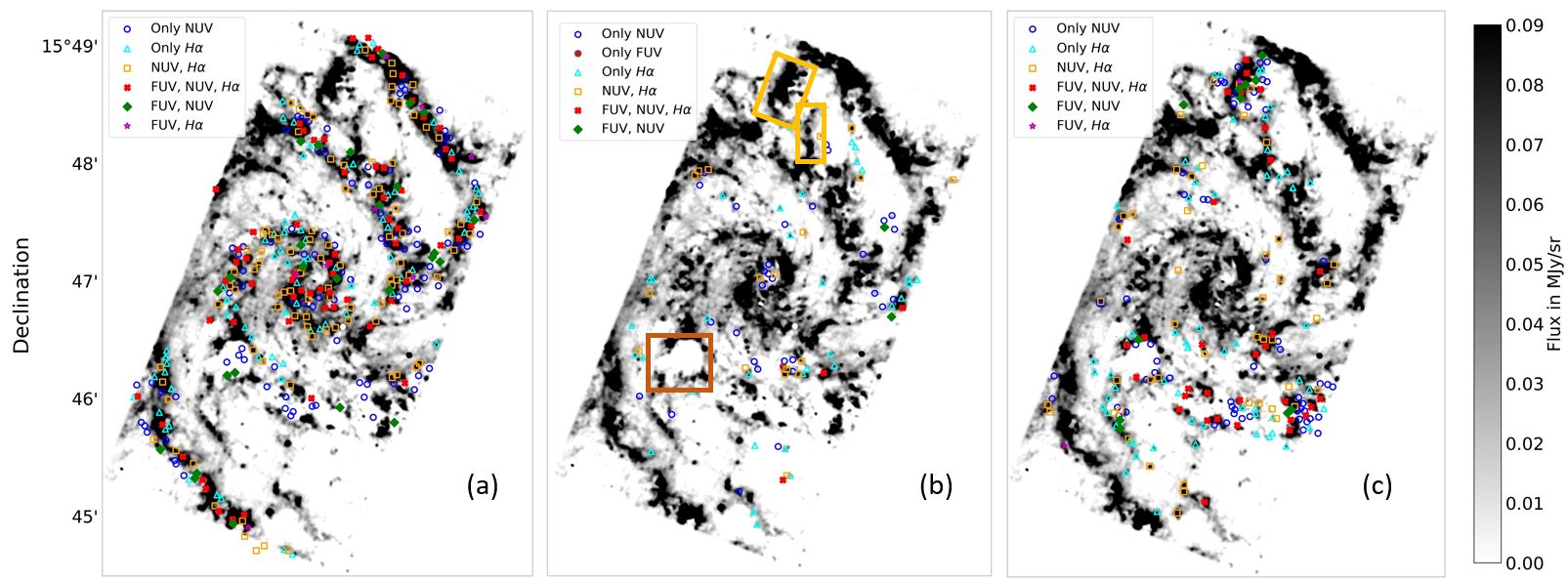}
\includegraphics[scale=0.435]{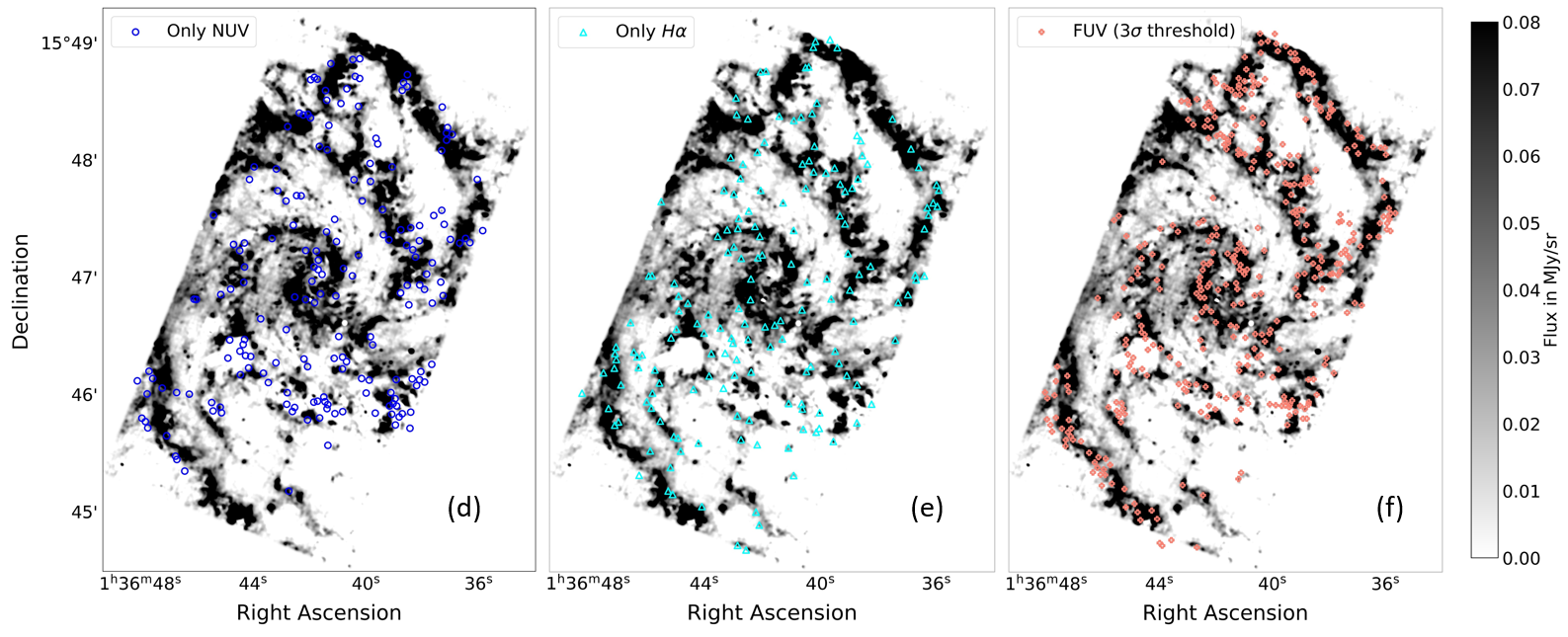}
\caption{(i) Association of 3.3 $\mu$m PAH emission with arm (a), interarm (b), and spur (c) SFCs. We also plotted the association of 3.3 $\mu$m PAH with NUV-only SFCs(d), H$\alpha$-only SFCs(e), and FUV emitting SFCs detected with a 3$\sigma$ threshold (f). The SFCs are overlaid on the continuum-subtracted JWST F335M image. Yellow rectangles in Figure (b) show an example of the spurs, and the brown rectangle indicates the position of the phantom void in NGC 628. }
\label{fig 3: arm_spur_interarm}
\end{figure*}

In this Section, we investigate how the intensity of PAH emission varies across the arm, spur, and interarm region. For that reason, we overplotted the SFCs onto the PAH images. We note the spatial overlap of 3.3 $\mu$m PAH emissions with SFC's position across the regions in Figure \ref{fig 3: arm_spur_interarm}. We have also overplotted NUV-only (No FUV and H$\alpha$ counterparts), H$\alpha$-only (No FUV and NUV counterparts), and FUV-emitting (3 $\sigma$ threshold) SFCs on the F335M image (bottom panel of Figure \ref{fig 3: arm_spur_interarm}).

\begin{figure*}
\begin{flushleft}
\hspace{3cm}(i)
\hspace{5.3cm} (ii) 
\hspace{5.7cm} (iii)\\
\end{flushleft}

\includegraphics[scale=0.33]{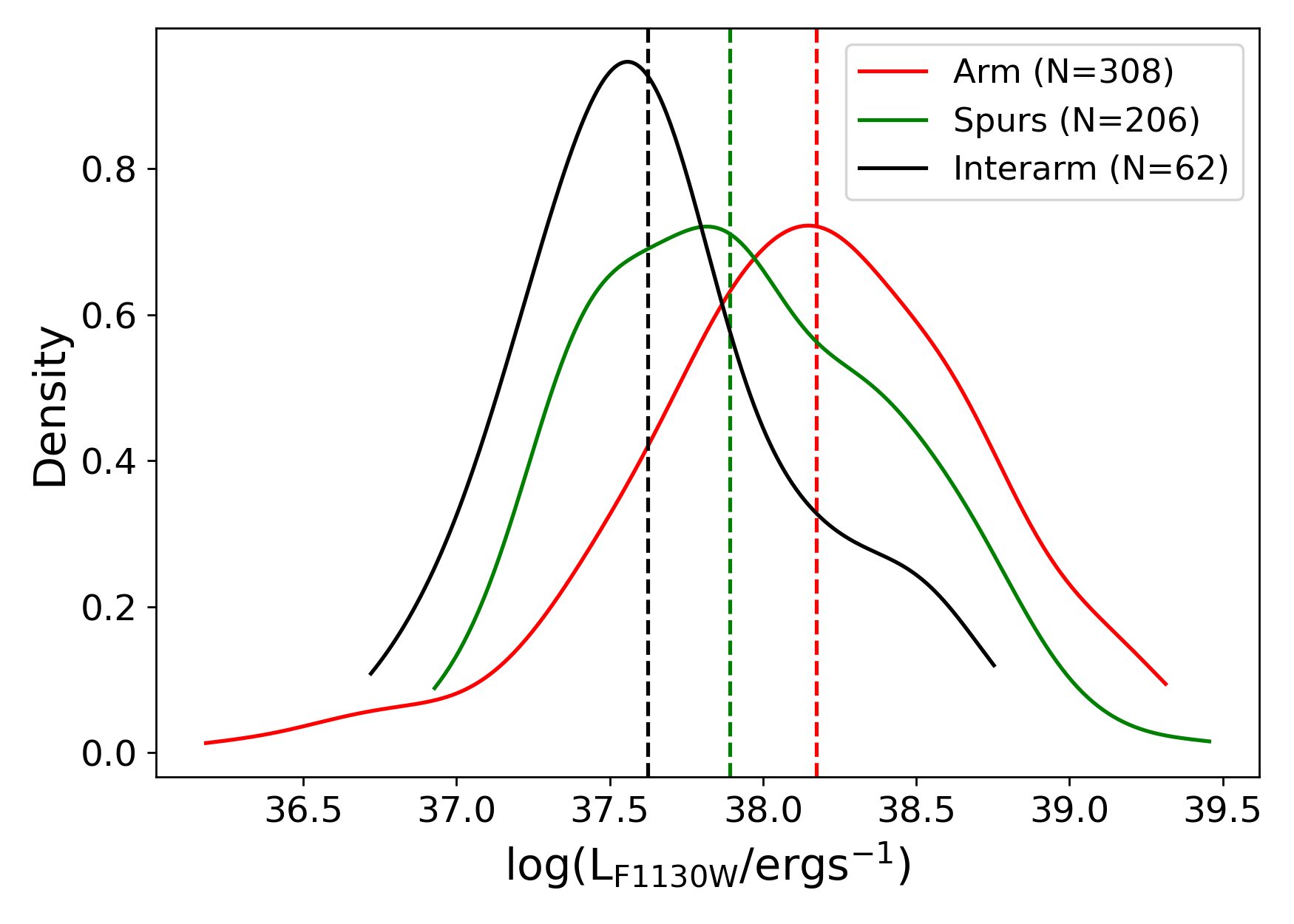}
\includegraphics[scale=0.33]{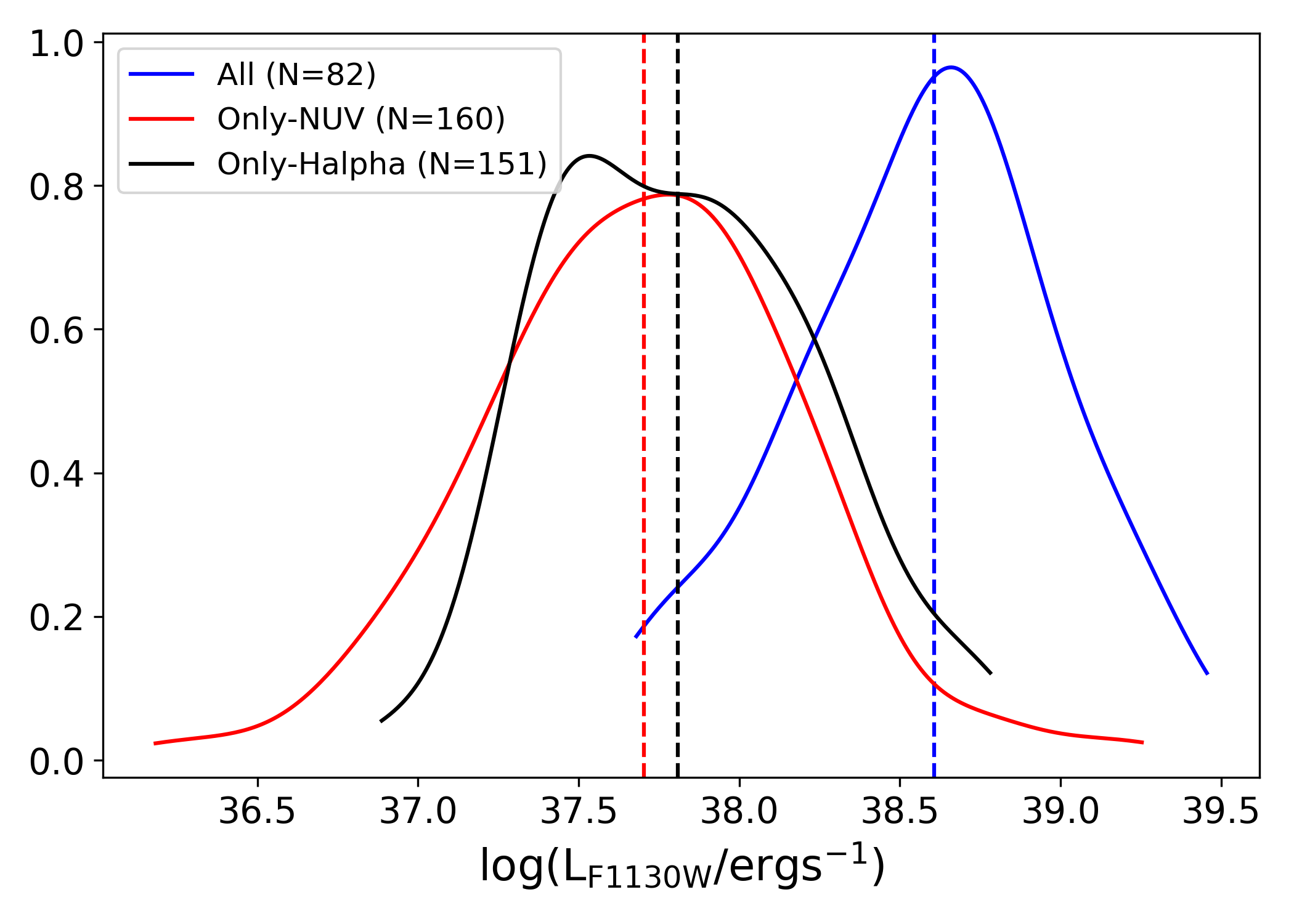}
\includegraphics[scale=0.33]{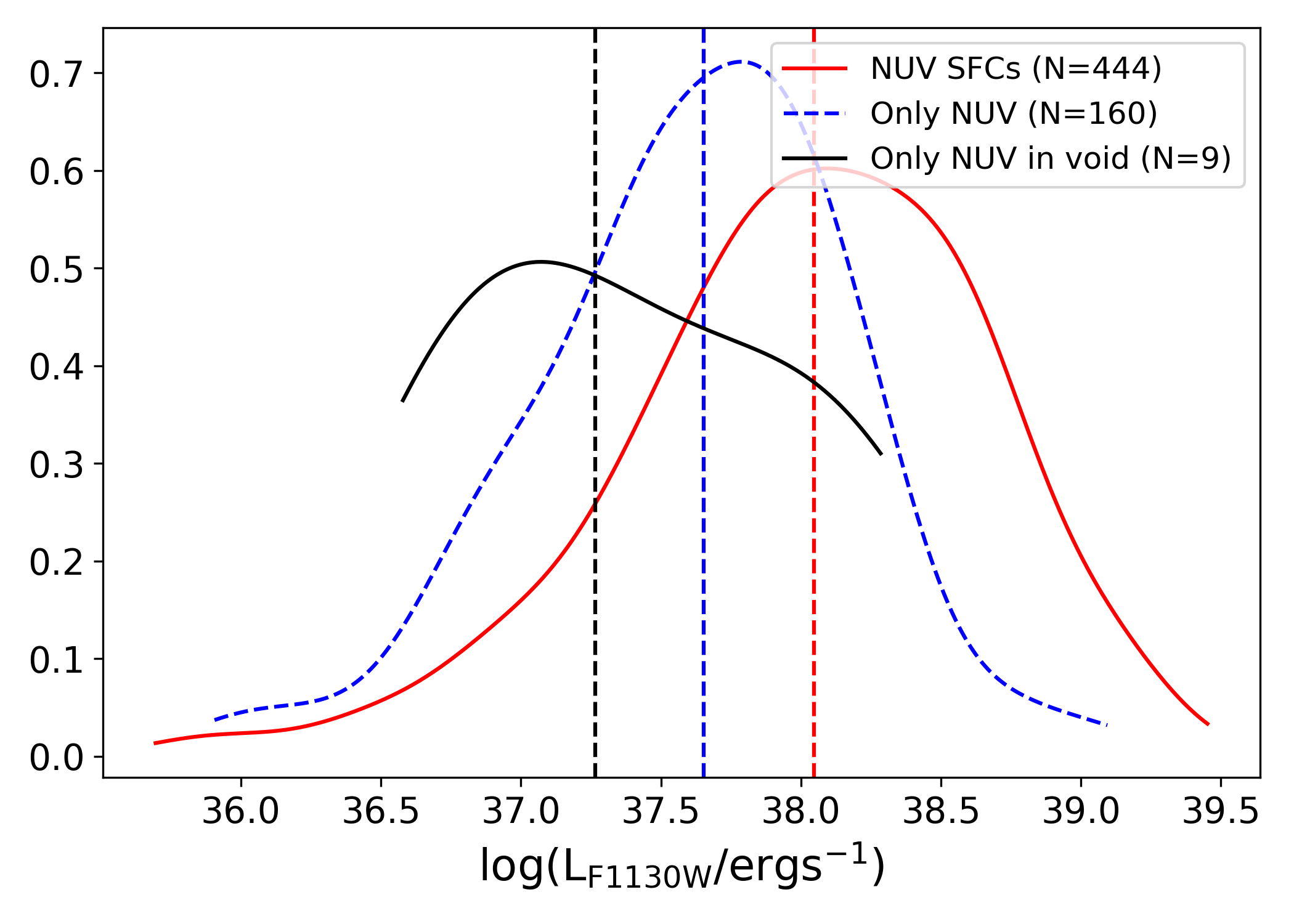}
\caption{The KDE plots representing the 11.3 $\mu$m PAH luminosity distribution (i) for SFCs in different regions, (ii) for SFCs emitting in all bands (UV and $H\alpha$), only-NUV and only-H$\alpha$ emissions, (iii) for SFCs emitting in NUV, only-NUV and only-NUV SFCs in the void region. The vertical lines represent the median values.}
\label{fig: KDE}
\end{figure*}

Visually, we find that arm and spur SFCs overlap with PAH emission, except for those that fall within bubbles and superbubbles, as seen in Figure \ref{fig 3: arm_spur_interarm}(d). However, around half of the interarm SFCs (a total of 62 SFCs) do not show a bright PAH counterpart. We plotted the Kernel density estimate (KDE) to examine the 11.3 $\mu$m luminosity distribution for the SFCs in arm, spurs, and interarm regions, as shown in Figure \ref{fig: KDE}(i). We clearly see that the median luminosity of the interarm SFCs (log(L$\mathrm{_{F1130W}/ergs^{-1}}$)=37.8) is lower than that of the arms (log(L$\mathrm{_{F1130W}/ergs^{-1}}$)=38.3) and spurs (log(L$\mathrm{_{F1130W}/ergs^{-1}}$)=38). However, the lower luminosity limit of arm SFCs is distributed below log(L$\mathrm{_{F1130W}/ergs^{-1}}$)=36.5. 

We also checked the KDE of 11.3 $\mu$m luminosity for SFCs with emission in all bands, only-NUV and only-H$\alpha$. The SFCs with emission in all bands clearly have very strong PAH counterparts. All of these SFCs with emission in all bands (82 SFCs) are at the higher end of the luminosity distribution.

The median values of these distributions depend on how we have classified SFCs by location (arms and spurs). The figure in Appendix \ref{sec:appendix} shows 11.3 $\mu$m PAH luminosity distribution of arm SFCs (top figure) and spur SFCs over the F1130W JWST band. In both figures, we can see that the luminosity distribution is nearly uniform. Hence, a slight shift in the boundary does not affect the trend, arm $\geq$ spurs $>$ interarm.

\begin{figure}
\includegraphics[scale=0.33]{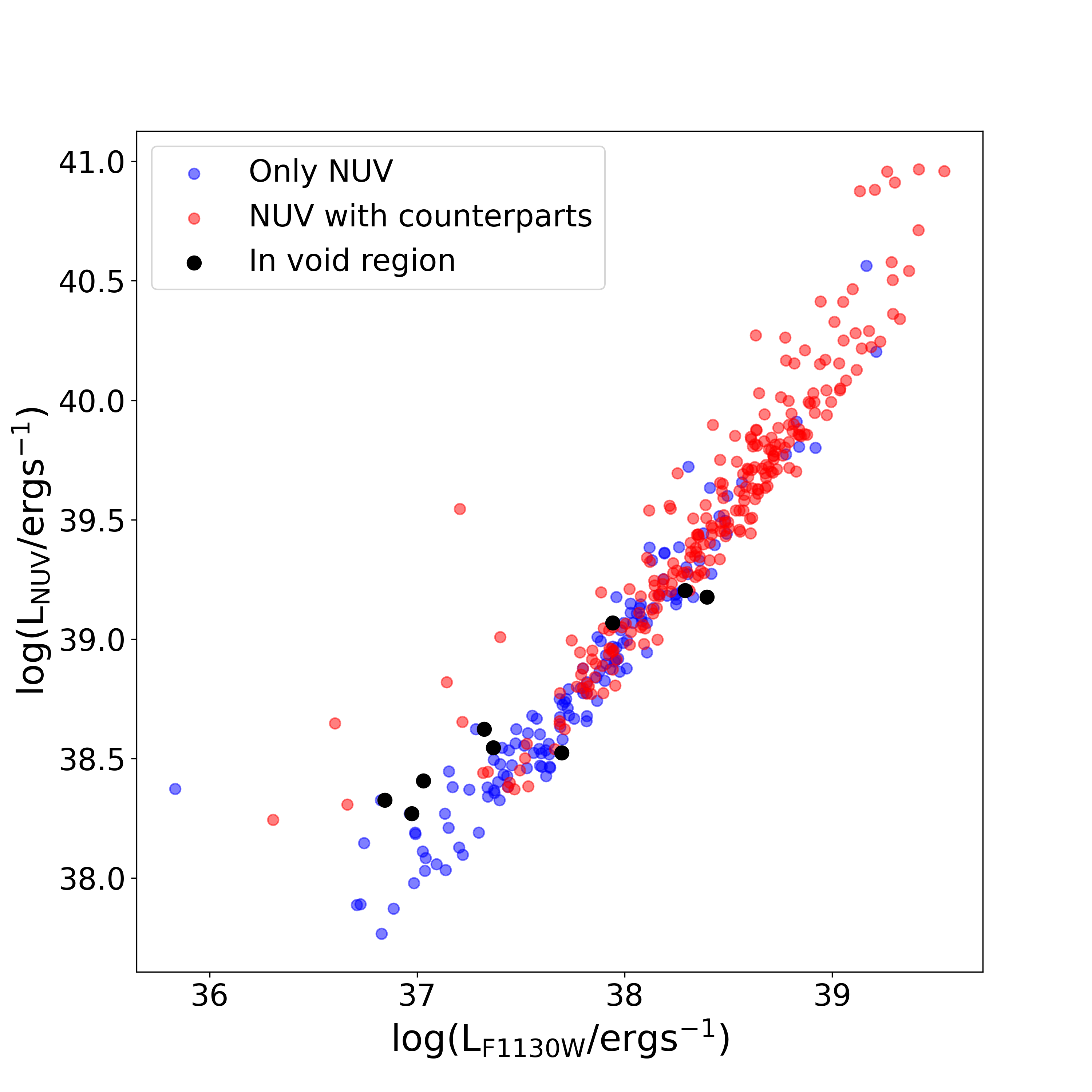}
\caption{Correlation of extinction corrected NUV luminosity of SFCs with NUV-only emission and NUV with counterparts (H$\alpha$, FUV) with the 11.3 $\mu$m PAH luminosity.}
\label{fig: NUV_correlation}
\end{figure}

Over 60 \% of SFCs within bubbles and superbubbles show only NUV emission, particularly in the southern arm, which hosts the phantom void. Therefore, we plotted the KDE of the 11.3 $\mu$m luminosity for SFCs with NUV emission (with and without counterparts), only NUV emission, and SFCs with only NUV emission within the phantom void region as shown in Figure \ref{fig: KDE}(iii). Here, we observed that SFCs with only NUV show a distribution shifted towards a lower PAH luminosity. We also plotted PAH and NUV luminosity correlations for SFCs with NUV emission, as shown in Figure \ref{fig: NUV_correlation}. The correlation between PAH and NUV luminosity reveals that NUV-only SFCs are at the lower end of both luminosity distributions, including those within the void region.

We also find that the galaxy centre has very faint PAH emission in all bands, with the 3.3\,$\mu$m being even further suppressed compared to the 7.7\,$\mu$m and 11.3\,$\mu$m emission. This indicates a dearth of small PAHs in the centre of this galaxy. We found that the SFCs around the centre are either NUV-only or show NUV and H$\alpha$ emission. The presence of H$\alpha$ emission indicates some recent star formation, or it could be emission from diffuse ionised gas. 

As described in Section \ref{Data}, SFCs were extracted by applying a constant threshold (5$\sigma$) in all three bands. Although the detection threshold was identical, the intrinsic depths of the images differ. Hence, to see if there are any faint FUV and H$\alpha$ emitting SFCs in the bubble and superbubble regions, we applied a 3$\sigma$ threshold. This does not alter the intrinsic depth of the FUV image, however corresponds to a formal 3$\sigma$ limiting magnitude of $m_{\mathrm{FUV}} = 23.4$. Using this less conservative selection, we identify 463 SFC candidates, which is significantly more than the 195 obtained using the  5$\sigma$ cutoff. The formal 3$\sigma$ FUV limit is comparable to the 5$\sigma$ depth of the NUV image. Since lowering the detection threshold can introduce spurious sources, we performed an empirical contamination test by running the identical source detection procedure on the inverted FUV image. We detect no sources in the negative image, corresponding to an estimated contamination fraction of zero. We therefore conclude that the additional detections introduced by the 3$\sigma$ threshold are unlikely to be dominated by noise. However, even after decreasing the threshold, we find that 4 out of 8 FUV-emitting SFCs are associated with NUV-emitting SFCs in the phantom void (at 5$\sigma$, only 1). Since other NUV-only SFCs lie at the faint end of the NUV luminosity distribution, their non-detection in FUV is consistent with the relative shallowness of the FUV image and the expected FUV–NUV colours of low-luminosity stellar populations. This likely indicates a relative deficiency of the most massive stars. A few FUV SFCs are located between the phantom void and two connecting bubbles, as shown in Figure \ref{fig 3: arm_spur_interarm}(f).


\begin{figure}
\includegraphics[scale=0.37]{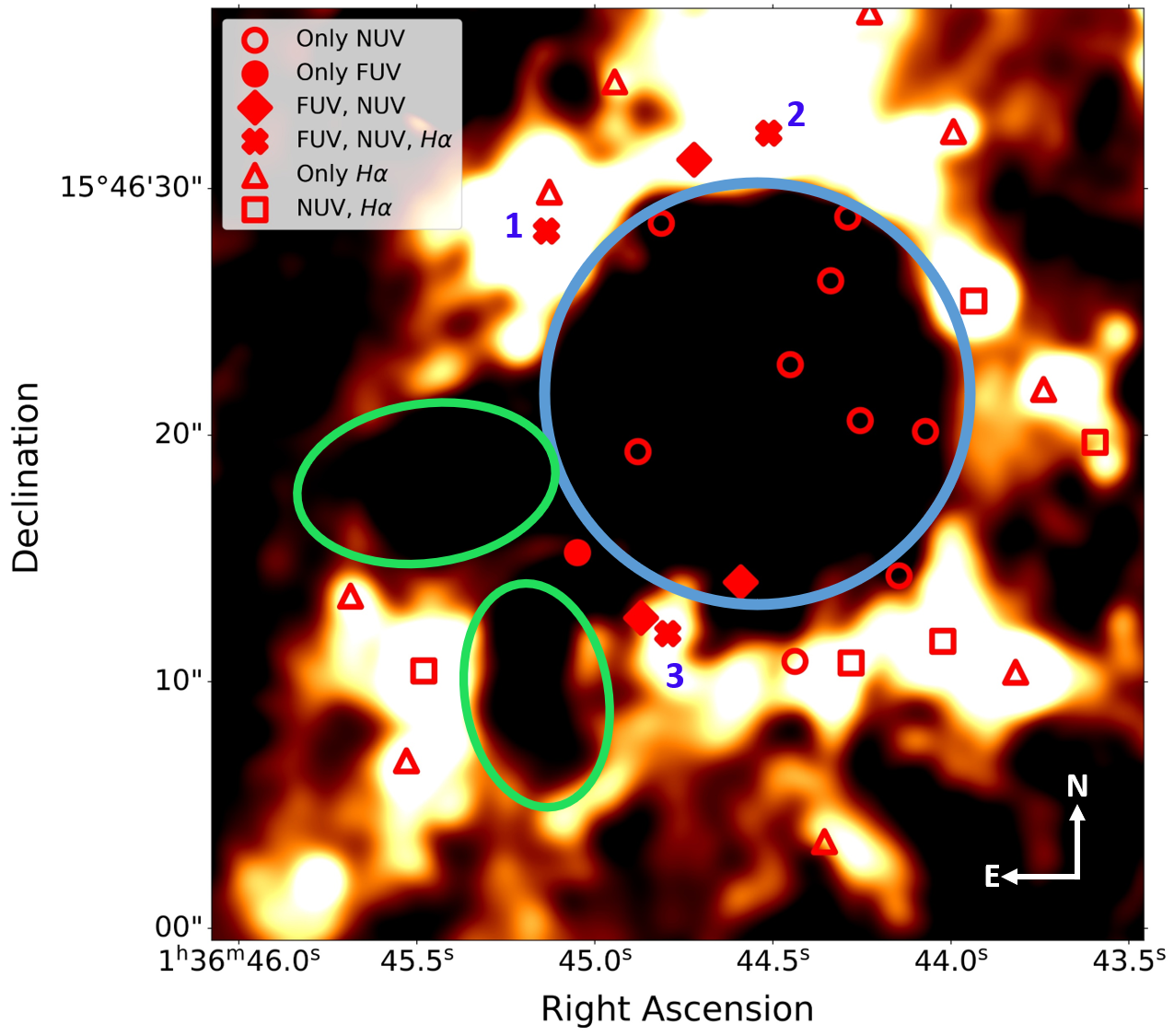}
\caption{The phantom void region of NGC 628. The markers are overlaid on the continuum-subtracted JWST F335M image. The blue superbubble is the phantom void, and the two green ellipses represent the connecting bubbles. The red markers represent the position of SFCs emitting at different wavelengths. The SFCs with both UV and H$\alpha$ emission are numbered.}
\label{fig: void}
\end{figure}

\subsection{The variation of PAH band ratios with the IMF for the SFCs} \label{The IMF and PAH correlation in SFCs}

The determination of the IMF indices follows A\&D25, where it was assumed that H$\alpha$ emission originates solely from O-type stars. Using this assumption, an approximate number of O stars was estimated. The number of B stars was then estimated from UV emission. From the O star to massive B stars ($M_{*} > 10 M_{\odot}$; B0V type stars) ratio, we derived the IMF index $\alpha_1$, and from the ratio of massive B stars to less massive B stars ($10 M_{\odot} \geq M_{*} \geq 3 M_{\odot}$), the IMF index $\alpha_2$ was estimated. The $\alpha_1$ IMF index represents the present-day IMF (relatively high-mass IMF), and $\alpha_2$, derived from the ratio of B stars, can provide a better estimate of the stellar IMF. In A\&D25, $\alpha_2$ showed strong trends with properties such as stellar mass and gas mass density. There was also a prominent correlation of  $\alpha_2$ with UV colour and star formation rate density. Hence, in this section, we use these IMF values to examine how the properties of PAH molecules change under different IMF conditions (e.g., when SFCs are top-light or top-heavy).

Both statistical and systematic uncertainties are associated with the estimation of these IMF indices. Statistical uncertainties are observed due to the stochastic sampling of O stars in the multiple realisations and photometric uncertainties in the measured fluxes. Systematic uncertainties arise from assumptions about the stellar types contributing to the H$\alpha$ emission (e.g., O8 only, O7–O9, and O3–O9 populations), and considering different O star combinations results in variations in the derived IMF $\alpha_1$ slope of up to $\sim$ 1. This reflects the strong sensitivity of the inferred $\alpha$ to the assumed massive-star population. The values, especially $\alpha_1$, depend on the mass limits we take. Although $\alpha_2$ depends on $\alpha_1$, both statistical and systematic uncertainties are very small compared to $\alpha_1$. Hence, $\alpha_2$ is a more robust indicator of the approximate IMF value.

As explained in Section \ref{sec:intro}, the relative strength of the PAH band ratios gives us information about the size and charge of the PAH molecules in that environment. An increase in the F335M/F1130W ratio likely indicates that the fraction of small PAH molecules is increasing. Similarly, the fraction of ionised molecules increases with an increase in F770W/F1130W. Recent studies, such as \cite{2021MNRAS.504.5287R} and \cite{2024MNRAS.532.1598R}, use diagnostic grids of PAH band ratios derived using theoretically computed PAH spectra by varying the numbers of carbon atoms, charge, and intensity of exposed radiation fields. These studies reveal the number of carbon atoms in the PAH molecule, the charge of the ISM, and information on what kind of radiation field the molecules are exposed to. Our study uses some of these results to explore the IMF variation with PAH band ratios.

\begin{figure*}
\includegraphics[scale=0.52]{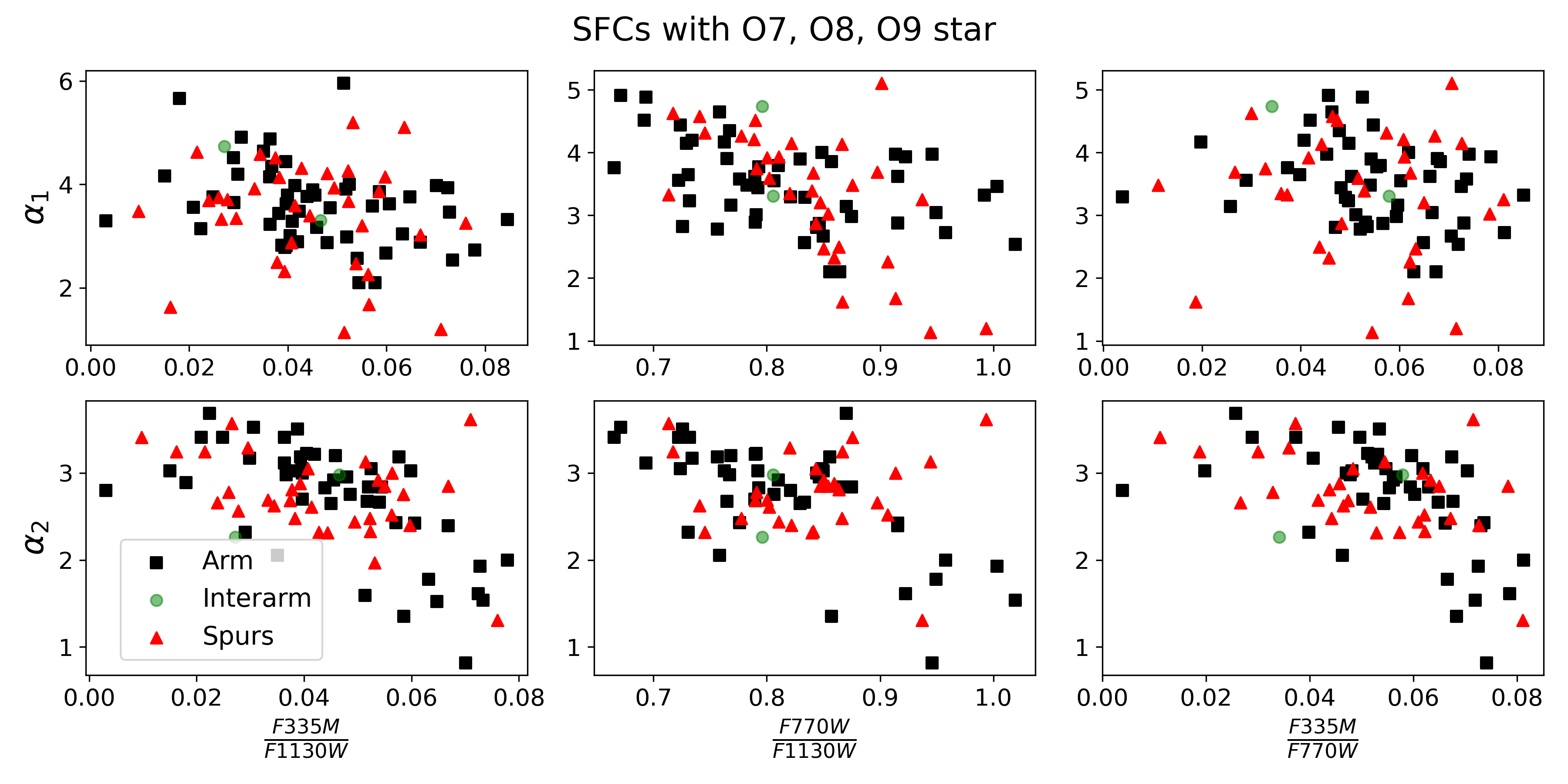}
\includegraphics[scale=0.52]{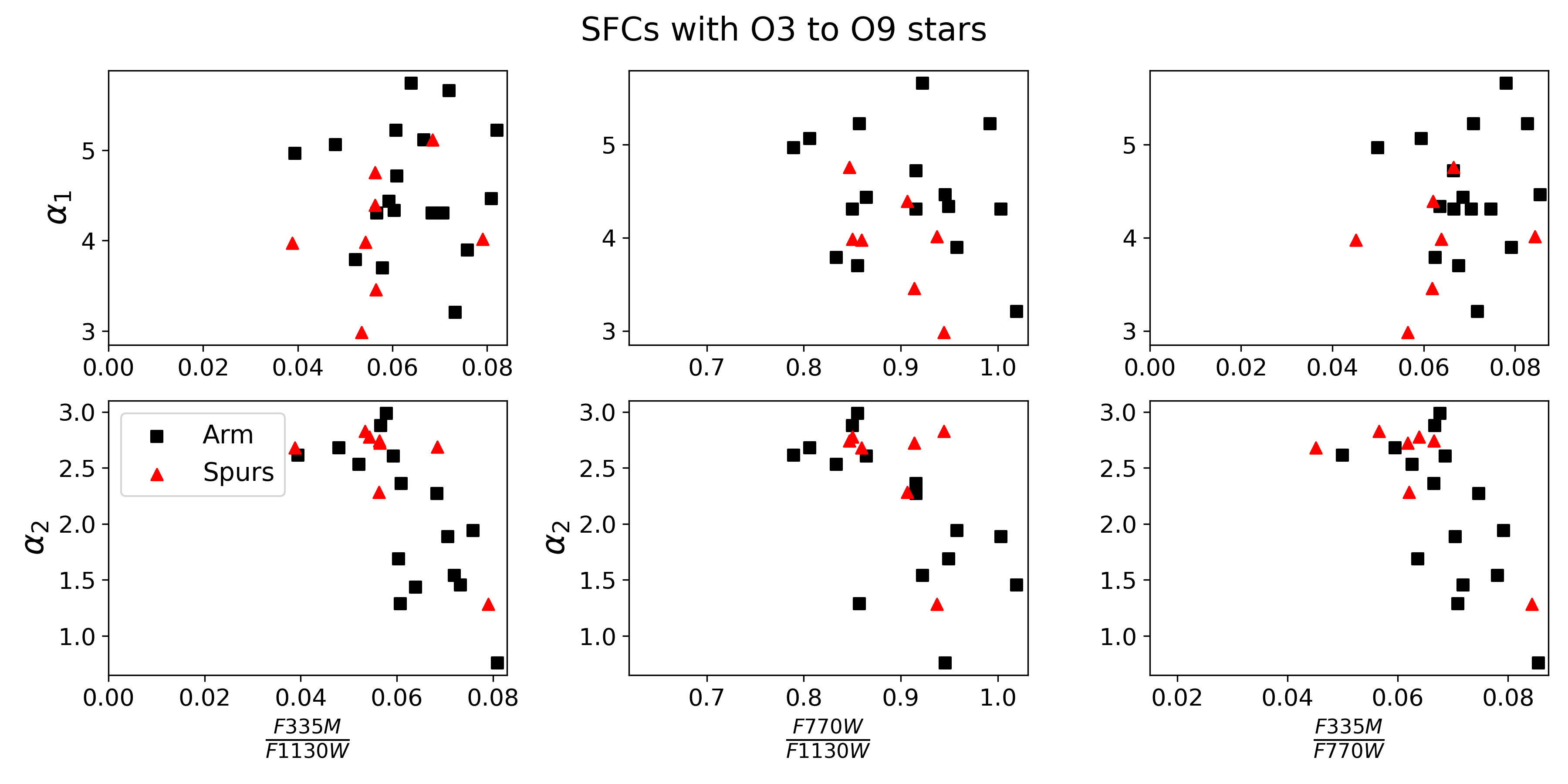}

\caption{Variation of IMF index ($\alpha_{1}$ and $\alpha_{2}$) with the PAH band ratios - F335M/F1130W, F770W/F1130W, and F335M/F770W (from left). Black square, red triangle and green circles represents arm, spur and interarm SFCs respectively.}
\label{fig:IMF}
\end{figure*}

We plotted $\alpha_1$ and $\alpha_2$ as a function of PAH band ratios (F335M/F1130W, F770W/F1130W, F335M/F770W) (Figure~\ref{fig:IMF}). We find no significant trend with $\alpha_1$, which remains relatively flat across the range of PAH band ratios. In contrast, $\alpha_2$ exhibits a noticeable trend with PAH band ratios. For arm SFCs, both F335M/F1130W and F770W/F1130W increase as the IMF becomes more top-heavy. However, $\alpha_2$ remains relatively flat for spur SFCs, clumps around values of 2.5-3.5 over a broad range of PAH ratios. The increasing PAH band ratios in arm SFCs are consistent with a relatively enhanced contribution from smaller and ionised PAH molecules, likely driven by stronger radiation fields associated with more top-heavy IMFs. Nonetheless, spur SFCs that exhibit a top-heavy IMF in the $\alpha_1$ PAH relation also tend to show an enhanced contribution from smaller and ionised PAH molecules.

We retained only SFCs with higher PAH ratios for the case where we populated SFCs with O3 to O9 type stars, as these correlate with high H$\alpha$ luminosity and can host massive stars (O stars up to 100 M$_{\odot}$). Hence, this shows that SFCs with higher H$\alpha$ luminosity exhibit a relatively enhanced contribution from smaller and ionised PAH molecules. However, Figure \ref{fig:IMF} shows that higher H$\alpha$ luminosity does not necessarily imply a flatter IMF.

Besides the correlation of IMF with PAH ratios, we investigated the IMF of SFCs in the shell region around the phantom void. Unfortunately, we have only three SFCs with H$\alpha$, FUV, and NUV emission. We named these SFCs SFC 1, 2, and 3, as shown in Figure \ref{fig: void}. The SFCs 1 and 2 have $\alpha_1$ (present-day mass function) value flatter than SFC 3, which is near the connecting elongated bubble. For the case where O7-O9V stars are populated, SFCs 1 and 2 have $\alpha_1$ values of 3.17$^{+0.5}_{-0.4}$ and 2.48$^{+0.6}_{-0.5}$, respectively, while SFC 3 has an $\alpha_1$ value of 4.57$^{+0.4}_{-0.3}$. Although the values differ across all three cases, we see the same trend. Hence, at the present day, SFC 1 and 2 host more massive stars than SFC 3. The mean $\alpha_2$ for three SFCs is 2.79 $\pm$ 0.09 for all the cases. We discuss the IMF variation with PAH band ratios, and the IMF of SFCs in the shell region around the phantom void, in more detail  in Section \ref{sec:discussion}.

\subsection{The variation of PAH band ratios with the colour for the SFCs} \label{The colour and PAH correlation in SFCs}

\begin{figure*}
\includegraphics[scale=0.52]{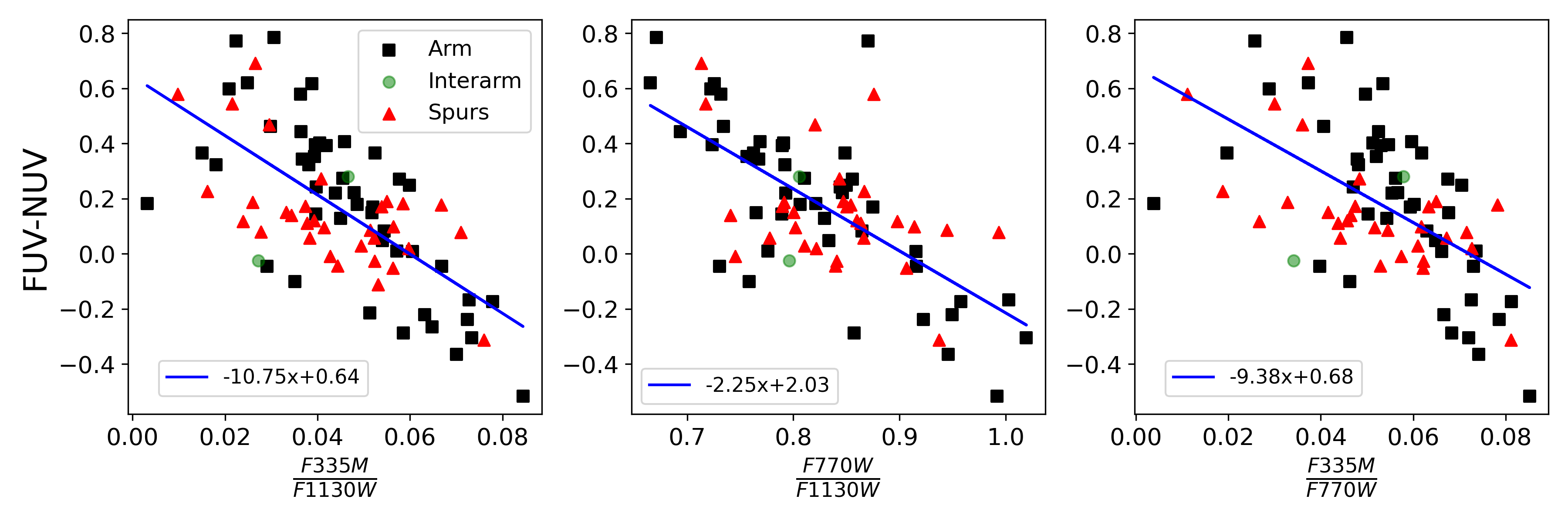}
\caption{Variation of FUV-NUV with the PAH band ratios - F335M/F1130W, F770W/F1130W, and F335M/F770W (from left).}
\label{fig: colour}
\end{figure*} 

A prominent correlation was observed for $\alpha_2$ with UV colour FUV-NUV in A\&D25. It followed a quadratic fit, and FUV-NUV was red for the SFCs with top-light IMF and blue for SFCs with top-heavy IMF, indicating more recent star formation in the latter. Therefore, we plotted FUV-NUV as a function of PAH band ratios (F335M/F1130W, F770W/F1130W, F335M/F770W) (see Figure \ref{fig: colour}) for the case with a larger number of SFCs. We fitted linear regression for each ratio as given below.

\begin{equation} \label{eq3}
\mathrm{(FUV-NUV)=Ck+D}
\end{equation}

Where C and D are fitting parameters given in the Table \ref{table2}. k is the PAH band flux ratio. As expected, we observe the general trend of F335M/F1130W, F770W/F1130W, and F335M/F770W increasing as SFCs get bluer. However, in Figure \ref{fig: colour}, especially for ratios F335M/F1130W and F335M/F770W, we found that more spur SFCs are below the fit than arm in the range $0.1>$ FUV-NUV$>-1$. This indicates that for the same colour, the relative contribution from smaller molecules is higher for arms SFCs than for spurs.

\begin{table}
\caption{Fitting parameters for Equation \ref{eq3}.}
\begin{center}
\begin{tabular}{|c|cccc|}
\hline

k   &C &D&$\sigma$   & $\rho$    \\
\hline
$F330M/F1130W$    & -14.02 $\pm$ 1.71        & 0.67 $\pm$ 0.06& 0.19         & -0.64                   \\
$F770W/F1130W$     &-2.95 $\pm$ 0.34       & 2.07 $\pm$ 0.21 & 0.19          & -0.63              \\
$F330M/F770W$ & -9.38 $\pm$ 1.51          & 0.68 $\pm$ 0.08& 0.22         & -0.58             \\
\hline

\end{tabular}
\end{center}
\label{table2}
\centering
 \begin{minipage}{80 mm}
    \textit{Note:} C and D are the fitting parameters for the Equation \ref{eq3}, which represents the correlation between FUV-NUV and PAH band ratios as shown in the Figure \ref{fig: colour}. $\sigma$ and $\rho$ represent the standard deviation and Spearman's correlation coefficient, respectively.
    \end{minipage}%
\end{table}

\section{Discussion}\label{sec:discussion}

Although PAH emission traces star formation, its strength depends on the nature of the source exciting the molecules and the local environment. For example, the median PAH luminosity is lower in interarm regions than in spiral arms and spurs, where the star formation efficiency remains lower than in the arms, even though molecular hydrogen gas is present \citep{Foyle_2010, Kreckel_2016}. Besides, studies have reported a central deficit of PAH emission in unbarred galaxies, attributed to the lack of continuous gas inflow typically driven by a bar \citep{Regan_2006}. Consistent with this, we find a compact peak of PAH emission at the nucleus of NGC~628, surrounded by a region of relatively faint PAH emission extending over $\sim$0.1 kpc. Such limited gas inflow may suppress star formation, as indicated by the presence of NUV emission from an evolved stellar population and the absence of FUV emission associated with ongoing star formation (Figure~\ref{fiG:multi-wavelength_images}).

In addition to the interarm and nuclear regions, bubbles and superbubbles in NGC 628 also exhibit a deficit of PAH emission \citep{Barnes_2023, Mayya2023}. Within these regions, we detected several NUV-only emitting SFCs. Lowering the FUV emission threshold increases the total number of FUV-emitting SFCs by a factor of 2.4, with a fourfold increase in the Phantom void (8 SFCs). The FUV-emitting SFCs within bubbles have minimal overlap with NUV-only emitting SFCs and are primarily located along the shell regions. The presence of NUV-only emitting SFCs suggests the presence of relatively evolved stellar populations and other intermediate-age stars. A few massive stars, such as O-type and massive B stars (as inferred from the FUV), are also found in these regions and may contribute to the formation and powering of these bubbles and superbubbles, while the most massive stars may have already evolved. This is consistent with \cite{Barnes_2023}, who reported several NUV emission peaks within the Phantom void, along with compact stellar emission sources from HST and AstroSat NUV observations. They also found diffuse H$\alpha$ within the void and associated it with ionisation by an older generation of stars. This explains the absence of distinct H$\alpha$ SFCs in these regions. However, the detection of FUV and the absence of H$\alpha$ SFCs suggest that the ionised gas has been expelled, likely due to feedback from massive stars or SNe.

Interestingly, bubbles in the northern arm have few NUV-only-emitting SFCs and lack other star formation tracers, suggesting they are older than those in the southern arm and show no recent star formation. This may be due to stellar feedback from evolved massive stars that has cleared the cold gas needed for star formation. Studies such as \cite{10.1093/mnras/stt1019, 10.1093/mnras/stt1970} have also confirmed the drastic difference in the inner structures between the spiral arms of NGC 628, proposing that a shock wave along the northern arm creates high pressure and suppresses star formation, while a regular magnetic field stabilises star formation along the southern arm. Alternatively, these holes may arise from galactic dynamical processes, such as spur or feathering instabilities and shear driven by the galaxy's rotation \citep{Watkins_2023}.  

The prominent superbubble in the southern arm, the phantom void, has a radius of around 600 pc \citep{Watkins_2023}. Two additional bubbles appear to have merged with it (green ellipses in Figure \ref{fig: void}). Such superbubbles likely form from the merging of smaller bubbles and are powered by stellar feedback from massive stars and successive SNe. Given their short lifetimes, low-mass massive stars are expected to remain within them \citep{Mayya2023, Jiménez_2024}. Therefore, we examined SFCs within the phantom void and found low luminosity NUV-only emitting SFCs in it (See Figure \ref{fig: NUV_correlation}). The faint FUV-only-emitting SFCs (detected only after reducing the threshold) are detected between the phantom void and the adjacent bubbles, while the connecting bubbles themselves show no detectable emission in these tracers. It is likely that they are older structures currently powered by massive stars located between them and the phantom void. 

There is also a clear age-dependent spatial segregation in the stellar population. \cite{Mayya2023} observed the youngest population at larger radii up to a radius of 450 pc and is moving away from the centre as the bubble expands. As in theirs, we also see that FUV-emitting SFCs are far from the bubble centre and the older population, suggesting star formation triggered in the swept-up material. However, a small number of O and B stars formed within the bubble could explain the detection of faint SFCs.

There is clear evidence of star formation in the shell of the phantom void (as seen in Figure \ref{fig: void}) due to positive feedback consistent with \cite{Mayya2023}. Figure \ref{fig: void} shows that most of the SFCs in the shell have NUV and H$\alpha$ emission, and there are very few FUV-emitting SFCs. However, we detected a few additional FUV SFCs when we lowered the threshold. Since H$\alpha$ emission confirms the presence of young massive stars, the relative deficit of FUV detections may therefore reflect both the shallowness of the FUV imaging and possible attenuation of FUV emission by dust. However, PAH emission is very bright throughout this region, indicating enough PAH molecules and radiation sources to excite the PAHs.

Out of three SFCs with emission detected in both UV and H$\alpha$, SFCs 1 and 2 exhibit flatter IMF slopes compared to SFC 3. This difference may be related to the formation and dynamical evolution of the structure. As shown in Figure \ref{fig: void}, the phantom void may have formed through the merging of bubbles and/or been influenced by galactic shear. The elongated morphology of the connecting bubbles indicates the influence of shear forces, while the phantom void near SFCs 1 and 2 (northeast of the superbubble) remains nearly circular, implying weaker shear in that region. Furthermore, \cite{Barnes_2023} identified supernova remnants mainly in the northwestern shell of the superbubble, which may contribute to its continued expansion. The flatter IMF slopes observed in SFCs 1 and 2 could therefore result from higher gas densities in the swept-up shell of the expanding superbubble. In contrast, the molecular gas near SFC 3 may have experienced stronger shear, reducing the gas density and the potential for forming massive stars. Consistent with this interpretation, the ALMA map in Figure \ref{fiG:multi-wavelength_images} shows little molecular gas in that region.

\begin{figure*}
\includegraphics[scale=0.57]{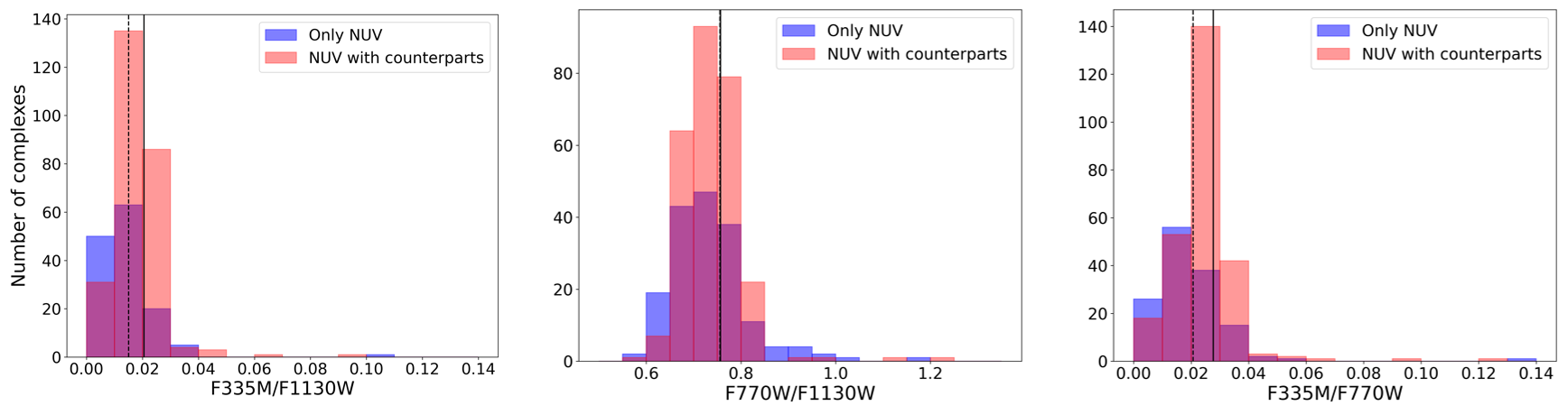}
\caption{The distribution of PAH band ratios (F335M/F1130W, F770W/F1130W, and F335M/F770W) for only NUV-emitting SFCs and NUV-emitting SFCs with counterparts in UV and/or H$\alpha$.}
\label{fig: onlynuv_prop}
\end{figure*}

To obtain information on the molecule size and the ionisation state of the PAH molecules in different environments, we compared PAH band ratios (F335M/F1130W, F770W/F1130W, and F335M/F770W) between NUV-only emitting SFCs and those emitting in NUV that also have counterparts in at least H$\alpha$ or FUV. This is to examine if there is any difference in PAH molecule properties, as shown in Figure \ref{fig: onlynuv_prop}. We found no significant difference in the median F770W/F1130W ratio; therefore, there is no difference in the charge of PAH molecules between NUV-emitting SFCs with or without counterparts. However, the median F335W/F1130W and F335M/F770W ratio for the NUV emitting SFCs with counterparts is slightly higher than for the NUV-only emitting SFCs, implying a higher fraction of small molecules in the NUV emitting SFCs with counterparts. If we say that NUV-emitting SFCs with counterparts in other bands host massive stars, whereas others do not, then here we see that the presence of massive stars slightly influences the size of molecules in the ISM, not their charge. 

We also checked the IMF and UV colour trends against PAH band ratios for arms and spurs. As seen in A\&D25, only the IMF index $\alpha_{2}$, obtained from the B-star ratio, correlates with PAH band ratios. The trend for cases where we populated H$\alpha$ SFCs with O7, O8, and O9, and also with only O8 stars, is similar. Hence, we only present the former in Figure \ref{fig:IMF}. Unlike previous observations in A\&D25, here we observe differences in the trends for the arm and spurs. We find that spurs have only a few SFCs with top-heavy IMF ($\alpha_{2}<2.3$) compared to arm SFCs. Consequently, the spurs display a flat trend, mostly clumped between 2 and 4, while arms show a decreasing trend in IMF with increasing PAH ratios. The lack of SFCs with a top-heavy IMF in the spurs could be due to low molecular gas relative to the arm. Interestingly, we also found that, for SFCs with the same UV colour, the fraction of small PAHs is higher in the arms than in the spurs. According to \cite{Schinnerer_2017}, the star formation is biased towards spurs rather than arms in galaxy M51, and suggested that the young clusters in spurs are larger and more massive than what we see in the arms. IMF index $\alpha_{2}$ in A\&D25 is derived from the ratios of massive to less massive B stars, whereas $\alpha_{1}$ is from the ratios of the number of O stars to massive B stars. In Figure \ref{fig:IMF}, we see that many spur SFCs are top-heavy if we consider $\alpha_{1}$, in contrast to $\alpha_{2}$. Hence, there might be more massive young stars in spurs than in arms, as mentioned in \cite{Schinnerer_2017}. The study has also shown that star formation in spurs is independent of the spiral arm. However, it is preferred towards the arm within the spurs.

Previous JWST studies of NGC 628 have shown that the star-forming regions or HII regions with a high star formation rate density exhibit elevated values of F335M/F1130W, F770W/F1130W, and F335M/F770W, which are commonly associated with enhanced contributions from small and ionised PAHs \citep{2023ApJ...944L..12C, 2023ApJ...944L..16E, ujjwal}. As in previous studies, we also observed that as the IMF for arm SFCs becomes top-heavy and blue (indicating recent star formation), the PAH band ratios F335M/F1130W, F770W/F1130W and F335M/F770W increase, consistent with enhanced excitation and ionisation of PAHs in stronger UV radiation fields. The SFCs with top-heavy IMF host more massive stars and hence higher UV luminosity. The intense UV radiation density from the compact PDRs in these regions can dominate the particle density, affecting recombination and ionisation rates. This leads to an increase in the number of charged particles in these regions \citep{2021ApJ...908..238M, 2024MNRAS.532.1598R}. There will also be an increase in the fraction of shocked gas to total gas in such a high UV radiation region, which can increase the relative contribution of ionised PAHs to the observed emission \citep{2023ApJ...944L..16E, 2023ApJ...944L..12C}. In contrast, theoretically, small molecules will not survive longer in such an extreme environment \citep{2010A&A...510A..37M}. However, numerical models show that shocks in dense clouds can cause gas-grain collisions, grain-grain collisions, and large-grain fragmentation, which in turn increases the fraction of small molecules \citep{1996ApJ...469..740J, 2011A&A...527A.123G}. It is also known that stronger UV radiation fields increase PAH temperatures, leading to enhanced emission at shorter wavelengths \citep{2021ApJ...917....3D}. Recent studies suggest that variations in these ratios reflect changes in the local radiation field more strongly than changes in the intrinsic PAH size distribution \citep{Baron_2024, 2025ApJ...991...76H}.

Studies have shown that the properties and emission characteristics of PAH molecules vary with the age of stellar clusters and associations. \cite{Whitmore_2025} reported a rapid decline in PAH emission and infrared dust emission with increasing age, which they attributed to dust grain destruction and stellar feedback processes. This provides a natural explanation for the lower PAH luminosities observed in the NUV-only emitting SFCs (Figure \ref{fig: KDE} (ii)). Furthermore, \cite{2025AJ....169..133D} found that the F335M/F1130W and F335M/F770W flux ratios decrease with increasing age. This behaviour is expected because older regions are characterised by softer radiation fields, which reduce PAH excitation and maintain lower dust temperatures \citep{2021ApJ...917....3D}. Our results, shown in Figures \ref{fig: colour} and \ref{fig: onlynuv_prop}, are consistent with this picture, as NUV-emitting SFCs exhibit redder UV colours and lack associated FUV and H$\alpha$ emission, both of which are indicative of more evolved stellar populations.

\cite{2022MNRAS.509.1303W}  demonstrated that the inner disk of NGC 628 shows minimal metallicity variation (8.45$<$ 12+log(O/H)$<$8.6). Therefore, properties of PAHs in different regions of the inner disk should not be affected by metallicity. However, trends can vary if there is a contamination by continuum. Though the F335M is continuum subtracted, as discussed in  Section \ref{Data}, \cite{Sandstrom_2023} has addressed that the aliphatic PAH at 3.4 $\mu$m and plateau at 3.47 $\mu$m might also contribute slightly to the detected 3.3 $\mu$m after continuum subtraction. However, their strength are too low \cite[8$\%$,][]{Lai_2020} compared to the 3.3 $\mu$m PAH emission to have a significant effect on the observed trends. Even with the F770W band, approximately 80$\%$ of the emissions are from  7.7 $\mu$m PAHs, and 70$\%$ of the emission seen in the filter F1130W is due to 11.3 $\mu$m PAH molecules \citep{Whitcomb_2023}. The continuum-subtraction prescription of \cite{2025ApJ...983...79D} applied to the F770W and F1130W filters yields PAH flux estimates with typical accuracies of $\sim$7--8\%. Therefore, continuum contamination is unlikely to be the primary driver of the observed regional variations in PAH emission.

Several star cluster catalogues exist for NGC 628 that resolve clusters on scales of a few tens of parsecs, such as Legacy ExtraGalactic Ultraviolet Survey (LEGUS; \cite{2015AJ....149...51C}) and PHANGS–HST \cite[]{Lee2022,2022MNRAS.509.4094T, Thilker_2025}. These catalogues identify compact star clusters and associations using high-resolution HST imaging. In contrast, due to UVIT's lower spatial resolution, our analysis detects larger stellar associations (OB associations and larger complexes). While using HST-based cluster catalogues would probe smaller spatial scales and may introduce some variation in the relations, the overall SFC–PAH trends are expected to remain broadly consistent with those found in this work.

 By comparing PAH emissions with other star formation indicators at a resolved scale of 57 pc, we aimed to gain insights into the properties of these SFCs and their surrounding environments. We also observed variations in star formation properties across different regions (arms and spurs) in NGC 628 at such small scales. This highlights the significance of examining star formation on a local level.

\section{Summary} \label{sec:Summary}

In this work, we used PAH maps (F335M, F770W, and F1130W) from JWST observations of NGC 628. We also utilised the catalogue of star formation properties of SFCs of NGC 628 from A\&D25, which includes H$\alpha$ and UV (FUV and NUV) luminosities from  MUSE and UVIT, respectively. We used these luminosities to get a correlation with PAH at a scale of 57 pc. We also analysed the PAH emission at the different regions (arm, spurs, interarms, and bubbles) of NGC 628, and the effect of the environment on PAH emission. From the A\&D25 catalogue, we used the IMF of the SFCs and derived some important trends with the PAH band ratios (F335M/F1130W, F770W/F1130W, and F335M/F770W) to understand the PAH size and ionised state in arms and spurs. We summarise the main results of our study below,

1. The PAH luminosities show a strong correlation (Spearman's correlation coefficient, $\rho > 0.73 $) with FUV, NUV, and H$\alpha$ luminosities for the extracted SFCs, even at the scale of SFCs (size $>$ 57 pc). 

2. The 7.7 and 11.3 $\mu$m PAH luminosities exhibit a flat linear fit compared to the linear fit with the 3.3 $\mu$m luminosity, suggesting that the 3.3 $\mu$m feature is more sensitive to variations in recent star formation.

3. The PAH emission is generally well associated with SFCs located in the spiral arms and spurs. In contrast, SFCs located within bubbles and superbubbles, including both the predominantly NUV-only SFCs and the few FUV-emitting SFCs, generally lack associated PAH emission. The predominance of NUV-only SFCs in these regions suggests that they are mainly populated by low-mass B stars, A stars, and a few evolved stars. The presence of a small number of FUV-emitting SFCs indicates that some O and massive B stars are still present, potentially contributing to the ongoing expansion of bubbles and superbubbles.

4. PAH molecules in only NUV-emitting SFCs and in NUV-emitting SFCs with counterparts in at least H$\alpha$ or FUV exhibit similar charge states, as indicated by their comparable median F770W/F1130W ratios. However, the median F335M/F1130W and F335M/F770W ratios are higher in NUV-emitting SFCs with counterparts than in only NUV-emitting SFCs. This suggests either enhanced emission from small PAH molecules or more efficient excitation of the 3.3 $\mu$m PAH feature in the former population. Conversely, the lower ratios in only NUV-emitting SFCs may be consistent with softer radiation fields that reduce PAH excitation.

5. There is a clear age-dependent spatial segregation in the stellar population in the phantom void, consistent with the previous study. The youngest population emitting FUV is at larger radii, far from the bubble's centre, and the older population. This segregation suggests that the recent star formation must have been triggered in the material swept by the expanding bubble. 

6. Two SFCs located in the shell region of the phantom void show flatter IMF slopes compared to an SFC near the elongated bubble. This difference may arise from gas accumulation driven by stellar feedback in the shell region, whereas shear in the elongated bubble may hinder the formation of more massive stars.

7. IMF index $\alpha_2$ derived from the ratio of B stars and UV colour ($\mathrm{FUV-NUV}$), shows a decreasing trend with PAH band ratios (F335M/F1130W, F770W/F1130W, and F335M/F770W). These elevated ratios are commonly associated with enhanced contributions from small and ionised PAHs, although recent studies suggest they may primarily reflect variations in the local radiation field. For a given $\mathrm{FUV-NUV}$ colour, arm SFCs exhibit systematically higher PAH band ratios than spur SFCs, indicating stronger PAH excitation and/or different ISM environments in the arm regions.

8. Spur SFCSs have a flat trend with PAH band ratios compared to the arm SFCs. This is because of fewer spur SFCs with a top-heavy IMF.

\section{acknowledgments}
We thank the anonymous referee for their careful reading of the manuscript and for their constructive comments and suggestions, which significantly improved the clarity and quality of this work. DR and MD acknowledge the support of "The Royal Society Yusuf Hamied International Exchange Award”, 
U.K. MD also gratefully acknowledges the support of the Department of Science and Technology (DST) grant DST/WIDUSHI-A/PM/PM/2023/25(G) for this research. JY acknowledges the support from the Spanish Ministry of Science and Innovation, project PID2022-136598NBC31 (ESTALLIDOS8) by MCIN/AEI/10.13039/501100011033.
This publication uses data from UVIT, which is part of the AstroSat mission of the Indian Space Research Organisation (ISRO) and is archived at the Indian Space Science Data Centre (ISSDC). We gratefully thank all the members of various teams for supporting the project from the early stages of design to launch and observations in orbit. 

This work is based on observations made with the NASA/ESA/CSA JWST. The data were obtained from the Mikulski Archive for Space Telescopes at the Space Telescope Science Institute, which is operated by the Association of Universities for Research in Astronomy, Inc., under NASA contract NAS 5-03127. The observations are associated with the JWST program 2107.  Based on observations collected at the European Southern Observatory under ESO programs 094.C-0623 (PI: Kreckel), 095.C-0473, 098.C-0484 (PI: Blanc), and 1100.B-0651 (PHANGS-MUSE; PI: Schinnerer), as well as 094.B-0321 (MAGNUM; PI: Marconi), 099.B-0242, 0100.B-0116, 098.B-0551 (MAD; PI: Carollo), and 097.B-0640 (TIMER; PI: Gadotti).  

\section*{Data Availability}

All the UVIT data used in this paper are publicly available at \url{https://astrobrowse.issdc.gov.in/astro_archive/archive/Home.jsp.}

The specific JWST observation analysed can be accessed via \url{https://doi.org/10.17909/76c4-r832}.



\bibliographystyle{mnras}
\bibliography{example} 

@ARTICLE{1996A&AS..117..393B,
       author = {{Bertin}, E. and {Arnouts}, S.},
        title = "{SExtractor: Software for source extraction.}",
      journal = {\aaps},
         year = "1996",
        month = "Jun",
       volume = {117},
        pages = {393-404},
          doi = {10.1051/aas:1996164},
       adsurl = {https://ui.adsabs.harvard.edu/abs/1996A&AS..117..393B}
}

@ARTICLE{2007ApJS..173..538T,
       author = {{Thilker}, David A. and {Bianchi}, Luciana and {Meurer}, Gerhardt and {Gil de Paz}, Armando and {Boissier}, Samuel and {Madore}, Barry F. and {Boselli}, Alessandro and {Ferguson}, Annette M.~N. and {Mu{\~n}oz-Mateos}, Juan Carlos and {Madsen}, Greg J. and {Hameed}, Salman and {Overzier}, Roderik A. and {Forster}, Karl and {Friedman}, Peter G. and {Martin}, D. Christopher and {Morrissey}, Patrick and {Neff}, Susan G. and {Schiminovich}, David and {Seibert}, Mark and {Small}, Todd and {Wyder}, Ted K. and {Donas}, Jos{\'e} and {Heckman}, Timothy M. and {Lee}, Young-Wook and {Milliard}, Bruno and {Rich}, R. Michael and {Szalay}, Alex S. and {Welsh}, Barry Y. and {Yi}, Sukyoung K.},
        title = "{A Search for Extended Ultraviolet Disk (XUV-Disk) Galaxies in the Local Universe}",
      journal = {\apjs},
         year = 2007,
        month = dec,
       volume = {173},
       number = {2},
        pages = {538-571},
          doi = {10.1086/523853},
archivePrefix = {arXiv},
       eprint = {0712.3555},
 primaryClass = {astro-ph},
       adsurl = {https://ui.adsabs.harvard.edu/abs/2007ApJS..173..538T}
}

@ARTICLE{2009ApJ...703.1672K,
       author = {{Kennicutt}, Robert C., Jr. and {Hao}, Cai-Na and {Calzetti}, Daniela and {Moustakas}, John and {Dale}, Daniel A. and {Bendo}, George and {Engelbracht}, Charles W. and {Johnson}, Benjamin D. and {Lee}, Janice C.},
        title = "{Dust-corrected Star Formation Rates of Galaxies. I. Combinations of H{\ensuremath{\alpha}} and Infrared Tracers}",
      journal = {\apj},
         year = 2009,
        month = oct,
       volume = {703},
       number = {2},
        pages = {1672-1695},
          doi = {10.1088/0004-637X/703/2/1672},
archivePrefix = {arXiv},
       eprint = {0908.0203},
 primaryClass = {astro-ph.CO},
       adsurl = {https://ui.adsabs.harvard.edu/abs/2009ApJ...703.1672K}
}

@article{Hao_2011,
doi = {10.1088/0004-637X/741/2/124},
url = {https://dx.doi.org/10.1088/0004-637X/741/2/124},
year = {2011},
month = {oct},
publisher = {The American Astronomical Society},
volume = {741},
number = {2},
pages = {124},
author = {Cai-Na Hao and Robert C. Kennicutt and Benjamin D. Johnson and Daniela Calzetti and Daniel A. Dale and John Moustakas},
title = {DUST-CORRECTED STAR FORMATION RATES OF GALAXIES. II. COMBINATIONS OF ULTRAVIOLET AND INFRARED TRACERS},
journal = {\apj}
}

@article{Calzetti_2007,
doi = {10.1086/520082},
url = {https://dx.doi.org/10.1086/520082},
year = {2007},
month = {sep},
publisher = {The American Astronomical Society},
volume = {666},
number = {2},
pages = {870},
author = {D. Calzetti and R. C. Kennicutt and C. W. Engelbracht and C. Leitherer and B. T. Draine and L. Kewley and J. Moustakas and M. Sosey and D. A. Dale and K. D. Gordon and G. X. Helou and D. J. Hollenbach and L. Armus and G. Bendo and C. Bot and B. Buckalew and T. Jarrett and A. Li and M. Meyer and E. J. Murphy and M. Prescott and M. W. Regan and G. H. Rieke and H. Roussel and K. Sheth and J. D. T. Smith and M. D. Thornley and F. Walter},
title = {The Calibration of Mid-Infrared Star Formation Rate Indicators*},
journal = {\apj}
}

@article{10.1093/mnras/staa3668,
    author = {Anand, Gagandeep S and Lee, Janice C and Van Dyk, Schuyler D and Leroy, Adam K and Rosolowsky, Erik and Schinnerer, Eva and Larson, Kirsten and Kourkchi, Ehsan and Kreckel, Kathryn and Scheuermann, Fabian and Rizzi, Luca and Thilker, David and Tully, R Brent and Bigiel, Frank and Blanc, Guillermo A and Boquien, Médéric and Chandar, Rupali and Dale, Daniel and Emsellem, Eric and Deger, Sinan and Glover, Simon C O and Grasha, Kathryn and Groves, Brent and S. Klessen, Ralf and Kruijssen, J M Diederik and Querejeta, Miguel and Sánchez-Blázquez, Patricia and Schruba, Andreas and Turner, Jordan and Ubeda, Leonardo and Williams, Thomas G and Whitmore, Brad},
    title = "{Distances to PHANGS galaxies: New tip of the red giant branch measurements and adopted distances}",
    journal = {\mnras},
    volume = {501},
    number = {3},
    pages = {3621-3639},
    year = {2020},
    month = {11},
    issn = {0035-8711},
    doi = {10.1093/mnras/staa3668},
    url = {https://doi.org/10.1093/mnras/staa3668},
    eprint = {https://academic.oup.com/mnras/article-pdf/501/3/3621/35699965/staa3668.pdf},
}

@article{Lang_2020,
doi = {10.3847/1538-4357/ab9953},
url = {https://dx.doi.org/10.3847/1538-4357/ab9953},
year = {2020},
month = {jul},
publisher = {The American Astronomical Society},
volume = {897},
number = {2},
pages = {122},
author = {Philipp Lang and Sharon E. Meidt and Erik Rosolowsky and Joseph Nofech and Eva Schinnerer and Adam K. Leroy and Eric Emsellem and Ismael Pessa and Simon C. O. Glover and Brent Groves and Annie Hughes and J. M. Diederik Kruijssen and Miguel Querejeta and Andreas Schruba and Frank Bigiel and Guillermo A. Blanc and Mélanie Chevance and Dario Colombo and Christopher Faesi and Jonathan D. Henshaw and Cinthya N. Herrera and Daizhong Liu and Jérôme Pety and Johannes Puschnig and Toshiki Saito and Jiayi Sun and Antonio Usero},
title = {PHANGS CO Kinematics: Disk Orientations and Rotation Curves at 150 pc Resolution},
journal = {\apj}
}

@ARTICLE{2019ApJS..244...24L,
       author = {{Leroy}, Adam K. and {Sandstrom}, Karin M. and {Lang}, Dustin and {Lewis}, Alexia and {Salim}, Samir and {Behrens}, Erica A. and {Chastenet}, J{\'e}r{\'e}my and {Chiang}, I-Da and {Gallagher}, Molly J. and {Kessler}, Sarah and {Utomo}, Dyas},
        title = "{A z = 0 Multiwavelength Galaxy Synthesis. I. A WISE and GALEX Atlas of Local Galaxies}",
      journal = {\apjs},
         year = 2019,
        month = oct,
       volume = {244},
       number = {2},
          eid = {24},
        pages = {24},
          doi = {10.3847/1538-4365/ab3925},
archivePrefix = {arXiv},
       eprint = {1910.13470},
 primaryClass = {astro-ph.GA},
       adsurl = {https://ui.adsabs.harvard.edu/abs/2019ApJS..244...24L}
}

@article{Kreckel_2016,
doi = {10.3847/0004-637X/827/2/103},
url = {https://dx.doi.org/10.3847/0004-637X/827/2/103},
year = {2016},
month = {aug},
publisher = {The American Astronomical Society},
volume = {827},
number = {2},
pages = {103},
author = {K. Kreckel and G. A. Blanc and E. Schinnerer and B. Groves and A. Adamo and A. Hughes and S. Meidt},
title = {CHARACTERIZING SPIRAL ARM AND INTERARM STAR FORMATION},
journal = {\apj}
}

@article{Yadav_2021,
doi = {10.3847/1538-4357/abf8c1},
url = {https://dx.doi.org/10.3847/1538-4357/abf8c1},
year = {2021},
month = {jun},
publisher = {The American Astronomical Society},
volume = {914},
number = {1},
pages = {54},
author = {Jyoti Yadav and Mousumi Das and Narendra Nath Patra and K. S. Dwarakanath and P. T. Rahna and Stacy S. McGaugh and James Schombert and Jayant Murthy},
title = {Comparing the Inner and Outer Star-forming Complexes in the Nearby Spiral Galaxies NGC 628, NGC 5457, and NGC 6946 Using UVIT Observations},
journal = {\apj}
}

@article{Foyle_2010,
doi = {10.1088/0004-637X/725/1/534},
url = {https://dx.doi.org/10.1088/0004-637X/725/1/534},
year = {2010},
month = {nov},
publisher = {The American Astronomical Society},
volume = {725},
number = {1},
pages = {534},
author = {K. Foyle and H.-W. Rix and F. Walter and A. K. Leroy},
title = {ARM AND INTERARM STAR FORMATION IN SPIRAL GALAXIES},
journal = {\apj}
}

@article{Leroy_2021a,
doi = {10.3847/1538-4365/abec80},
url = {https://dx.doi.org/10.3847/1538-4365/abec80},
year = {2021},
month = {jul},
publisher = {The American Astronomical Society},
volume = {255},
number = {1},
pages = {19},
author = {Adam K. Leroy and Annie Hughes and Daizhong Liu and Jérôme Pety and Erik Rosolowsky and Toshiki Saito and Eva Schinnerer and Andreas Schruba and Antonio Usero and Christopher M. Faesi and Cinthya N. Herrera and Mélanie Chevance and Alexander P. S. Hygate and Amanda A. Kepley and Eric W. Koch and Miguel Querejeta and Kazimierz Sliwa and David Will and Christine D. Wilson and Gagandeep S. Anand and Ashley Barnes and Francesco Belfiore and Ivana Bešlić and Frank Bigiel and Guillermo A. Blanc and Alberto D. Bolatto and Médéric Boquien and Yixian Cao and Rupali Chandar and Jérémy Chastenet and I-Da Chiang and Enrico Congiu and Daniel A. Dale and Sinan Deger and Jakob S. den Brok and Cosima Eibensteiner and Eric Emsellem and Axel García-Rodríguez and Simon C. O. Glover and Kathryn Grasha and Brent Groves and Jonathan D. Henshaw and María J. Jiménez Donaire and Jaeyeon Kim and Ralf S. Klessen and Kathryn Kreckel and J. M. Diederik Kruijssen and Kirsten L. Larson and Janice C. Lee and Ness Mayker and Rebecca McElroy and Sharon E. Meidt and Angus Mok and Hsi-An Pan and Johannes Puschnig and Alessandro Razza and Patricia Sánchez-Bl’azquez and Karin M. Sandstrom and Francesco Santoro and Amy Sardone and Fabian Scheuermann and Jiayi Sun and David A. Thilker and Jordan A. Turner and Leonardo Ubeda and Dyas Utomo and Elizabeth J. Watkins and Thomas G. Williams},
title = {PHANGS–ALMA Data Processing and Pipeline},
journal = {The Astrophysical Journal Supplement Series}
}

@article{ Eric,
	author = {{Emsellem, Eric} and {Schinnerer, Eva} and {Santoro, Francesco} and {Belfiore, Francesco} and {Pessa, Ismael} and {McElroy, Rebecca} and {Blanc, Guillermo A.} and {Congiu, Enrico} and {Groves, Brent} and {Ho, I-Ting} and {Kreckel, Kathryn} and {Razza, Alessandro} and {Sanchez-Blazquez, Patricia} and {Egorov, Oleg} and {Faesi, Chris} and {Klessen, Ralf S.} and {Leroy, Adam K.} and {Meidt, Sharon} and {Querejeta, Miguel} and {Rosolowsky, Erik} and {Scheuermann, Fabian} and {Anand, Gagandeep S.} and {Barnes, Ashley T.} and {Bešlić, Ivana} and {Bigiel, Frank} and {Boquien, Médéric} and {Cao, Yixian} and {Chevance, Mélanie} and {Dale, Daniel A.} and {Eibensteiner, Cosima} and {Glover, Simon C. O.} and {Grasha, Kathryn} and {Henshaw, Jonathan D.} and {Hughes, Annie} and {Koch, Eric W.} and {Kruijssen, J. M. Diederik} and {Lee, Janice} and {Liu, Daizhong} and {Pan, Hsi-An} and {Pety, Jérôme} and {Saito, Toshiki} and {Sandstrom, Karin M.} and {Schruba, Andreas} and {Sun, Jiayi} and {Thilker, David A.} and {Usero, Antonio} and {Watkins, Elizabeth J.} and {Williams, Thomas G.}},
	title = {The PHANGS-MUSE survey - Probing the chemo-dynamical evolution of disc galaxies},
	DOI= "10.1051/0004-6361/202141727",
	url= "https://doi.org/10.1051/0004-6361/202141727",
	journal = {A\&A},
	year = 2022,
	volume = 659,
	pages = "A191",
}

@article{Lee_2023,
doi = {10.3847/2041-8213/acaaae},
url = {https://dx.doi.org/10.3847/2041-8213/acaaae},
year = {2023},
month = {feb},
publisher = {The American Astronomical Society},
volume = {944},
number = {2},
pages = {L17},
author = {Lee, Janice C. and Sandstrom, Karin M. and Leroy, Adam K. and Thilker, David A. and Schinnerer, Eva and Rosolowsky, Erik and Larson, Kirsten L. and Egorov, Oleg V. and Williams, Thomas G. and Schmidt, Judy and Emsellem, Eric and Anand, Gagandeep S. and Barnes, Ashley T. and Belfiore, Francesco and Bešlić, Ivana and Bigiel, Frank and Blanc, Guillermo A. and Bolatto, Alberto D. and Boquien, Médéric and Brok, Jakob den and Cao, Yixian and Chandar, Rupali and Chastenet, Jérémy and Chevance, Mélanie and Chiang, I-Da and Congiu, Enrico and Dale, Daniel A. and Deger, Sinan and Eibensteiner, Cosima and Faesi, Christopher M. and Glover, Simon C. O. and Grasha, Kathryn and Groves, Brent and Hassani, Hamid and Henny, Kiana F. and Henshaw, Jonathan D. and Hoyer, Nils and Hughes, Annie and Jeffreson, Sarah and Jiménez-Donaire, María J. and Kim, Jaeyeon and Kim, Hwihyun and Klessen, Ralf S. and Koch, Eric W. and Kreckel, Kathryn and Kruijssen, J. M. Diederik and Li, Jing and Liu, Daizhong and Lopez, Laura A. and Maschmann, Daniel and Chen, Ness Mayker and Meidt, Sharon E. and Murphy, Eric J. and Neumann, Justus and Neumayer, Nadine and Pan, Hsi-An and Pessa, Ismael and Pety, Jérôme and Querejeta, Miguel and Pinna, Francesca and Rodríguez, M. Jimena and Saito, Toshiki and Sánchez-Blázquez, Patricia and Santoro, Francesco and Sardone, Amy and Smith, Rowan J. and Sormani, Mattia C. and Scheuermann, Fabian and Stuber, Sophia K. and Sutter, Jessica and Sun, Jiayi and Teng, Yu-Hsuan and Treß, Robin G. and Usero, Antonio and Watkins, Elizabeth J. and Whitmore, Bradley C. and Razza, Alessandro},
title = {The PHANGS–JWST Treasury Survey: Star Formation, Feedback, and Dust Physics at High Angular Resolution in Nearby GalaxieS},
journal = {ApJL}
}

@ARTICLE{2015AJ....149...51C,
       author = {{Calzetti}, D. and {Lee}, J.~C. and {Sabbi}, E. and {Adamo}, A. and {Smith}, L.~J. and {Andrews}, J.~E. and {Ubeda}, L. and {Bright}, S.~N. and {Thilker}, D. and {Aloisi}, A. and {Brown}, T.~M. and {Chandar}, R. and {Christian}, C. and {Cignoni}, M. and {Clayton}, G.~C. and {da Silva}, R. and {de Mink}, S.~E. and {Dobbs}, C. and {Elmegreen}, B.~G. and {Elmegreen}, D.~M. and {Evans}, A.~S. and {Fumagalli}, M. and {Gallagher}, III, J.~S. and {Gouliermis}, D.~A. and {Grebel}, E.~K. and {Herrero}, A. and {Hunter}, D.~A. and {Johnson}, K.~E. and {Kennicutt}, R.~C. and {Kim}, H. and {Krumholz}, M.~R. and {Lennon}, D. and {Levay}, K. and {Martin}, C. and {Nair}, P. and {Nota}, A. and {{\"O}stlin}, G. and {Pellerin}, A. and {Prieto}, J. and {Regan}, M.~W. and {Ryon}, J.~E. and {Schaerer}, D. and {Schiminovich}, D. and {Tosi}, M. and {Van Dyk}, S.~D. and {Walterbos}, R. and {Whitmore}, B.~C. and {Wofford}, A.},
        title = "{Legacy Extragalactic UV Survey (LEGUS) With the Hubble Space Telescope. I. Survey Description}",
      journal = {\aj},
         year = 2015,
        month = feb,
       volume = {149},
       number = {2},
          eid = {51},
        pages = {51},
          doi = {10.1088/0004-6256/149/2/51},
archivePrefix = {arXiv},
       eprint = {1410.7456},
 primaryClass = {astro-ph.GA},
       adsurl = {https://ui.adsabs.harvard.edu/abs/2015AJ....149...51C}
}

@article{ujjwal,
	author = {{Ujjwal, Krishnan} and {Kartha, Sreeja S.} and {Akhil, Krishna R.} and {Mathew, Blesson} and {Subramanian, Smitha} and {Sudheesh, T. P.} and {Thomas, Robin}},
	title = {Disentangling the association of PAH molecules with star formation - Insights from the James Webb Space Telescope and from the UltraViolet Imaging Telescope⋆},
	DOI= "10.1051/0004-6361/202347620",
	url= "https://doi.org/10.1051/0004-6361/202347620",
	journal = {A\&A},
	year = 2024,
	volume = 684,
	pages = "A71",
}

@article{Sandstrom_2023,
doi = {10.3847/2041-8213/acb0cf},
url = {https://dx.doi.org/10.3847/2041-8213/acb0cf},
year = {2023},
month = {feb},
publisher = {The American Astronomical Society},
volume = {944},
number = {2},
pages = {L7},
author = {Sandstrom, Karin M. and Chastenet, Jérémy and Sutter, Jessica and Leroy, Adam K. and Egorov, Oleg V. and Williams, Thomas G. and Bolatto, Alberto D. and Boquien, Médéric and Cao, Yixian and Dale, Daniel A. and Lee, Janice C. and Rosolowsky, Erik and Schinnerer, Eva and Barnes, Ashley. T. and Belfiore, Francesco and Bigiel, F. and Chevance, Mélanie and Grasha, Kathryn and Groves, Brent and Hassani, Hamid and Hughes, Annie and Klessen, Ralf S. and Kruijssen, J. M. Diederik and Larson, Kirsten L. and Liu, Daizhong and Lopez, Laura A. and Meidt, Sharon E. and Murphy, Eric J. and Sormani, Mattia C. and Thilker, David A. and Watkins, Elizabeth J.},
title = {PHANGS–JWST First Results: Mapping the 3.3 μm Polycyclic Aromatic Hydrocarbon Vibrational Band in Nearby Galaxies with NIRCam Medium Bands},
journal = {\apjl}
}

@article{Barnes_2023,
   title={PHANGS–JWST First Results: Multiwavelength View of Feedback-driven Bubbles (the Phantom Voids) across NGC 628},
   volume={944},
   ISSN={2041-8213},
   url={http://dx.doi.org/10.3847/2041-8213/aca7b9},
   DOI={10.3847/2041-8213/aca7b9},
   number={2},
   journal={\apjl},
   publisher={American Astronomical Society},
   author={Barnes, Ashley. T. and Watkins, Elizabeth J. and Meidt, Sharon E. and Kreckel, Kathryn and Sormani, Mattia C. and Treß, Robin G. and Glover, Simon C. O. and Bigiel, Frank and Chandar, Rupali and Emsellem, Eric and Lee, Janice C. and Leroy, Adam K. and Sandstrom, Karin M. and Schinnerer, Eva and Rosolowsky, Erik and Belfiore, Francesco and Blanc, Guillermo A. and Boquien, Médéric and Brok, Jakob den and Cao, Yixian and Chevance, Mélanie and Dale, Daniel A. and Egorov, Oleg V. and Eibensteiner, Cosima and Grasha, Kathryn and Groves, Brent and Hassani, Hamid and Henshaw, Jonathan D. and Jeffreson, Sarah and Jiménez-Donaire, María J. and Keller, Benjamin W. and Klessen, Ralf S. and Koch, Eric W. and Kruijssen, J. M. Diederik and Larson, Kirsten L. and Li, Jing and Liu, Daizhong and Lopez, Laura A. and Murphy, Eric J. and Neumann, Lukas and Pety, Jérôme and Pinna, Francesca and Querejeta, Miguel and Renaud, Florent and Saito, Toshiki and Sarbadhicary, Sumit K. and Sardone, Amy and Smith, Rowan J. and Stuber, Sophia K. and Sun, Jiayi and Thilker, David A. and Usero, Antonio and Whitmore, Bradley C. and Williams, Thomas G.},
   year={2023},
   month=feb, pages={L22} }

@ARTICLE{2024MNRAS.532.1598R,
       author = {{Rigopoulou}, D. and {Donnan}, F.~R. and {Garc{\'\i}a-Bernete}, I. and {Pereira-Santaella}, M. and {Alonso-Herrero}, A. and {Davies}, R. and {Hunt}, L.~K. and {Roche}, P.~F. and {Shimizu}, T.},
        title = "{Polycyclic aromatic hydrocarbon emission in galaxies as seen with JWST}",
      journal = {\mnras},
         year = 2024,
        month = aug,
       volume = {532},
       number = {2},
        pages = {1598-1611},
          doi = {10.1093/mnras/stae1535},
archivePrefix = {arXiv},
       eprint = {2406.11415},
 primaryClass = {astro-ph.GA},
       adsurl = {https://ui.adsabs.harvard.edu/abs/2024MNRAS.532.1598R}
}

@article{Lai_2020,
doi = {10.3847/1538-4357/abc002},
url = {https://dx.doi.org/10.3847/1538-4357/abc002},
year = {2020},
month = {dec},
publisher = {The American Astronomical Society},
volume = {905},
number = {1},
pages = {55},
author = {Lai, Thomas S.-Y. and Smith, J. D. T. and Baba, Shunsuke and Spoon, Henrik W. W. and Imanishi, Masatoshi},
title = {All the PAHs: An AKARI–Spitzer Cross-archival Spectroscopic Survey of Aromatic Emission in Galaxies},
journal = {\apj}
}

@ARTICLE{2004ApJ...613..986P,
       author = {{Peeters}, E. and {Spoon}, H.~W.~W. and {Tielens}, A.~G.~G.~M.},
        title = "{Polycyclic Aromatic Hydrocarbons as a Tracer of Star Formation?}",
      journal = {\apj},
         year = 2004,
        month = oct,
       volume = {613},
       number = {2},
        pages = {986-1003},
          doi = {10.1086/423237},
archivePrefix = {arXiv},
       eprint = {astro-ph/0406183},
 primaryClass = {astro-ph},
       adsurl = {https://ui.adsabs.harvard.edu/abs/2004ApJ...613..986P}
}

@article{Kennicutt_1998,
   title={STAR FORMATION IN GALAXIES ALONG THE HUBBLE SEQUENCE},
   volume={36},
   ISSN={1545-4282},
   url={http://dx.doi.org/10.1146/annurev.astro.36.1.189},
   DOI={10.1146/annurev.astro.36.1.189},
   number={1},
   journal={\araa},
   publisher={Annual Reviews},
   author={Kennicutt, Robert C.},
   year={1998},
   month=sep, pages={189–231} }

@ARTICLE{1989ApJS...71..733A,
       author = {{Allamandola}, L.~J. and {Tielens}, A.~G.~G.~M. and {Barker}, J.~R.},
        title = "{Interstellar Polycyclic Aromatic Hydrocarbons: The Infrared Emission Bands, the Excitation/Emission Mechanism, and the Astrophysical Implications}",
      journal = {\apjs},
         year = 1989,
        month = dec,
       volume = {71},
        pages = {733},
          doi = {10.1086/191396},
       adsurl = {https://ui.adsabs.harvard.edu/abs/1989ApJS...71..733A}
}

@article{Smith_2007,
doi = {10.1086/510549},
url = {https://dx.doi.org/10.1086/510549},
year = {2007},
month = {feb},
publisher = {},
volume = {656},
number = {2},
pages = {770},
author = {Smith, J. D. T. and Draine, B. T. and Dale, D. A. and Moustakas, J. and Kennicutt, Jr., R. C. and Helou, G. and Armus, L. and Roussel, H. and Sheth, K. and Bendo, G. J. and Buckalew, B. A. and Calzetti, D. and Engelbracht, C. W. and Gordon, K. D. and Hollenbach, D. J. and Li, A. and Malhotra, S. and Murphy, E. J. and Walter, F.},
title = {The Mid-Infrared Spectrum of Star-forming Galaxies: Global Properties of Polycyclic Aromatic Hydrocarbon Emission},
journal = {\apj}
}

@ARTICLE{2020NatAs...4..339L,
       author = {{Li}, Aigen},
        title = "{Spitzer's perspective of polycyclic aromatic hydrocarbons in galaxies}",
      journal = {Nature Astronomy},
         year = 2020,
        month = mar,
       volume = {4},
        pages = {339-351},
          doi = {10.1038/s41550-020-1051-1},
archivePrefix = {arXiv},
       eprint = {2003.10489},
 primaryClass = {astro-ph.GA},
       adsurl = {https://ui.adsabs.harvard.edu/abs/2020NatAs...4..339L}
}

@ARTICLE{1984A&A...137L...5L,
       author = {{Leger}, A. and {Puget}, J.~L.},
        title = "{Identification of the Unidentified Infrared Emission Features of Interstellar Dust}",
      journal = {\aap},
         year = 1984,
        month = aug,
       volume = {137},
        pages = {L5-L8},
       adsurl = {https://ui.adsabs.harvard.edu/abs/1984A&A...137L...5L}
}

@ARTICLE{2008ARA&A..46..289T,
       author = {{Tielens}, A.~G.~G.~M.},
        title = "{Interstellar polycyclic aromatic hydrocarbon molecules.}",
      journal = {\araa},
         year = 2008,
        month = sep,
       volume = {46},
        pages = {289-337},
          doi = {10.1146/annurev.astro.46.060407.145211},
       adsurl = {https://ui.adsabs.harvard.edu/abs/2008ARA&A..46..289T}
}

@article{Tielens1993,
 ISSN = {00368075, 10959203},
 URL = {http://www.jstor.org/stable/2885669},
 author = {A. G. G. M. Tielens and M. M. Meixner and P. P. van der Werf and J. Bregman and J. A. Tauber and J. Stutzki and D. Rank},
 journal = {Science},
 number = {5130},
 pages = {86--89},
 publisher = {American Association for the Advancement of Science},
 title = {Anatomy of the Photodissociation Region in the Orion Bar},
 urldate = {2025-06-01},
 volume = {262},
 year = {1993}
}

@article{Allamandola_1999,
doi = {10.1086/311843},
url = {https://dx.doi.org/10.1086/311843},
year = {1999},
month = {jan},
publisher = {},
volume = {511},
number = {2},
pages = {L115},
author = {Allamandola, L. J. and Hudgins, D. M. and Sandford, S. A.},
title = {Modeling the Unidentified Infrared Emission with Combinations of Polycyclic Aromatic Hydrocarbons},
journal = {\apj}
}

@ARTICLE{2021ApJ...917....3D,
       author = {{Draine}, B.~T. and {Li}, Aigen and {Hensley}, Brandon S. and {Hunt}, L.~K. and {Sandstrom}, K. and {Smith}, J. -D.~T.},
        title = "{Excitation of Polycyclic Aromatic Hydrocarbon Emission: Dependence on Size Distribution, Ionization, and Starlight Spectrum and Intensity}",
      journal = {\apj},
         year = 2021,
        month = aug,
       volume = {917},
       number = {1},
          eid = {3},
        pages = {3},
          doi = {10.3847/1538-4357/abff51},
archivePrefix = {arXiv},
       eprint = {2011.07046},
 primaryClass = {astro-ph.GA},
       adsurl = {https://ui.adsabs.harvard.edu/abs/2021ApJ...917....3D}
}

@ARTICLE{2022ApJ...931...38M,
       author = {{Maragkoudakis}, A. and {Boersma}, C. and {Temi}, P. and {Bregman}, J.~D. and {Allamandola}, L.~J.},
        title = "{Linking Characteristics of the Polycyclic Aromatic Hydrocarbon Population with Galaxy Properties: A Quantitative Approach Using the NASA Ames PAH IR Spectroscopic Database}",
      journal = {\apj},
         year = 2022,
        month = may,
       volume = {931},
       number = {1},
          eid = {38},
        pages = {38},
          doi = {10.3847/1538-4357/ac666f},
archivePrefix = {arXiv},
       eprint = {2204.05292},
 primaryClass = {astro-ph.GA},
       adsurl = {https://ui.adsabs.harvard.edu/abs/2022ApJ...931...38M}
}

@ARTICLE{2021MNRAS.504.5287R,
       author = {{Rigopoulou}, D. and {Barale}, M. and {Clary}, D.~C. and {Shan}, X. and {Alonso-Herrero}, A. and {Garc{\'\i}a-Bernete}, I. and {Hunt}, L. and {Kerkeni}, B. and {Pereira-Santaella}, M. and {Roche}, P.~F.},
        title = "{The properties of polycyclic aromatic hydrocarbons in galaxies: constraints on PAH sizes, charge and radiation fields}",
      journal = {\mnras},
         year = 2021,
        month = jul,
       volume = {504},
       number = {4},
        pages = {5287-5300},
          doi = {10.1093/mnras/stab959},
archivePrefix = {arXiv},
       eprint = {2011.10114},
 primaryClass = {astro-ph.GA},
       adsurl = {https://ui.adsabs.harvard.edu/abs/2021MNRAS.504.5287R}
}

@ARTICLE{2023ApJ...944L..12C,
       author = {{Chastenet}, J{\'e}r{\'e}my and {Sutter}, Jessica and {Sandstrom}, Karin and {Belfiore}, Francesco and {Egorov}, Oleg V. and {Larson}, Kirsten L. and {Leroy}, Adam K. and {Liu}, Daizhong and {Rosolowsky}, Erik and {Thilker}, David A. and {Watkins}, Elizabeth J. and {Williams}, Thomas G. and {Barnes}, Ashley. T. and {Bigiel}, F. and {Boquien}, M{\'e}d{\'e}ric and {Chevance}, M{\'e}lanie and {Dale}, Daniel A. and {Kruijssen}, J.~M. Diederik and {Emsellem}, Eric and {Grasha}, Kathryn and {Groves}, Brent and {Hassani}, Hamid and {Hughes}, Annie and {Kreckel}, Kathryn and {Meidt}, Sharon E. and {Pan}, Hsi-An and {Querejeta}, Miguel and {Schinnerer}, Eva and {Whitcomb}, Cory M.},
        title = "{PHANGS-JWST First Results: Measuring Polycyclic Aromatic Hydrocarbon Properties across the Multiphase Interstellar Medium}",
      journal = {\apjl},
         year = 2023,
        month = feb,
       volume = {944},
       number = {2},
          eid = {L12},
        pages = {L12},
          doi = {10.3847/2041-8213/acac94},
       adsurl = {https://ui.adsabs.harvard.edu/abs/2023ApJ...944L..12C}
}

@article{Rigopoulou_1999,
doi = {10.1086/301146},
url = {https://dx.doi.org/10.1086/301146},
year = {1999},
month = {dec},
publisher = {},
volume = {118},
number = {6},
pages = {2625},
author = {Rigopoulou, D. and Spoon, H. W. W. and Genzel, R. and Lutz, D. and Moorwood, A. F. M. and Tran, Q. D.},
title = {A Large Mid-Infrared Spectroscopic and Near-Infrared
Imaging Survey of Ultraluminous Infrared Galaxies: Their Nature and
Evolution* **},
journal = {\aj}
}

@article{Kaneda_2008,
doi = {10.1086/590243},
url = {https://dx.doi.org/10.1086/590243},
year = {2008},
month = {sep},
publisher = {},
volume = {684},
number = {1},
pages = {270},
author = {Kaneda, H. and Onaka, T. and Sakon, I. and Kitayama, T. and Okada, Y. and Suzuki, T.},
title = {Properties of Polycyclic Aromatic Hydrocarbons in Local Elliptical Galaxies Revealed by the Infrared Spectrograph on Spitzer},
journal = {\apj}
}

@ARTICLE{2010ApJ...721.1090V,
       author = {{Vega}, O. and {Bressan}, A. and {Panuzzo}, P. and {Rampazzo}, R. and {Clemens}, M. and {Granato}, G.~L. and {Buson}, L. and {Silva}, L. and {Zeilinger}, W.~W.},
        title = "{Unusual PAH Emission in Nearby Early-type Galaxies: A Signature of an Intermediate-age Stellar Population?}",
      journal = {\apj},
         year = 2010,
        month = oct,
       volume = {721},
       number = {2},
        pages = {1090-1104},
          doi = {10.1088/0004-637X/721/2/1090},
archivePrefix = {arXiv},
       eprint = {1008.0009},
 primaryClass = {astro-ph.CO},
       adsurl = {https://ui.adsabs.harvard.edu/abs/2010ApJ...721.1090V}
}

@ARTICLE{2012A&A...541A..10Y,
       author = {{Yamagishi}, M. and {Kaneda}, H. and {Ishihara}, D. and {Kondo}, T. and {Onaka}, T. and {Suzuki}, T. and {Minh}, Y.~C.},
        title = "{AKARI near-infrared spectroscopy of the aromatic and aliphatic hydrocarbon emission features in the galactic superwind of M 82}",
      journal = {\aap},
         year = 2012,
        month = may,
       volume = {541},
          eid = {A10},
        pages = {A10},
          doi = {10.1051/0004-6361/201218904},
archivePrefix = {arXiv},
       eprint = {1203.2794},
 primaryClass = {astro-ph.GA},
       adsurl = {https://ui.adsabs.harvard.edu/abs/2012A&A...541A..10Y}
}

@article{ García2022,
	author = {{García-Bernete, I.} and {Rigopoulou, D.} and {Alonso-Herrero, A.} and {Donnan, F. R.} and {Roche, P. F.} and {Pereira-Santaella, M.} and {Labiano, A.} and {Peralta de Arriba, L.} and {Izumi, T.} and {Ramos Almeida, C.} and {Shimizu, T.} and {Hönig, S.} and {García-Burillo, S.} and {Rosario, D. J.} and {Ward, M. J.} and {Bellocchi, E.} and {Hicks, E. K. S.} and {Fuller, L.} and {Packham, C.}},
	title = {A high angular resolution view of the PAH emission in Seyfert galaxies using JWST/MRS data},
	DOI= "10.1051/0004-6361/202244806",
	url= "https://doi.org/10.1051/0004-6361/202244806",
	journal = {A\&A},
	year = 2022,
	volume = 666,
	pages = "L5",
}

@article{ garcia2024,
	author = {{García-Bernete, I.} and {Alonso-Herrero, A.} and {Rigopoulou, D.} and {Pereira-Santaella, M.} and {Shimizu, T.} and {Davies, R.} and {Donnan, F. R.} and {Roche, P. F.} and {González-Martín, O.} and {Ramos Almeida, C.} and {Bellocchi, E.} and {Boorman, P.} and {Combes, F.} and {Efstathiou, A.} and {Esparza-Arredondo, D.} and {García-Burillo, S.} and {González-Alfonso, E.} and {Hicks, E. K. S.} and {Hönig, S.} and {Labiano, A.} and {Levenson, N. A.} and {López-Rodríguez, E.} and {Ricci, C.} and {Packham, C.} and {Rouan, D.} and {Stalevski, M.} and {Ward, M. J.}},
	title = {The Galaxy Activity, Torus, and Outflow Survey (GATOS) - III. Revealing the inner icy structure in local active galactic nuclei},
	DOI= "10.1051/0004-6361/202348266",
	url= "https://doi.org/10.1051/0004-6361/202348266",
	journal = {A\&A},
	year = 2024,
	volume = 681,
	pages = "L7",
}

@article{Armus_2023,
doi = {10.3847/2041-8213/acac66},
url = {https://dx.doi.org/10.3847/2041-8213/acac66},
year = {2023},
month = {jan},
publisher = {The American Astronomical Society},
volume = {942},
number = {2},
pages = {L37},
author = {Armus, L. and Lai, T. and U, V. and Larson, K. L. and Diaz-Santos, T. and Evans, A. S. and Malkan, M. A. and Rich, J. and Medling, A. M. and Law, D. R. and Inami, H. and Muller-Sanchez, F. and Charmandaris, V. and Werf, P. van der and Stierwalt, S. and Linden, S. and Privon, G. C. and Barcos-Muñoz, L. and Hayward, C. and Song, Y. and Appleton, P. and Aalto, S. and Bohn, T. and Böker, T. and Brown, M. J. I. and Finnerty, L. and Howell, J. and Iwasawa, K. and Kemper, F. and Marshall, J. and Mazzarella, J. M. and McKinney, J. and Murphy, E. J. and Sanders, D. and Surace, J.},
title = {GOALS-JWST: Mid-infrared Spectroscopy of the Nucleus of NGC 7469},
journal = {\apjl}
}

@article{madden2006,
	author = {{Madden, S. C.} and {Galliano, F.} and {Jones, A. P.} and {Sauvage, M.}},
	title = {ISM properties in low-metallicity environments - I. Mid-infrared spectra of dwarf galaxies},
	DOI= "10.1051/0004-6361:20053890",
	url= "https://doi.org/10.1051/0004-6361:20053890",
	journal = {A\&A},
	year = 2006,
	volume = 446,
	number = 3,
	pages = "877-896",
}

@ARTICLE{2024A&A...690A..89S,
       author = {{Shivaei}, Irene and {Alberts}, Stacey and {Florian}, Michael and {Rieke}, George and {Wuyts}, Stijn and {Bodansky}, Sarah and {Bunker}, Andrew J. and {Cameron}, Alex J. and {Curti}, Mirko and {D'Eugenio}, Francesco and {Dudzevi{\v{c}}i{\={u}}t{\.{e}}}, Ugn{\.{e}} and {Ji}, Zhiyuan and {Johnson}, Benjamin D. and {Kramarenko}, Ivan and {Lyu}, Jianwei and {Matthee}, Jorryt and {Morrison}, Jane and {Naidu}, Rohan and {P{\'e}rez-Gonz{\'a}lez}, Pablo G. and {Reddy}, Naveen and {Robertson}, Brant and {Sun}, Yang and {Tacchella}, Sandro and {Whitaker}, Katherine and {Williams}, Christina C. and {Willmer}, Christopher N.~A. and {Witstok}, Joris and {Xiao}, Mengyuan and {Zhu}, Yongda},
        title = "{A new census of dust and polycyclic aromatic hydrocarbons at z = 0.7{\textendash}2 with JWST MIRI}",
      journal = {\aap},
         year = 2024,
        month = oct,
       volume = {690},
          eid = {A89},
        pages = {A89},
          doi = {10.1051/0004-6361/202449579},
archivePrefix = {arXiv},
       eprint = {2402.07989},
 primaryClass = {astro-ph.GA},
       adsurl = {https://ui.adsabs.harvard.edu/abs/2024A&A...690A..89S}
}

@ARTICLE{2023ApJ...950....7S,
       author = {{Shen}, Lu and {Papovich}, Casey and {Yang}, Guang and {Matharu}, Jasleen and {Wang}, Xin and {Magnelli}, Benjamin and {Elbaz}, David and {Jogee}, Shardha and {Alavi}, Anahita and {Arrabal Haro}, Pablo and {Backhaus}, Bren E. and {Bagley}, Micaela B. and {Bell}, Eric F. and {Bisigello}, Laura and {Calabr{\`o}}, Antonello and {Cooper}, M.~C. and {Costantin}, Luca and {Daddi}, Emanuele and {Dickinson}, Mark and {Finkelstein}, Steven L. and {Fujimoto}, Seiji and {Giavalisco}, Mauro and {Grogin}, Norman A. and {Guo}, Yuchen and {Holwerda}, Benne W. and {Kartaltepe}, Jeyhan S. and {Koekemoer}, Anton M. and {Kurczynski}, Peter and {Lucas}, Ray A. and {P{\'e}rez-Gonz{\'a}lez}, Pablo G. and {Pirzkal}, Nor and {Prichard}, Laura and {Rafelski}, Marc and {Ronayne}, Kaila and {Simons}, Raymond C. and {Sunnquist}, Ben and {Teplitz}, Harry I. and {Trump}, Jonathan R. and {Weiner}, Benjamin J. and {Windhorst}, Rogier A. and {Yung}, L.~Y. Aaron},
        title = "{CEERS: Spatially Resolved UV and Mid-infrared Star Formation in Galaxies at 0.2 < z < 2.5: The Picture from the Hubble and James Webb Space Telescopes}",
      journal = {\apj},
         year = 2023,
        month = jun,
       volume = {950},
       number = {1},
          eid = {7},
        pages = {7},
          doi = {10.3847/1538-4357/acc944},
archivePrefix = {arXiv},
       eprint = {2301.05727},
 primaryClass = {astro-ph.GA},
       adsurl = {https://ui.adsabs.harvard.edu/abs/2023ApJ...950....7S}
}

@article{Chastenet_2023,
doi = {10.3847/2041-8213/acadd7},
url = {https://dx.doi.org/10.3847/2041-8213/acadd7},
year = {2023},
month = {feb},
publisher = {The American Astronomical Society},
volume = {944},
number = {2},
pages = {L11},
author = {Chastenet, Jérémy and Sutter, Jessica and Sandstrom, Karin and Belfiore, Francesco and Egorov, Oleg V. and Larson, Kirsten L. and Leroy, Adam K. and Liu, Daizhong and Rosolowsky, Erik and Thilker, David A. and Watkins, Elizabeth J. and Williams, Thomas G. and Barnes, Ashley. T. and Bigiel, Frank and Boquien, Médéric and Chevance, Mélanie and Chiang, I-Da and Dale, Daniel A. and Kruijssen, J. M. Diederik and Emsellem, Eric and Grasha, Kathryn and Groves, Brent and Hassani, Hamid and Hughes, Annie and Kreckel, Kathryn and Meidt, Sharon E. and Rickards Vaught, Ryan J. and Sardone, Amy and Schinnerer, Eva},
title = {PHANGS–JWST First Results: Variations in PAH Fraction as a Function of ISM Phase and Metallicity},
journal = {\apjl}
}

@article{Sutter_2024,
doi = {10.3847/1538-4357/ad54bd},
url = {https://dx.doi.org/10.3847/1538-4357/ad54bd},
year = {2024},
month = {aug},
publisher = {The American Astronomical Society},
volume = {971},
number = {2},
pages = {178},
author = {Sutter, Jessica and Sandstrom, Karin and Chastenet, Jérémy and Leroy, Adam K. and Koch, Eric W. and Williams, Thomas G. and Chown, Ryan and Belfiore, Francesco and Bigiel, Frank and Boquien, Médéric and Cao, Yixian and Chevance, Mélanie and Dale, Daniel A. and Egorov, Oleg V. and Glover, Simon C. O. and Groves, Brent and Klessen, Ralf S. and Kreckel, Kathryn and Larson, Kirsten L. and Oakes, Elias K. and Pathak, Debosmita and Ramambason, Lise and Rosolowsky, Erik and Watkins, Elizabeth J.},
title = {The Fraction of Dust Mass in the Form of Polycyclic Aromatic Hydrocarbons on 10–50 pc Scales in Nearby Galaxies},
journal = {\apj}
}

@article{Watkins_2023,
doi = {10.3847/2041-8213/aca6e4},
url = {https://dx.doi.org/10.3847/2041-8213/aca6e4},
year = {2023},
month = {feb},
publisher = {The American Astronomical Society},
volume = {944},
number = {2},
pages = {L24},
author = {Watkins, Elizabeth J. and Barnes, Ashley T. and Henny, Kiana and Kim, Hwihyun and Kreckel, Kathryn and Meidt, Sharon E. and Klessen, Ralf S. and Glover, Simon C. O. and Williams, Thomas G. and Keller, Benjamin W. and Leroy, Adam K. and Rosolowsky, Erik and Lee, Janice C. and Anand, Gagandeep S. and Belfiore, Francesco and Bigiel, Frank and Blanc, Guillermo A. and Boquien, Médéric and Cao, Yixian and Chandar, Rupali and Chen, Ness Mayker and Chevance, Mélanie and Congiu, Enrico and Dale, Daniel A. and Deger, Sinan and Egorov, Oleg V. and Emsellem, Eric and Faesi, Christopher M. and Grasha, Kathryn and Groves, Brent and Hassani, Hamid and Henshaw, Jonathan D. and Herrera, Cinthya and Hughes, Annie and Jeffreson, Sarah and Jiménez-Donaire, María J. and Koch, Eric W. and Kruijssen, J. M. Diederik and Larson, Kirsten L. and Liu, Daizhong and Lopez, Laura A. and Pessa, Ismael and Pety, Jérôme and Querejeta, Miguel and Saito, Toshiki and Sandstrom, Karin and Scheuermann, Fabian and Schinnerer, Eva and Sormani, Mattia C. and Stuber, Sophia K. and Thilker, David A. and Usero, Antonio and Whitmore, Bradley C.},
title = {PHANGS–JWST First Results: A Statistical View on Bubble Evolution in NGC 628},
journal = {\apjl}
}

@article{Thilker_2023,
doi = {10.3847/2041-8213/acaeac},
url = {https://dx.doi.org/10.3847/2041-8213/acaeac},
year = {2023},
month = {feb},
publisher = {The American Astronomical Society},
volume = {944},
number = {2},
pages = {L13},
author = {Thilker, David A. and Lee, Janice C. and Deger, Sinan and Barnes, Ashley T. and Bigiel, Frank and Boquien, Médéric and Cao, Yixian and Chevance, Mélanie and Dale, Daniel A. and Egorov, Oleg V. and Glover, Simon C. O. and Grasha, Kathryn and Henshaw, Jonathan D. and Klessen, Ralf S. and Koch, Eric and Kruijssen, J. M. Diederik and Leroy, Adam K. and Lessing, Ryan A. and Meidt, Sharon E. and Pinna, Francesca and Querejeta, Miguel and Rosolowsky, Erik and Sandstrom, Karin M. and Schinnerer, Eva and Smith, Rowan J. and Watkins, Elizabeth J. and Williams, Thomas G. and Anand, Gagandeep S. and Belfiore, Francesco and Blanc, Guillermo A. and Chandar, Rupali and Congiu, Enrico and Emsellem, Eric and Groves, Brent and Kreckel, Kathryn and Larson, Kirsten L. and Liu, Daizhong and Pessa, Ismael and Whitmore, Bradley C.},
title = {PHANGS–JWST First Results: The Dust Filament Network of NGC 628 and Its Relation to Star Formation Activity},
journal = {\apjl}
}

@article{Dale_2023,
doi = {10.3847/2041-8213/aca769},
url = {https://dx.doi.org/10.3847/2041-8213/aca769},
year = {2023},
month = {feb},
publisher = {The American Astronomical Society},
volume = {944},
number = {2},
pages = {L23},
author = {Dale, Daniel A. and Boquien, Médéric and Barnes, Ashley T. and Belfiore, Francesco and Bigiel, Frank and Cao, Yixian and Chandar, Rupali and Chastenet, Jérémy and Chevance, Mélanie and Deger, Sinan and Egorov, Oleg V. and Grasha, Kathryn and Groves, Brent and Hassani, Hamid and Henny, Kiana F. and Klessen, Ralf S. and Kreckel, Kathryn and Kruijssen, J. M. Diederik and Larson, Kirsten L. and Lee, Janice C. and Leroy, Adam K. and Liu, Daizhong and Murphy, Eric J. and Rosolowsky, Erik and Sandstrom, Karin and Schinnerer, Eva and Sutter, Jessica and Thilker, David A. and Watkins, Elizabeth J. and Whitmore, Bradley C. and Williams, Thomas G.},
title = {PHANGS–JWST First Results: The Influence of Stellar Clusters on Polycyclic Aromatic Hydrocarbons in Nearby Galaxies},
journal = {\apjl}
}

@article{Shipley_2016,
doi = {10.3847/0004-637X/818/1/60},
url = {https://dx.doi.org/10.3847/0004-637X/818/1/60},
year = {2016},
month = {feb},
publisher = {The American Astronomical Society},
volume = {818},
number = {1},
pages = {60},
author = {Shipley, Heath V. and Papovich, Casey and Rieke, George H. and Brown, Michael J. I. and Moustakas, John},
title = {A NEW STAR FORMATION RATE CALIBRATION FROM POLYCYCLIC AROMATIC HYDROCARBON EMISSION FEATURES AND APPLICATION TO HIGH-REDSHIFT GALAXIES},
journal = {\apj}
}

@article{Williams_2022,
doi = {10.3847/2041-8213/aca674},
url = {https://dx.doi.org/10.3847/2041-8213/aca674},
year = {2022},
month = {dec},
publisher = {The American Astronomical Society},
volume = {941},
number = {2},
pages = {L27},
author = {Williams, Thomas G. and Sun, Jiayi and Barnes, Ashley T. and Schinnerer, Eva and Henshaw, Jonathan D. and Meidt, Sharon E. and Querejeta, Miguel and Watkins, Elizabeth J. and Bigiel, Frank and Blanc, Guillermo A. and Boquien, Médéric and Cao, Yixian and Chevance, Mélanie and Egorov, Oleg V. and Emsellem, Eric and Glover, Simon C. O. and Grasha, Kathryn and Hassani, Hamid and Jeffreson, Sarah and Jiménez-Donaire, María J. and Kim, Jaeyeon and Klessen, Ralf S. and Kreckel, Kathryn and Kruijssen, J. M. Diederik and Larson, Kirsten L. and Leroy, Adam K. and Liu, Daizhong and Pessa, Ismael and Pety, Jérôme and Pinna, Francesca and Rosolowsky, Erik and Sandstrom, Karin M. and Smith, Rowan and Sormani, Mattia C. and Stuber, Sophia and Thilker, David A. and Whitmore, Bradley C.},
title = {PHANGS-JWST First Results: Spurring on Star Formation: JWST Reveals Localized Star Formation in a Spiral Arm Spur of NGC 628},
journal = {\apjl}
}

@article{Battisti_2015,
doi = {10.1088/0004-637X/800/2/143},
url = {https://dx.doi.org/10.1088/0004-637X/800/2/143},
year = {2015},
month = {feb},
publisher = {The American Astronomical Society},
volume = {800},
number = {2},
pages = {143},
author = {Battisti, A. J. and Calzetti, D. and Johnson, B. D. and Elbaz, D.},
title = {CONTINUOUS MID-INFRARED STAR FORMATION RATE INDICATORS: DIAGNOSTICS FOR 0 &lt; z &lt; 3 STAR-FORMING GALAXIES},
journal = {\apj}
}

@article{Belfiore ,
	author = {{Belfiore, Francesco} and {Leroy, Adam K.} and {Williams, Thomas G.} and {Barnes, Ashley T.} and {Bigiel, Frank} and {Boquien, Médéric} and {Cao, Yixian} and {Chastenet, Jérémy} and {Congiu, Enrico} and {Dale, Daniel A.} and {Egorov, Oleg V.} and {Eibensteiner, Cosima} and {Emsellem, Eric} and {Glover, Simon C. O.} and {Groves, Brent} and {Hassani, Hamid} and {Klessen, Ralf S.} and {Kreckel, Kathryn} and {Neumann, Lukas} and {Neumann, Justus} and {Querejeta, Miguel} and {Rosolowsky, Erik} and {Sanchez-Blazquez, Patricia} and {Sandstrom, Karin} and {Schinnerer, Eva} and {Sun, Jiayi} and {Sutter, Jessica} and {Watkins, Elizabeth J.}},
	title = {Calibrating mid-infrared emission as a tracer of obscured star formation on H II-region scales in the era of JWST},
	DOI= "10.1051/0004-6361/202347175",
	url= "https://doi.org/10.1051/0004-6361/202347175",
	journal = {A\&A},
	year = 2023,
	volume = 678,
	pages = "A129",
}

@article{Zhang_2021,
doi = {10.3847/1538-3881/abc693},
url = {https://dx.doi.org/10.3847/1538-3881/abc693},
year = {2020},
month = {dec},
publisher = {The American Astronomical Society},
volume = {161},
number = {1},
pages = {29},
author = {Zhang, Lulu and Ho, Luis C. and Xie, Yanxia},
title = {A Method to Extract Spatially Resolved Polycyclic Aromatic Hydrocarbon Emission from Spitzer Spectra: Application to M51},
journal = {\aj}
}

@article{Amrutha_2025,
doi = {10.3847/1538-4357/add68d},
url = {https://dx.doi.org/10.3847/1538-4357/add68d},
year = {2025},
month = {jun},
publisher = {The American Astronomical Society},
volume = {987},
number = {1},
pages = {11},
author = {Amrutha, S and Das, Mousumi},
title = {A Pilot Method to Determine the High Mass End of the Stellar Initial Mass Function in Galaxies Using UVIT, Hα-MUSE Observations and Applied to NGC 628},
journal = {\apj}
}

@article{Regan_2006,
doi = {10.1086/505382},
url = {https://dx.doi.org/10.1086/505382},
year = {2006},
month = {dec},
publisher = {},
volume = {652},
number = {2},
pages = {1112},
author = {Regan, Michael W. and Thornley, Michele D. and Vogel, Stuart N. and Sheth, Kartik and Draine, Bruce T. and Hollenbach, David J. and Meyer, Martin and Dale, Daniel A. and Engelbracht, Charles W. and Kennicutt, Robert C. and Armus, Lee and Buckalew, Brent and Calzetti, Daniela and Gordon, Karl D. and Helou, George and Leitherer, Claus and Malhotra, Sangeeta and Murphy, Eric and Rieke, George H. and Rieke, Marcia J. and Smith, J. D.},
title = {The Radial Distribution of the Interstellar Medium in Disk Galaxies: Evidence for Secular Evolution},
journal = {\apj}
}

@article{Mayya2023,
    author = {Mayya, Y D and Alzate, J A and Lomelí-Núez, L and Zaragoza-Cardiel, J and Gómez-González, V M A and Silich, S and Fernández-Arenas, D and Vega, O and Ovando, P A and Rodríguez, L H and Rosa-González, D and Luna, A and Zamora-Avilés, M and Rosales-Ortega, F},
    title = {The stellar population responsible for a kiloparsec-size superbubble seen in the JWST ‘phantom’ images of NGC 628},
    journal = {\mnras},
    volume = {521},
    number = {4},
    pages = {5492-5507},
    year = {2023},
    month = {03},
    issn = {0035-8711},
    doi = {10.1093/mnras/stad636},
    url = {https://doi.org/10.1093/mnras/stad636},
    eprint = {https://academic.oup.com/mnras/article-pdf/521/4/5492/49780418/stad636.pdf},
}

@ARTICLE{2023ApJ...944L..16E,
       author = {{Egorov}, Oleg V. and {Kreckel}, Kathryn and {Sandstrom}, Karin M. and {Leroy}, Adam K. and {Glover}, Simon C.~O. and {Groves}, Brent and {Kruijssen}, J.~M. Diederik and {Barnes}, Ashley. T. and {Belfiore}, Francesco and {Bigiel}, F. and {Blanc}, Guillermo A. and {Boquien}, M{\'e}d{\'e}ric and {Cao}, Yixian and {Chastenet}, J{\'e}r{\'e}my and {Chevance}, M{\'e}lanie and {Congiu}, Enrico and {Dale}, Daniel A. and {Emsellem}, Eric and {Grasha}, Kathryn and {Klessen}, Ralf S. and {Larson}, Kirsten L. and {Liu}, Daizhong and {Murphy}, Eric J. and {Pan}, Hsi-An and {Pessa}, Ismael and {Pety}, J{\'e}r{\^o}me and {Rosolowsky}, Erik and {Scheuermann}, Fabian and {Schinnerer}, Eva and {Sutter}, Jessica and {Thilker}, David A. and {Watkins}, Elizabeth J. and {Williams}, Thomas G.},
        title = "{PHANGS-JWST First Results: Destruction of the PAH Molecules in H II Regions Probed by JWST and MUSE}",
      journal = {\apjl},
         year = 2023,
        month = feb,
       volume = {944},
       number = {2},
          eid = {L16},
        pages = {L16},
          doi = {10.3847/2041-8213/acac92},
archivePrefix = {arXiv},
       eprint = {2212.09159},
 primaryClass = {astro-ph.GA},
       adsurl = {https://ui.adsabs.harvard.edu/abs/2023ApJ...944L..16E}
}

@ARTICLE{2021ApJ...908..238M,
       author = {{McKinney}, J. and {Armus}, L. and {Pope}, A. and {D{\'\i}az-Santos}, T. and {Charmandaris}, V. and {Inami}, H. and {Song}, Y. and {Evans}, A.~S.},
        title = "{Regulating Star Formation in Nearby Dusty Galaxies: Low Photoelectric Efficiencies in the Most Compact Systems}",
      journal = {\apj},
         year = 2021,
        month = feb,
       volume = {908},
       number = {2},
          eid = {238},
        pages = {238},
          doi = {10.3847/1538-4357/abd6f2},
archivePrefix = {arXiv},
       eprint = {2101.01182},
 primaryClass = {astro-ph.GA},
       adsurl = {https://ui.adsabs.harvard.edu/abs/2021ApJ...908..238M}
}

@ARTICLE{2010A&A...510A..37M,
       author = {{Micelotta}, E.~R. and {Jones}, A.~P. and {Tielens}, A.~G.~G.~M.},
        title = "{Polycyclic aromatic hydrocarbon processing in a hot gas}",
      journal = {\aap},
         year = 2010,
        month = feb,
       volume = {510},
          eid = {A37},
        pages = {A37},
          doi = {10.1051/0004-6361/200911683},
archivePrefix = {arXiv},
       eprint = {0912.1595},
 primaryClass = {astro-ph.GA},
       adsurl = {https://ui.adsabs.harvard.edu/abs/2010A&A...510A..37M}
}

@ARTICLE{1996ApJ...469..740J,
       author = {{Jones}, A.~P. and {Tielens}, A.~G.~G.~M. and {Hollenbach}, D.~J.},
        title = "{Grain Shattering in Shocks: The Interstellar Grain Size Distribution}",
      journal = {\apj},
         year = 1996,
        month = oct,
       volume = {469},
        pages = {740},
          doi = {10.1086/177823},
       adsurl = {https://ui.adsabs.harvard.edu/abs/1996ApJ...469..740J}
}

@ARTICLE{2011A&A...527A.123G,
       author = {{Guillet}, V. and {Pineau Des For{\^e}ts}, G. and {Jones}, A.~P.},
        title = "{Shocks in dense clouds. III. Dust processing and feedback effects in C-type shocks}",
      journal = {\aap},
         year = 2011,
        month = mar,
       volume = {527},
          eid = {A123},
        pages = {A123},
          doi = {10.1051/0004-6361/201015973},
       adsurl = {https://ui.adsabs.harvard.edu/abs/2011A&A...527A.123G}
}

@article{Schinnerer_2017,
doi = {10.3847/1538-4357/836/1/62},
url = {https://dx.doi.org/10.3847/1538-4357/836/1/62},
year = {2017},
month = {feb},
publisher = {The American Astronomical Society},
volume = {836},
number = {1},
pages = {62},
author = {Schinnerer, Eva and Meidt, Sharon E. and Colombo, Dario and Chandar, Rupali and Dobbs, Clare L. and García-Burillo, Santiago and Hughes, Annie and Leroy, Adam K. and Pety, Jérôme and Querejeta, Miguel and Kramer, Carsten and Schuster, Karl F.},
title = {The PdBI Arcsecond Whirlpool Survey (PAWS): The Role of Spiral Arms in Cloud and Star Formation},
journal = {\apj}
}

@article{10.1093/mnras/stt1019,
    author = {Gusev, A. S. and Efremov, Yu. N.},
    title = {Regular chains of star formation complexes in spiral arms of NGC 628},
    journal = {\mnras},
    volume = {434},
    number = {1},
    pages = {313-324},
    year = {2013},
    month = {06},
    issn = {0035-8711},
    doi = {10.1093/mnras/stt1019},
    url = {https://doi.org/10.1093/mnras/stt1019},
    eprint = {https://academic.oup.com/mnras/article-pdf/434/1/313/18499572/stt1019.pdf},
}

@article{10.1093/mnras/stt1970,
    author = {Gusev, A. S. and Egorov, O. V. and Sakhibov, F.},
    title = {Parameters of the brightest star formation regions in the two principal spiral arms of NGC 628},
    journal = {\mnras},
    volume = {437},
    number = {2},
    pages = {1337-1351},
    year = {2013},
    month = {11},
    issn = {0035-8711},
    doi = {10.1093/mnras/stt1970},
    url = {https://doi.org/10.1093/mnras/stt1970},
    eprint = {https://academic.oup.com/mnras/article-pdf/437/2/1337/3844282/stt1970.pdf},
}

@INCOLLECTION{calzetti.book.2013,
       author = {{Calzetti}, Daniela},
        title = "{Star Formation Rate Indicators}",
    booktitle = {Secular Evolution of Galaxies},
         year = 2013,
    publisher = {Cambridge University Press},
       editor = {{Falc{\'o}n-Barroso}, Jes{\'u}s and {Knapen}, Johan H.},
        pages = {419},
          doi = {10.48550/arXiv.1208.2997},
       adsurl = {https://ui.adsabs.harvard.edu/abs/2013seg..book..419C}
}

@article{Lai_2023,
doi = {10.3847/2041-8213/ad0387},
url = {https://dx.doi.org/10.3847/2041-8213/ad0387},
year = {2023},
month = {nov},
publisher = {The American Astronomical Society},
volume = {957},
number = {2},
pages = {L26},
author = {Lai, Thomas S.-Y. and Armus, Lee and Bianchin, Marina and Díaz-Santos, Tanio and Linden, Sean T. and Privon, George C. and Inami, Hanae and U, Vivian and Bohn, Thomas and Evans, Aaron S. and Larson, Kirsten L. and Hensley, Brandon S. and Smith, J.-D. T. and Malkan, Matthew A. and Song, Yiqing and Stierwalt, Sabrina and van der Werf, Paul P. and McKinney, Jed and Aalto, Susanne and Buiten, Victorine A. and Rich, Jeff and Charmandaris, Vassilis and Appleton, Philip and Barcos-Muñoz, Loreto and Böker, Torsten and Finnerty, Luke and Kader, Justin A. and Law, David R. and Medling, Anne M. and Brown, Michael J. I. and Hayward, Christopher C. and Howell, Justin and Iwasawa, Kazushi and Kemper, Francisca and Marshall, Jason and Mazzarella, Joseph M. and Müller-Sánchez, Francisco and Murphy, Eric J. and Sanders, David and Surace, Jason},
title = {GOALS-JWST: Small Neutral Grains and Enhanced 3.3 μm PAH Emission in the Seyfert Galaxy NGC 7469},
journal = {\apjl}
}

@article{10.1093/mnras/stac3729,
    author = {Donnan, F R and García-Bernete, I and Rigopoulou, D and Pereira-Santaella, M and Alonso-Herrero, A and Roche, P F and Hernán-Caballero, A and Spoon, H W W},
    title = {The obscured nucleus and shocked environment of VV 114E revealed by JWST/MIRI spectroscopy},
    journal = {\mnras},
    volume = {519},
    number = {3},
    pages = {3691-3705},
    year = {2022},
    month = {12},
    issn = {0035-8711},
    doi = {10.1093/mnras/stac3729},
    url = {https://doi.org/10.1093/mnras/stac3729},
    eprint = {https://academic.oup.com/mnras/article-pdf/519/3/3691/48600139/stac3729.pdf},
}

@ARTICLE{2023ApJ...944L...8S,
       author = {{Sandstrom}, Karin M. and {Koch}, Eric W. and {Leroy}, Adam K. and {Rosolowsky}, Erik and {Emsellem}, Eric and {Smith}, Rowan J. and {Egorov}, Oleg V. and {Williams}, Thomas G. and {Larson}, Kirsten L. and {Lee}, Janice C. and {Schinnerer}, Eva and {Thilker}, David A. and {Barnes}, Ashley T. and {Belfiore}, Francesco and {Bigiel}, F. and {Blanc}, Guillermo A. and {Bolatto}, Alberto D. and {Boquien}, M{\'e}d{\'e}ric and {Cao}, Yixian and {Chastenet}, J{\'e}r{\'e}my and {Chevance}, M{\'e}lanie and {Chiang}, I-Da and {Dale}, Daniel A. and {Faesi}, Christopher M. and {Glover}, Simon C.~O. and {Grasha}, Kathryn and {Groves}, Brent and {Hassani}, Hamid and {Henshaw}, Jonathan D. and {Hughes}, Annie and {Kim}, Jaeyeon and {Klessen}, Ralf S. and {Kreckel}, Kathryn and {Kruijssen}, J.~M. Diederik and {Lopez}, Laura A. and {Liu}, Daizhong and {Meidt}, Sharon E. and {Murphy}, Eric J. and {Pan}, Hsi-An and {Querejeta}, Miguel and {Saito}, Toshiki and {Sardone}, Amy and {Sormani}, Mattia C. and {Sutter}, Jessica and {Usero}, Antonio and {Watkins}, Elizabeth J.},
        title = "{PHANGS-JWST First Results: Tracing the Diffuse Interstellar Medium with JWST Imaging of Polycyclic Aromatic Hydrocarbon Emission in Nearby Galaxies}",
      journal = {\apjl},
         year = 2023,
        month = feb,
       volume = {944},
       number = {2},
          eid = {L8},
        pages = {L8},
          doi = {10.3847/2041-8213/aca972},
archivePrefix = {arXiv},
       eprint = {2212.11177},
 primaryClass = {astro-ph.GA},
       adsurl = {https://ui.adsabs.harvard.edu/abs/2023ApJ...944L...8S}
}

@ARTICLE{2025ApJ...983..137R,
       author = {{Rodr{\'\i}guez}, M. Jimena and {Lee}, Janice C. and {Indebetouw}, Remy and {Whitmore}, B.~C. and {Maschmann}, Daniel and {Williams}, Thomas G. and {Chandar}, Rupali and {Barnes}, A.~T. and {Gnedin}, Oleg Y. and {Sandstrom}, Karin M. and {Rosolowsky}, Erik and {Leroy}, Adam K. and {Thilker}, David A. and {Kim}, Hwihyun and {Sun}, Jiayi and {Klessen}, Ralf S. and {Groves}, Brent and {Wofford}, Aida and {Boquien}, M{\'e}d{\'e}ric and {Dale}, Daniel A. and {{\'U}beda}, Leonardo and {Larson}, Kirsten L. and {Grasha}, Kathryn and {Johnson}, Kelsey E. and {Levy}, Rebecca C. and {Bigiel}, Frank and {Hassani}, Hamid and {Sarbadhicary}, Sumit K.},
        title = "{Tracing the Earliest Stages of Star and Cluster Formation in 19 Nearby Galaxies with PHANGS-JWST and HST: Compact 3.3 {\ensuremath{\mu}}m Polycyclic Aromatic Hydrocarbon Emitters and Their Relation to the Optical Census of Star Clusters}",
      journal = {\apj},
         year = 2025,
        month = apr,
       volume = {983},
       number = {2},
          eid = {137},
        pages = {137},
          doi = {10.3847/1538-4357/adbb69},
archivePrefix = {arXiv},
       eprint = {2412.07862},
 primaryClass = {astro-ph.GA},
       adsurl = {https://ui.adsabs.harvard.edu/abs/2025ApJ...983..137R}
}

@ARTICLE{2024ApJ...971...32P,
       author = {{Pedrini}, Alex and {Adamo}, Angela and {Calzetti}, Daniela and {Bik}, Arjan and {Gregg}, Benjamin and {Linden}, Sean T. and {Bajaj}, Varun and {Ryon}, Jenna E. and {Ali}, Ahmad A. and {Bortolini}, Giacomo and {Correnti}, Matteo and {Elmegreen}, Bruce G. and {Elmegreen}, Debra Meloy and {Gallagher}, John S. and {Grasha}, Kathryn and {Gutermuth}, Robert A. and {Johnson}, Kelsey E. and {Melinder}, Jens and {Messa}, Matteo and {{\"O}stlin}, G{\"o}ran and {Sabbi}, Elena and {Smith}, Linda J. and {Tosi}, Monica and {Faustino Vieira}, Helena},
        title = "{FEAST: Feedback in Emerging extragAlactic Star ClusTers: JWST Spots Polycyclic Aromatic Hydrocarbon Destruction in NGC 628 during the Emerging Phase of Star Formation}",
      journal = {\apj},
         year = 2024,
        month = aug,
       volume = {971},
       number = {1},
          eid = {32},
        pages = {32},
          doi = {10.3847/1538-4357/ad534d},
archivePrefix = {arXiv},
       eprint = {2406.01666},
 primaryClass = {astro-ph.GA},
       adsurl = {https://ui.adsabs.harvard.edu/abs/2024ApJ...971...32P}
}

@article{Whitcomb_2023,
doi = {10.3847/2515-5172/acc073},
url = {https://doi.org/10.3847/2515-5172/acc073},
year = {2023},
month = {mar},
publisher = {The American Astronomical Society},
volume = {7},
number = {3},
pages = {38},
author = {Whitcomb, Cory M. and Sandstrom, Karin and Smith, John-David T.},
title = {JWST-MIRI Synthetic Photometry Composition using 460 Spitzer-IRS Spectra of Nearby Galaxies},
journal = {Research Notes of the AAS}
}

@ARTICLE{2001ApJ...554..778L,
       author = {{Li}, Aigen and {Draine}, B.~T.},
        title = "{Infrared Emission from Interstellar Dust. II. The Diffuse Interstellar Medium}",
      journal = {\apj},
         year = 2001,
        month = jun,
       volume = {554},
       number = {2},
        pages = {778-802},
          doi = {10.1086/323147},
archivePrefix = {arXiv},
       eprint = {astro-ph/0011319},
 primaryClass = {astro-ph},
       adsurl = {https://ui.adsabs.harvard.edu/abs/2001ApJ...554..778L}
}

@article{Lyu_2025,
doi = {10.3847/1538-4357/add538},
url = {https://doi.org/10.3847/1538-4357/add538},
year = {2025},
month = {jun},
publisher = {The American Astronomical Society},
volume = {986},
number = {2},
pages = {156},
author = {Lyu, Jianwei and Yang, Xuejuan and Li, Aigen and Sun, Fengwu and Rieke, George H. and Alberts, Stacey and Shivaei, Irene},
title = {Unveiling the Aromatic and Aliphatic Universe at Redshifts z ∼ 0.2–0.5 with JWST NIRCam/WFSS},
journal = {The Astrophysical Journal}
}

@article{Lai2022,
   title={GOALS-JWST: Tracing AGN Feedback on the Star-forming Interstellar Medium in NGC 7469},
   volume={941},
   ISSN={2041-8213},
   url={http://dx.doi.org/10.3847/2041-8213/ac9ebf},
   DOI={10.3847/2041-8213/ac9ebf},
   number={2},
   journal={The Astrophysical Journal Letters},
   publisher={American Astronomical Society},
   author={Lai , Thomas S.-Y. and Armus, Lee and U, Vivian and Díaz-Santos, Tanio and Larson, Kirsten L. and Evans, Aaron and Malkan, Matthew A. and Appleton, Philip and Rich, Jeff and Müller-Sánchez, Francisco and Inami, Hanae and Bohn, Thomas and McKinney, Jed and Finnerty, Luke and Law, David R. and Linden, Sean T. and Medling, Anne M. and Privon, George C. and Song, Yiqing and Stierwalt, Sabrina and van der Werf, Paul P. and Barcos-Muñoz, Loreto and Smith, J. D. T. and Togi, Aditya and Aalto, Susanne and Böker, Torsten and Charmandaris, Vassilis and Howell, Justin and Iwasawa, Kazushi and Kemper, Francisca and Mazzarella, Joseph M. and Murphy, Eric J. and Brown, Michael J. I. and Hayward, Christopher C. and Marshall, Jason and Sanders, David and Surace, Jason},
   year={2022},
   month=dec, pages={L36} }

@ARTICLE{2025arXiv251007365M,
       author = {{McKinney}, Jed and {Eleazer}, Miriam and {Pope}, Alexandra and {Sajina}, Anna and {Alberts}, Stacey and {Stone}, Meredith and {Sajkov}, Leonid and {Vanicek}, Virginia and {Kirkpatrick}, Allison and {Lai}, Thomas and {Casey}, Caitlin M. and {Armus}, Lee and {Diaz-Santos}, Tanio and {Korkus}, Andrew and {Cooper}, Olivia and {House}, Lindsay R. and {Akins}, Hollis and {Lambrides}, Erini and {Long}, Arianna and {Yan}, Lin},
        title = "{A JWST MIRI LRS Survey of 37 Massive Star-Forming Galaxies and AGN at Cosmic Noon -- Overview and First Results}",
      journal = {arXiv e-prints},
         year = 2025,
        month = oct,
          eid = {arXiv:2510.07365},
        pages = {arXiv:2510.07365},
          doi = {10.48550/arXiv.2510.07365},
archivePrefix = {arXiv},
       eprint = {2510.07365},
 primaryClass = {astro-ph.GA},
       adsurl = {https://ui.adsabs.harvard.edu/abs/2025arXiv251007365M}
}

@article{Young_2023,
doi = {10.3847/2041-8213/ad07e1},
url = {https://doi.org/10.3847/2041-8213/ad07e1},
year = {2023},
month = {nov},
publisher = {The American Astronomical Society},
volume = {958},
number = {1},
pages = {L5},
author = {Young, Jason and Pope, Alexandra and Sajina, Anna and Yan, Lin and Gonçalves, Thiago S and Eleazer, Miriam and Alberts, Stacey and Armus, Lee and Bonato, Matteo and Dale, Daniel A. and Farrah, Duncan and Ferkinhoff, Carl and Hayward, Christopher C. and McKinney, Jed and Murphy, Eric J. and Nesvadba, Nicole and Ogle, Patrick and Sajkov, Leonid and Veilleux, Sylvain},
title = {Halfway to the Peak: Spatially Resolved Star Formation and Kinematics in a z = 0.54 Dusty Galaxy with JWST/MIRI},
journal = {The Astrophysical Journal Letters}
}

@ARTICLE{2024ApJ...971..115G,
       author = {{Gregg}, Benjamin and {Calzetti}, Daniela and {Adamo}, Angela and {Bajaj}, Varun and {Ryon}, Jenna E. and {Linden}, Sean T. and {Correnti}, Matteo and {Cignoni}, Michele and {Messa}, Matteo and {Sabbi}, Elena and {Gallagher}, John S. and {Grasha}, Kathryn and {Pedrini}, Alex and {Gutermuth}, Robert A. and {Melinder}, Jens and {Kotulla}, Ralf and {P{\'e}rez}, Gustavo and {Krumholz}, Mark R. and {Bik}, Arjan and {{\"O}stlin}, G{\"o}ran and {Johnson}, Kelsey E. and {Bortolini}, Giacomo and {Smith}, Linda J. and {Tosi}, Monica and {Maji}, Subhransu and {Faustino Vieira}, Helena},
        title = "{Feedback in Emerging Extragalactic Star Clusters, FEAST: The Relation between 3.3 {\ensuremath{\mu}}m Polycyclic Aromatic Hydrocarbon Emission and Star Formation Rate Traced by Ionized Gas in NGC 628}",
      journal = {\apj},
         year = 2024,
        month = aug,
       volume = {971},
       number = {1},
          eid = {115},
        pages = {115},
          doi = {10.3847/1538-4357/ad54b4},
archivePrefix = {arXiv},
       eprint = {2405.09667},
 primaryClass = {astro-ph.GA},
       adsurl = {https://ui.adsabs.harvard.edu/abs/2024ApJ...971..115G}
}

@ARTICLE{2026ApJ...997...20G,
       author = {{Gregg}, Benjamin and {Calzetti}, Daniela and {Adamo}, Angela and {Pedrini}, Alex and {Linden}, Sean T. and {Bajaj}, Varun and {Ryon}, Jenna E. and {Bik}, Arjan and {Bortolini}, Giacomo and {Correnti}, Matteo and {Draine}, Bruce T. and {Elmegreen}, Bruce G. and {Faustino Vieira}, Helena and {Gallagher}, John S. and {Grasha}, Kathryn and {Johnson}, Kelsey E. and {Lai}, Thomas S.-Y. and {Messa}, Matteo and {{\"O}stlin}, G{\"o}ran and {Smith}, Linda J. and {Tosi}, Monica},
        title = "{The Calibration of Short-wavelength Polycyclic Aromatic Hydrocarbon Emission as Star Formation Rate Indicators with JWST}",
      journal = {\apj},
         year = 2026,
        month = jan,
       volume = {997},
       number = {1},
          eid = {20},
        pages = {20},
          doi = {10.3847/1538-4357/ae1ca8},
archivePrefix = {arXiv},
       eprint = {2511.06481},
 primaryClass = {astro-ph.GA},
       adsurl = {https://ui.adsabs.harvard.edu/abs/2026ApJ...997...20G}
}

@article{Jiménez_2024,
doi = {10.3847/1538-4357/ad0cb8},
url = {https://doi.org/10.3847/1538-4357/ad0cb8},
year = {2023},
month = {dec},
publisher = {The American Astronomical Society},
volume = {960},
number = {1},
pages = {81},
author = {Jiménez, S. and Silich, S. and Mayya, Y. D. and Zaragoza-Cardiel, J.},
title = {What Holes in the Gas Distribution of Nearly Face-on Galaxies Can Tell Us about the Host Disk Parameters: The Case of the NGC 628 Southeast Superbubble},
journal = {The Astrophysical Journal}
}

@article{Whitmore_2025,
doi = {10.3847/1538-4357/adb3a2},
url = {https://doi.org/10.3847/1538-4357/adb3a2},
year = {2025},
month = {mar},
publisher = {The American Astronomical Society},
volume = {982},
number = {1},
pages = {50},
author = {Whitmore, Bradley C. and Chandar, Rupali and Lee, Janice C. and Henny, Kiana F. and Rodríguez, M. Jimena and Baron, Dalya and Bigiel, F. and Boquien, Médéric and Chevance, Mélanie and Chown, Ryan and Dale, Daniel A. and Floyd, Matthew and Grasha, Kathryn and Glover, Simon C. O. and Gnedin, Oleg and Hassani, Hamid and Indebetouw, Remy and Kapoor, Anand Utsav and Larson, Kirsten L. and Leroy, Adam K. and Maschmann, Daniel and Scheuermann, Fabian and Sutter, Jessica and Schinnerer, Eva and Sarbadhicary, Sumit K. and Thilker, David A. and Williams, Thomas G. and Wofford, Aida},
title = {Empirical SED Templates for Star Clusters Observed with HST and JWST: No Strong PAH or IR Dust Emission after 5 Myr},
journal = {The Astrophysical Journal}
}

@ARTICLE{2022MNRAS.509.4094T,
       author = {{Thilker}, David A. and {Whitmore}, Bradley C. and {Lee}, Janice C. and {Deger}, Sinan and {Chandar}, Rupali and {Larson}, Kirsten L. and {Hannon}, Stephen and {Ubeda}, Leonardo and {Dale}, Daniel A. and {Glover}, Simon C.~O. and {Grasha}, Kathryn and {Klessen}, Ralf S. and {Kruijssen}, J.~M. Diederik and {Rosolowsky}, Erik and {Schruba}, Andreas and {White}, Richard L. and {Williams}, Thomas G.},
        title = "{PHANGS-HST: new methods for star cluster identification in nearby galaxies}",
      journal = {\mnras},
         year = 2022,
        month = jan,
       volume = {509},
       number = {3},
        pages = {4094-4127},
          doi = {10.1093/mnras/stab3183},
archivePrefix = {arXiv},
       eprint = {2106.13366},
 primaryClass = {astro-ph.GA},
       adsurl = {https://ui.adsabs.harvard.edu/abs/2022MNRAS.509.4094T}
}

@article{Thilker_2025,
doi = {10.3847/1538-4365/addabb},
url = {https://doi.org/10.3847/1538-4365/addabb},
year = {2025},
month = {aug},
publisher = {The American Astronomical Society},
volume = {280},
number = {1},
pages = {1},
author = {Thilker, David A. and Lee, Janice C. and Whitmore, Bradley C. and Maschmann, Daniel and Henny, Kiana and Chandar, Rupali and Dale, Daniel A. and Deger, Sinan and Boquien, Médéric and Wofford, Aida and Úbeda, Leonardo and Razza, Alessandro and Barnes, Ashley T. and Belfiore, Francesco and Bigiel, Frank and Grasha, Kathryn and Groves, Brent and Kim, Hwihyun and Klessen, Ralf S. and Neumann, Justus and Pinna, Francesca and Rodríguez, M. Jimena and Rosolowsky, Erik and Schinnerer, Eva and Williams, Thomas G.},
title = {PHANGS-HST Catalogs for ∼100,000 Star Clusters and Compact Associations in 38 Galaxies. II. Physical Properties from Decision-tree-based Spectral Energy Distribution Fitting of NUV-U-B-V-I Photometry with Categorical Priors Set by Hα Emission, Cluster Morphology, and Other Auxiliary Information},
journal = {The Astrophysical Journal Supplement Series}
}

@ARTICLE{Rieke2015PASP..127..584R,
       author = {{Rieke}, G.~H. and {Wright}, G.~S. and {B{\"o}ker}, T. and {Bouwman}, J. and {Colina}, L. and {Glasse}, Alistair and {Gordon}, K.~D. and {Greene}, T.~P. and {G{\"u}del}, Manuel and {Henning}, Th. and {Justtanont}, K. and {Lagage}, P. -O. and {Meixner}, M.~E. and {N{\o}rgaard-Nielsen}, H. -U. and {Ray}, T.~P. and {Ressler}, M.~E. and {van Dishoeck}, E.~F. and {Waelkens}, C.},
        title = "{The Mid-Infrared Instrument for the James Webb Space Telescope, I: Introduction}",
      journal = {\pasp},
         year = 2015,
        month = jul,
       volume = {127},
       number = {953},
        pages = {584},
          doi = {10.1086/682252},
archivePrefix = {arXiv},
       eprint = {1508.02294},
 primaryClass = {astro-ph.IM},
       adsurl = {https://ui.adsabs.harvard.edu/abs/2015PASP..127..584R}
}




\appendix
\section{Some extra material}\label{sec:appendix}

\begin{figure*}
\includegraphics[scale=0.43]{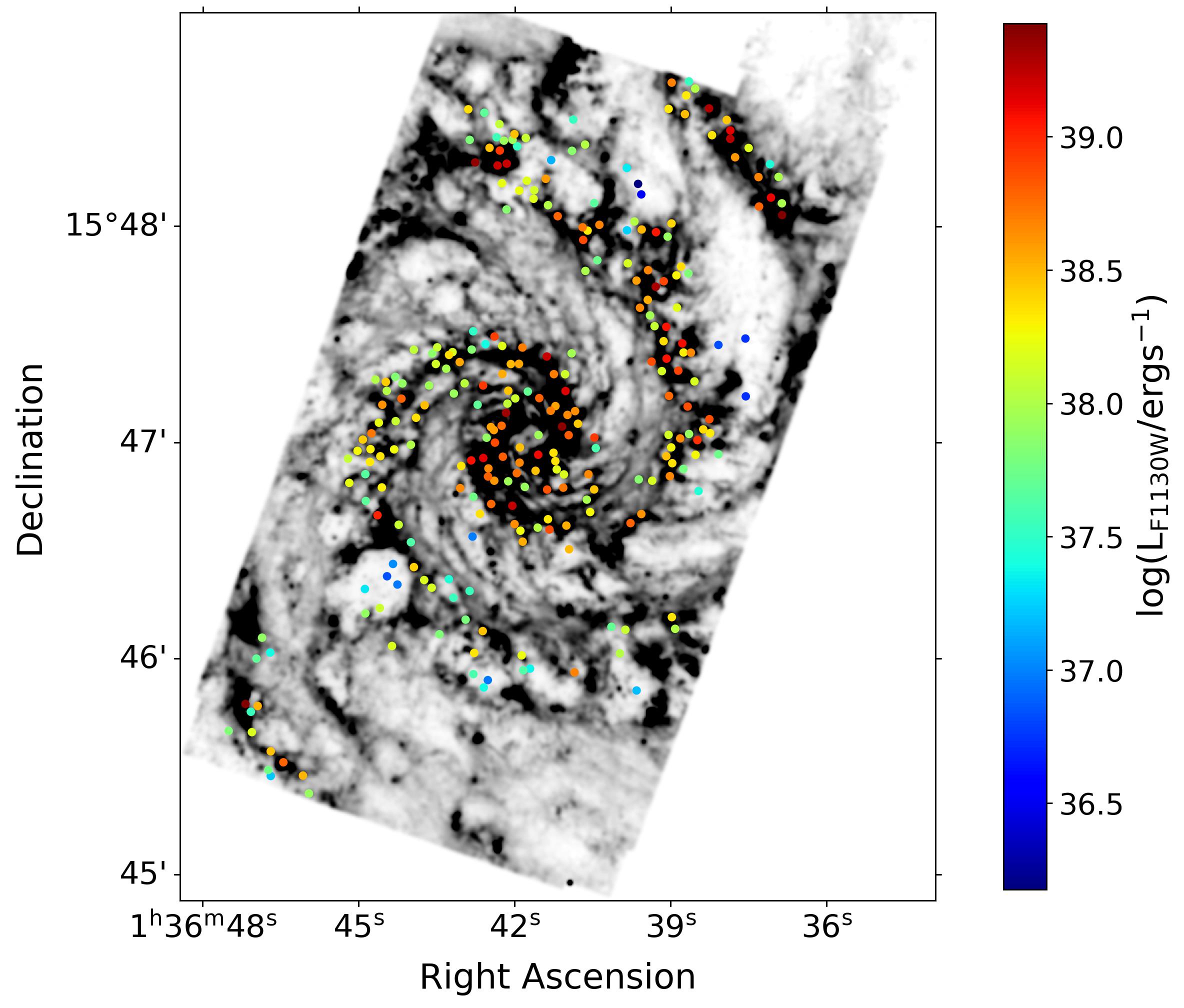}
\includegraphics[scale=0.435]{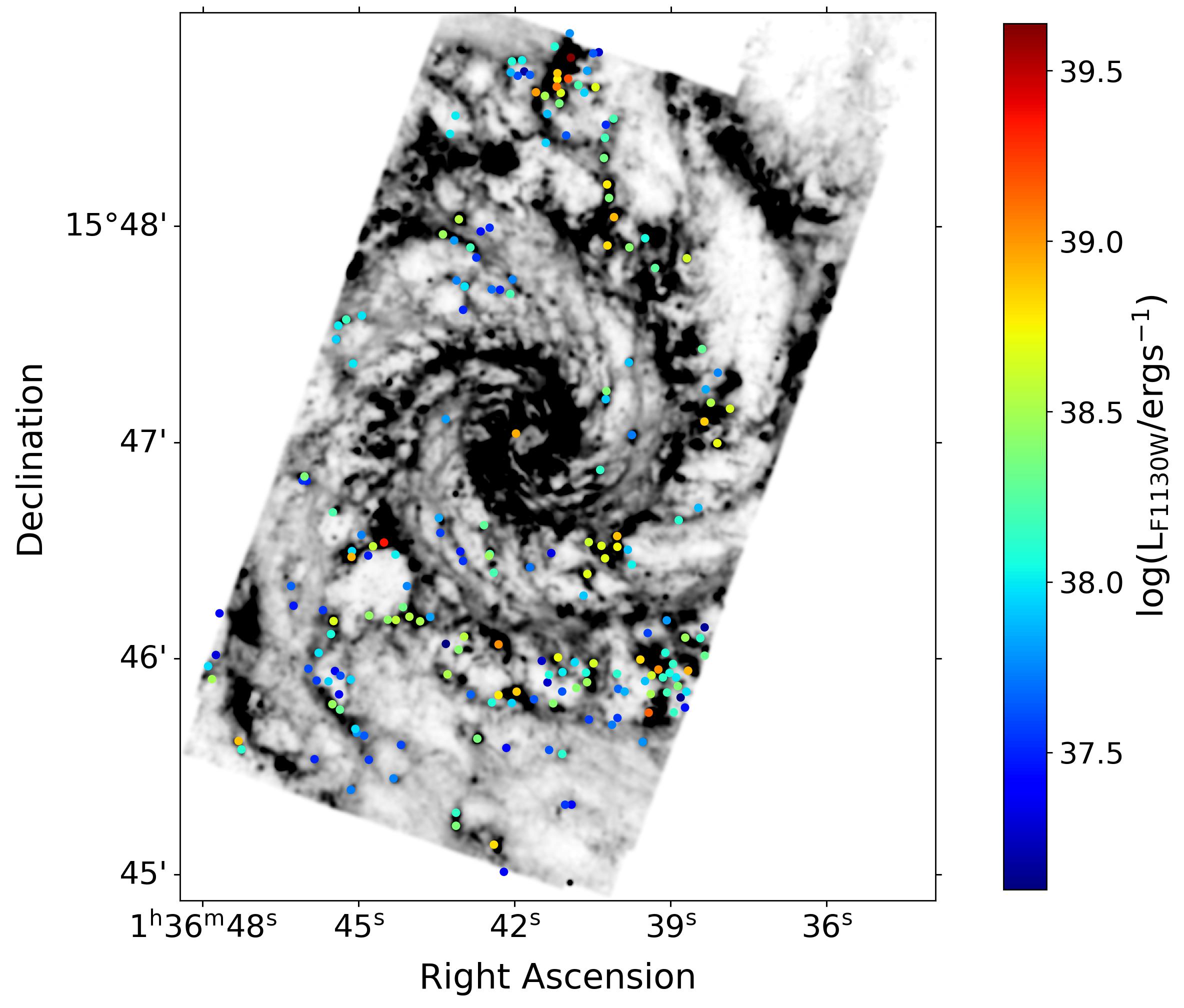}
\caption{11.3 $\mu$m PAH luminosity distribution of arm SFCs (top figure) and spur SFCs over F1130W JWST band. }
\label{fig A: scatter}
\end{figure*}

 The Figure \ref{fig A: scatter} shows 11.3 $\mu$m PAH luminosity distribution of arm SFCs (top figure) and spur SFCs over the F1130W JWST band.

\bsp	
\label{lastpage}
\end{document}